\documentclass[aps,pra,amsmath,amssymb,floatfix,twocolumn,amsmath,superscriptaddress,twocolumn,nofootinbib,tighten,letterpaper]{revtex4}
\usepackage{amsfonts, relsize, color}
\usepackage{graphics}
\usepackage{graphicx}
\usepackage{subfigure}
\usepackage{hyperref}
\usepackage{amssymb}
\usepackage{mathrsfs}
\usepackage{comment}
\usepackage{lscape}
\usepackage{multirow}
\usepackage{hhline}
\usepackage{multirow,tabularx}
\usepackage{array}
\newcolumntype{P}[1]{>{\centering\arraybackslash}m{#1}}

\usepackage[english]{babel}
\usepackage[utf8]{inputenc}
\usepackage{fancyhdr}

\newcommand{\be}{\begin{equation}}
\newcommand{\ee}{\end{equation}}
\begin{document}
\title{Global Phase Diagram of a Dirty Weyl Liquid and Emergent Superuniversality}

\author{Bitan Roy}
\affiliation{Condensed Matter Theory Center and Joint Quantum Institute, University of Maryland, College Park, Maryland 20742, USA}
\affiliation{Department of Physics and Astronomy, Rice University, Houston, Texas 77005, USA}
\affiliation{Max-Planck-Institut f$\ddot{\mbox{u}}$r Physik komplexer Systeme, N$\ddot{\mbox{o}}$thnitzer Str. 38, 01187 Dresden, Germany}

\author{Robert-Jan Slager}
\affiliation{Max-Planck-Institut f$\ddot{\mbox{u}}$r Physik komplexer Systeme, N$\ddot{\mbox{o}}$thnitzer Str. 38, 01187 Dresden, Germany}

\author{Vladimir Juri\v ci\' c}
\affiliation{Nordita, KTH Royal Institute of Technology and Stockholm University, Roslagstullsbacken 23, 10691 Stockholm, Sweden}

\date{\today}

\begin{abstract}
Pursuing complementary field-theoretic and numerical methods, we here paint the global phase diagram of a three-dimensional dirty Weyl system. The generalized Harris criterion, augmented by a perturbative renormalization-group (RG) analysis shows that weak disorder is an irrelevant perturbation at the Weyl semimetal(WSM)-insulator quantum critical point (QCP). But, a metallic phase sets in through a quantum phase transition (QPT) at strong disorder across a multicritical point (MCP). The field theoretic predictions for the correlation length exponent $\nu=2$ and dynamic scaling exponent $z=5/4$ at this MCP are in good agreement with the ones extracted numerically, yielding $\nu=1.98 \pm 0.10$ and $z=1.26 \pm 0.05$, from the scaling of the average density of states (DOS). Deep inside the WSM phase, generic disorder is also an irrelevant perturbation, while a metallic phase appears at strong disorder through a QPT. We here demonstrate that in the presence of generic, but strong disorder the WSM-metal QPT is ultimately always characterized by the exponents $\nu=1$ and $z=3/2$ (to one-loop order), originating from intra-node or chiral symmetric (e.g., regular and axial potential) disorder. We here anchor such emergent \emph{chiral superuniversality} through complementary RG calculations, controlled via $\epsilon$-expansions, and numerical analysis of average DOS across WSM-metal QPT. In addition, we also discuss a subsequent QPT (at even stronger disorder) of a Weyl metal into an Anderson insulator by numerically computing the typical DOS at zero energy. The scaling behavior of various physical observables, such as residue of quasiparticle pole, dynamic conductivity, specific heat, Gr$\ddot{\mbox{u}}$neisen ratio, inside various phases as well as across various QPTs in the global phase diagram of a dirty Weyl liquid are discussed.
\end{abstract}

\maketitle

\section{Introduction}

The complex energy landscape of electronic quantum-mechanical states in solid state compounds, commonly known as band structure, can display accidental or symmetry protected \emph{band touching} at isolated points in the Brillouin zone~\cite{herring, dornhaus, volovik, RyuTeo, tanmoy-RMP, kane-prb, balatsky, newfermions, slager2016}. In the vicinity of such diabolic points, low energy excitations can often be described as quasi-relativistic Dirac or Weyl fermions~\cite{dirac-1,dirac-2, Weyl}, which may provide an ideal platform for condensed matter realization of various peculiar phenomena, such as chiral anomaly, Casimir effect, and axionic electrodynamics \cite{burkov-review,rao-review, armitage-review}. Recently, three dimensional Weyl semimetals (WSMs) have attracted a lot of interest due to the growing evidence of their material realization~\cite{taas-1, taas-2, taas-3, nbas-1, tap-1, nbp-1, nbp-2, tas, borisenko, chiorescu}.

\begin{figure}[t!]
\includegraphics[scale=0.5]{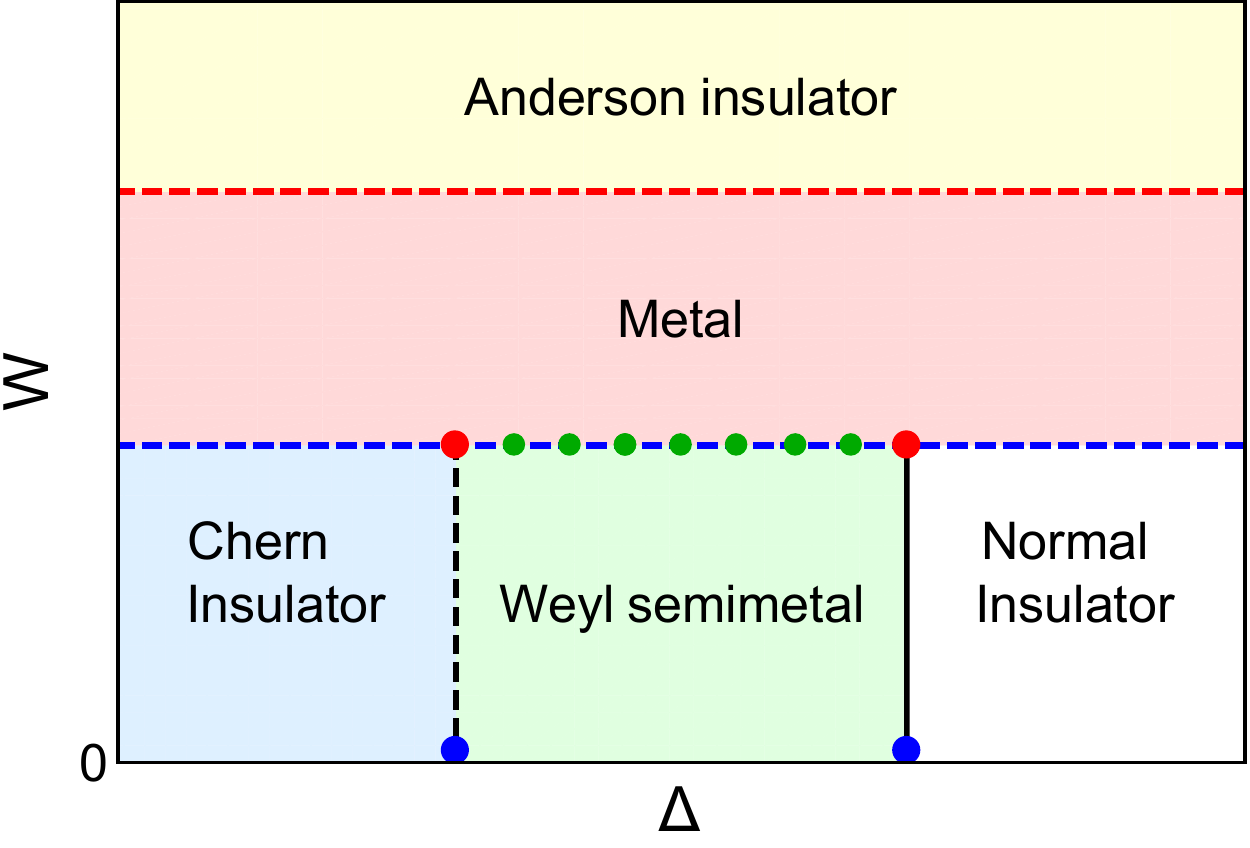}
\caption[]{A schematic phase diagram of a dirty Weyl semimetal. Here, $\Delta$ is a tuning parameter that drives quantum phase transition from Weyl semimetal to (Chern or normal) insulator in clean system [see Sec.~\ref{Weyl_Lattice}], and $W$ denotes the strength of disorder (the nature of which is not specified here). Semimetal-insulator quantum critical points are denoted by the blue dots. The red dots represent multicritical points, where an insulator, a metal and the Weyl semimetal meet [see Sec.~\ref{WSM-Ins-QPT}]. The string of green dots represents a line of quantum critical points through one of which (depending on the bare value of $\Delta$) the Weyl semimetal undergoes a quantum phase transition into a metallic phase [see Secs.~\ref{dirtyWSM_intro},~\ref{CSP_disorder},~\ref{numerics_analysis},~\ref{CSB_disorder}]. At stronger disorder the metallic phase undergoes a second quantum phase transition into the Anderson insulator phase [see Sec.~\ref{anderson}]. The shape of the phase boundaries is, however, non-universal. See, for example, Fig.~\ref{MCP_numerics_PD} for numerically obtained phase diagram from a lattice model.
}\label{Global_PD}
\end{figure}

A WSM, the prime example of a gapless topological phase of matter, is constituted by so called Weyl nodes that in the reciprocal space (Brillouin zone) act as the source and sinks of Abelian Berry curvature, and thus always appear in pairs~\cite{nielsen-ninomiya}. In a nutshell, the Abelian Berry flux enclosed by the system determines the integer topological invariant of a WSM and the degeneracy of topologically protected surface Fermi arcs. A question of fundamental and practical importance in this context concerns the stability of such gapless topological phase against impurities or disorder, inevitably present in real materials. Combining complementary field theoretic renormalization group (RG) calculations and a numerical analysis of the average density of states (DOS), we here study the role of randomness in various regimes of the phase diagram of a Weyl system to arrive at the \emph{global phase diagram}, schematically illustrated in Fig.~\ref{Global_PD}.

A WSM can be constructed by appropriately stacking two-dimensional layers of quantum anomalous Hall insulator (QAHI) in the momentum space along the $k_z$ direction, for example. Thus, by construction a WSM inherits the two dimensional integer topological invariant of constituting layers of QAHI, and the \emph{momentum space skyrmion number} of QAHI jumps by an integer amount across two Weyl nodes. As a result, the Weyl nodes serve as the sources and sinks for Abelian Berry curvature, and in a clean system WSM is sandwiched between a topological Chern and a trivial insulating phase, as shown in Fig.~\ref{Global_PD}. In an effective tight-binding model a WSM-insulator quantum phase transition (QPT), the blue dot in Fig.~\ref{Global_PD}, can be tuned by changing the effective hopping in the $k_z$ direction, as demonstrated in Sec.~\ref{Weyl_Lattice}. In this work we first assess the stability of such a clean semimetal-insulator quantum critical point (QCP) in the presence of generic randomness in the system, and arrive at the following conclusions:

1. By generalizing the Harris criterion~\cite{harris}, we find that WSM-insulator QCP is \emph{stable} against sufficiently weak, but otherwise generic disorder (see Sec.~\ref{WSM-Ins-QPT}). Such an outcome is further substantiated from the scaling analysis of disorder couplings, suggesting that any disorder is an \emph{irrelevant} perturbation at such a clean QCP.

2. From an appropriate $\epsilon$-expansion (see Sec.~\ref{WSM-Ins-QPT}), we demonstrate that a \emph{multicritical point} (MCP) emerges at stronger disorder, where the WSM, a band insulator (either Chern or trivial) and a metallic phase meet, the red dot in Fig.~\ref{Global_PD}. The critical semimetal residing at the phase boundary between a WSM and an insulator (along the black dashed line in Fig.~\ref{Global_PD}) then becomes unstable toward the formation of a compressible metal through such a MCP. The exponents capturing the instability of critical excitations toward the onset of a metal are: (a) correlation length exponent (CLE) $\nu=2$, and (b) dynamic scaling exponent (DSE) $z=5/4$ to the leading order in the $\epsilon$-expansion. These two exponents also determine the scaling behavior of physical observables across the anisotropic critical semimetal-metal QPT.

\begin{figure}[t!]
\includegraphics[width=4.2cm,height=4.0cm]{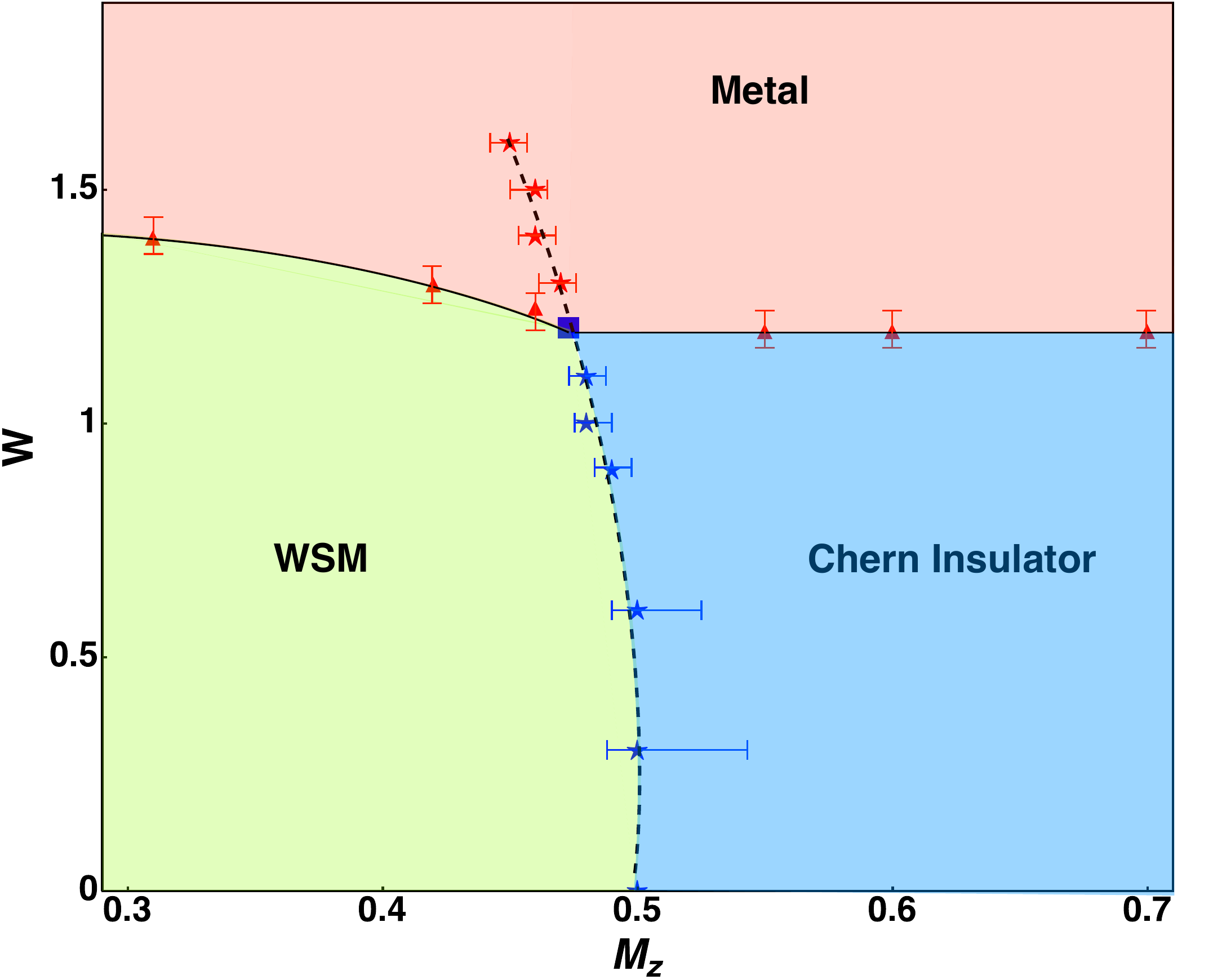}
\includegraphics[width=4.2cm,height=4.1cm]{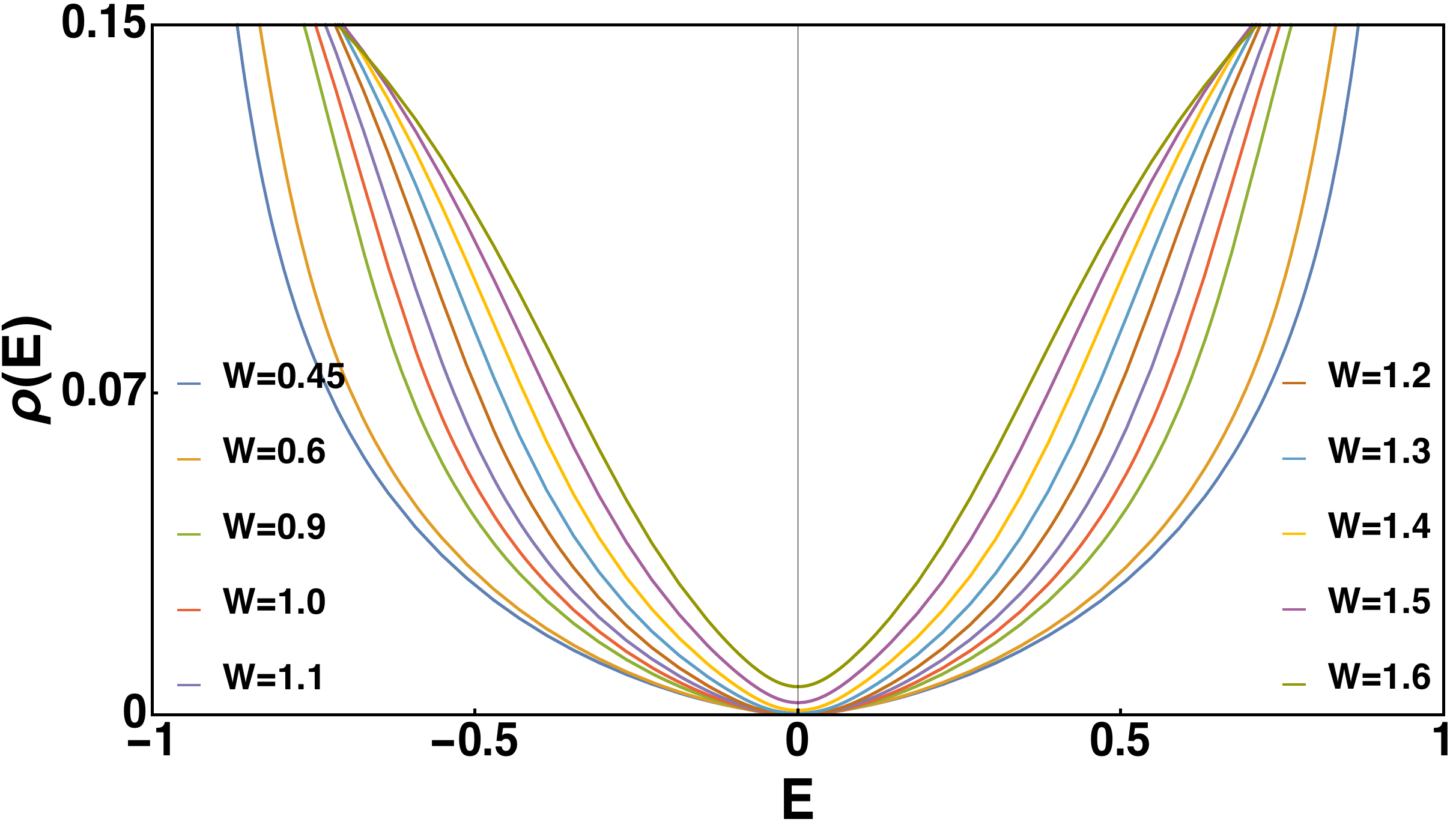}
\caption{ (Left) Numerically obtained phase diagram of a Weyl material residing in the proximity to the WSM-insulator QCP (blue dot for $W=0$) in the presence of random charge impurities ($W$). The black dashed line represents the phase boundary between these two phases, and blue square is the multi-critical point (MCP) where the WSM, a band insulator (Chern in the present situation) and a metal meet. The density of states at the phase semimetal-insulator phase boundary scales as $\varrho(E) \sim |E|^{1.5}$, as shown in the right panel [see Sec.~\ref{WSM-Ins-QPT} for details]. With increasing strength of disorder the direct transition between WSM and insulator gets avoided by an intervening metallic phase, where DOS at zero energy is finite [see the right panel]. The metallicity sets in through the MCP, where the DOS scales roughly as $\varrho(E) \sim |E|$. These findings are in qualitative agreement with the field theoretic predictions [see Fig.~\ref{WSM_Ins_Flow_PD}].
}~\label{MCP_numerics_PD}
\end{figure}

3. By following the scaling of DOS along the phase boundary (the black dashed line in Fig.~\ref{Global_PD}) between the WSM and insulator with increasing randomness in the system, we numerically extract $\nu$ and $z$ at the MCP across the critical semimetal-metal QPT [see Fig.~\ref{MCP_numerics_PD}]. Numerically extracted values of these two exponents are $\nu=1.98 \pm 0.10$ and $z=1.26 \pm 0.05$ [see Sec.~\ref{MCP_numerics_analysis}], which are in good agreement with our prediction from the leading order $\epsilon$-expansion (see Appendix~\ref{Append:data_analysis}, Table~\ref{Table:Data-analysis}).

We now turn our focus on the WSM phase (the green shaded region in Fig.~\ref{Global_PD}). The study of disorder effects in topological phases of matter has recently attracted a lot of attention, leading to a surge of analytical~\cite{fradkin, shindou, ominato, chakravarty, nomura-ryu, hosur, arovas, roy-dassarma, radzihovsky, altland, roy-dassarma-erratum, Syzranov-exponent, roy-dassarma-intdis, juricic-disorder, pallab-sudip2016, nandkishore, carpentier-1, radzihovsky-2, kim-moon, lars, shovkovy, gilbert, carpentier-2} and numerical~\cite{imura, herbut-disorder, brouwer-1, pixley-1, brouwer-2, pixley-2, ohtsuki, chen-song, roy-bera, hughes, pixley-3, roy-alavirad, pixley-4, takane2016, roy-Fermiarc} works. In particular, the focus has been concentrated on massless Dirac critical point separating two topologically distinct insulators (electrical or thermal), as well as inside a Dirac and Weyl semimetal phases. Even though the effects of generic disorders have been studied to some extent theoretically~\cite{chakravarty, roy-dassarma, roy-dassarma-intdis, juricic-disorder, pallab-sudip2016}, most of the numerical works solely focused on random charge impurities (for exception see Refs.~\cite{pixley-1, pixley-2}). By now there is both analytical and numerical evidence that chemical potential disorder when strong enough drives a QPT from the WSM to a diffusive metal, leaving its imprint on different observables, e.g., average DOS, specific heat and conductivity [see Sec.~\ref{physicalobservable}]. Deep inside the WSM phase, the system possesses various \emph{emergent} symmetries (see Table~\ref{table-disorder}), such as a \emph{continuous global chiral} $U(1)$ symmetry that is tied with the translational symmetry of a clean noninteracting WSM in the continuum limit~\cite{roy-sau}. In the absence of both inversion and time-reversal symmetries, the simplest realization of a WSM with only two Weyl nodes is susceptible to \emph{sixteen} possible sources of elastic scattering, displayed in Table~\ref{table-disorder}. They can be grouped in \emph{eight} classes, among which only four preserve the emergent global chiral symmetry (intranode scattering), while the remaining ones directly mix two Weyl nodes with opposite (left and right) chiralities (internode scattering)~\footnote{Throughout this paper, we will use chiral-symmetric and intra-node disorder synonymously. We also will use chiral-symmetry breaking and inter-node disorder synonymously. However, such classification is only germane for infinitesimal strength of randomness. At strong disorder all possible types of randomness are generated, leading to the notion of emergent superuniversality across the disorder-driven WSM-metal QPT.}. As we demonstrate in the paper, such characterization of disorders based on the chiral symmetry allows us to classify the WSM-metal QPTs (across one of the green dots shown in Fig.~\ref{Global_PD}) in the presence of generic disorder.

\begin{figure*}[t!]
\includegraphics[width=18.00cm, height=4.0cm]{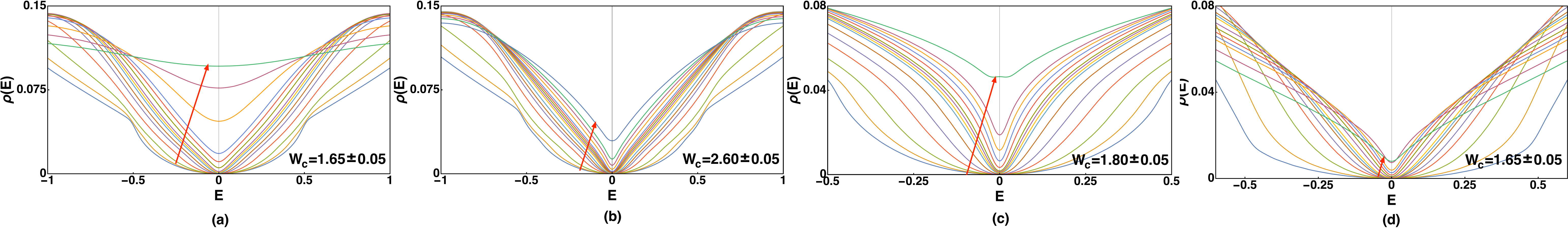}
\includegraphics[width=18.00cm, height=4.0cm]{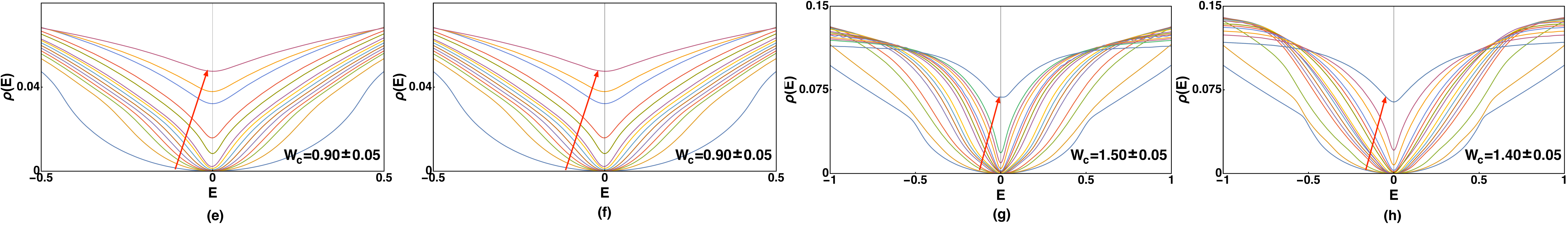}
\caption{ Scaling of numerically evaluated [using the kernel polynomial method~\cite{KPM-RMP}] average density of states (ADOS) in dirty Weyl semimetals (WSMs) in the presence of (a) potential, (b) axial, (c) axial current, (d) current, (e) spin-orbit (represented by temporal component of tensor), (f) axial magnetic (represented by spatial component of tensor), (g) scalar mass and (h) pseudo-scalar mass disorder for weak to strong disorder regime, in a cubic lattice of linear dimension $L=220$ [see Table~\ref{table-disorder} for definition and field theoretic nomenclature]. Notice that for weak enough disorder ADOS $\varrho(E) \sim |E|^2$ for $|E| \ll 1$. In the metallic phase, appearing for strong enough disorder, ADOS at zero energy $\varrho(0)$ becomes finite. Around a (non-universal) critical strength of disorder $W=W_c$ the ADOS scales as $\varrho(E) \sim |E|$ for $|E| \ll 1$. Since $\varrho(E) \sim |E|^{\frac{d}{z}-1}$, the dynamic scaling exponent $z \approx 1.5$  across the WSM-metal quantum phase transitions (QPTs), irrespective of the nature of the elastic scatterers. Disorders in panel (a)-(d) preserve the emergent global chiral symmetry and represent intranode scattering, while the remaining ones [(e)-(h)] break that symmetry and represent internode scattering. Numerically extracted critical exponents across WSM-metal QPTs and their comparison with the field theoretic predictions are displayed in Table~\ref{table-exponent} and Table~\ref{Tab:CSB_exponents}, suggesting an excellent agreement between these two methods and emergence of a \emph{superuniversality} across WSM-metal QPT. The strength of discorder increases monotonically in the direction of the \emph{red arrow} in each subfigure.
}~\label{DOS_Fan_General}
\end{figure*}

\begin{table}[t!]
\begin{tabular}{ |m{1.3cm}||P{1.45cm}|P{1.5cm}|P{1.5cm}||P{0.75cm}|P{0.75cm}|}
\hline
\multirow{2}{6em}{Disorder} &\multicolumn{3}{c||}{Numerical Analysis}&\multicolumn{2}{c|}{Field Theory}\\ \cline{2-6}
&$W_{c}$&$z$&$\nu$&$z$&$\nu$ \\ \hhline{|=#=|=|=#=|=|}
Potential & $1.65 \pm 0.05$ & $1.47 \pm 0.05$ & $1.00 \pm 0.08$ & $3/2$ & $1$ \\ \hline
Axial & $2.60 \pm 0.05$ & $1.47 \pm 0.05$ & $1.06 \pm 0.10$ & $3/2$ & $1$ \\ \hline
Magnetic & $1.80 \pm 0.05$ & $1.51 \pm 0.05$ & $1.03 \pm 0.10$ & $3/2$ & $1$ \\ \hline
Current & $1.65 \pm 0.05$ & $1.48 \pm 0.05$ & $1.02 \pm 0.09$ & $3/2$ & $1$\\ \hhline{|=#=|=|=#=|=|}
\hline
\end{tabular}
\caption{Comparison of numerically extracted values of dynamic scaling exponent ($z$) and correlation length exponent ($\nu$) across the WSM-metal QPT [takes place at $W=W_c$], with the ones obtained from the leading order $\epsilon$-expansions using field theoretic techniques. All four disorders preserve continuous global chiral symmetry of a WSM. This comparison strongly suggests that a WSM-metal transition driven by a CSP disorder is insensitive to the nature of elastic scatterers, thus motivating an emergent chiral superuniversality class of the QPTs, consult Sec.~\ref{CSP_disorder}. The fact that $z\approx 1.5$ for all types of disorder, reflects through \emph{almost} linear scaling of DOS around the WSM-metal QPT, see Fig.~\ref{DOS_Fan_General} (top panel). Here, error bars in $z$ and $\nu$ are ``\emph{fitting error bars}" (see Fig.~\ref{numeric_analysis_figure}). For detailed discussion see Appendix~\ref{Append:data_analysis} and Table~\ref{Table:Data-analysis}.
}~\label{table-exponent}
\end{table}

To motivate our theoretical analysis, we now discuss the possible microscopic origin of disorders in the Weyl materials. Furthermore, knowing this in future may facilitate a control over randomness in experiments on these materials. For example, chemical potential disorder can be controlled by modifying the concentration of random charge impurities. Random asymmetric shifts of chemical potential between the left and right chiral Weyl cones correspond to the \emph{axial potential disorder}. Therefore, in an inversion asymmetric WSM such disorder is always present. Magnetic disorder is yet another type of chiral symmetry preserving (CSP) disorder, and the strength of random magnetic scatterers can be efficiently tuned by systematically injecting magnetic ions in the system~\footnote{We here do not consider Kondo effect or Ruderman–Kittel–Kasuya–Yosida (RKKY) interaction.}. In contrast, all chiral symmetry breaking (CSB) disorders cause mixing of two Weyl nodes and in an effective model for WSMs they stem from various types of \emph{random bond} disorder that also cause random fluctuation of band-width (see Appendix~\ref{disorder_lattice}). Therefore, strength of CSB disorder may be tuned by applying \emph{inhomogeneous} pressure (hydrostatic or chemical) in the Weyl materials. Since the WSMs are found in strong spin-orbit coupled materials, a random spin-orbit coupling can be achieved when hopping (hybridization) between two orbitals with opposite parity acquires random spatial modulation. Yet another CSB but vector-like type of disorder is a random axial Zeeman coupling. Its source is the different $g$-factor of two hybridizing bands that touch at the Weyl point~\cite{model-TI, qi-anomaly, GR-field-theory}. Therefore, when magnetic impurities are injected in the system such disorder is naturally introduced, and depending on the relative strength of the $g$-factor in different bands, one can access regular (intranode) or axial (internode) random magnetic coupling. Finally, two different types of CSB \emph{mass} disorders that tend to gap out the Weyl points are represented by random charge- or spin-density-wave order, depending on the microscopic details \cite{li-roy2016}. These disorders correspond to random scalar and pseudo-scalar mass in the field theory language. Due to their presence, Weyl nodes are gapped out in each disorder configuration, but the sign of the gap is random from realization to realization, and in the thermodynamic limit the nodes remain gapless. To the best of our knowledge, it is currently unknown how to tune the strength of all individual sources of elastic scattering in real Weyl materials. \emph{Nevertheless, we elucidate how all possible disorders can be obtained from a simple effective tight-binding model on a cubic lattice for a WSM with two nodes (see Appendix~\ref{disorder_lattice}), allowing us to numerically investigate the effects of generic disorder in this system}.

\begin{table}[t!]
\begin{tabular}{|c||c|c|c|}
\hline
Disorder & $W_c$ & $z$ & $\nu$ \\
\hline \hline
spin-orbit & $0.90 \pm 0.05$ & $1.53 \pm 0.05$ & $1.01 \pm 0.10$             \\
\hline
axial magnetic & $0.90 \pm 0.05$ & $1.53 \pm 0.05$ & $0.99 \pm 0.12$        \\
\hline
Scalar mass & $1.50 \pm 0.05$ & $1.49 \pm 0.05$ & $0.99 \pm 0.12$             \\
\hline
Pseudo-scalar mass &  $1.40 \pm 0.05$ & $1.49 \pm 0.05$ & $1.01 \pm 0.11$      \\
\hline \hline
\end{tabular}
\caption{ Numerically extracted critical strength of disorder for WSM-metal QPT ($W_c$), dynamic scaling exponent ($z$) and correlation length exponent ($\nu$) in the presence of four individual disorder potentials that mix two Weyl nodes (non-chiral disorder), obtained from the scaling of average DOS. The fact that $z\approx 1.5$ for all types of disorder, reflects through \emph{almost} linear scaling of DOS around the WSM-metal QPT, see Fig.~\ref{DOS_Fan_General} (lower panel). For field theoretic analysis of internode scatterers or non-chiral disorder see Sec.~\ref{CSB_disorder}. Here, error bars in $z$ and $\nu$ are ``\emph{fitting error bars}" (see Fig.~\ref{numeric_analysis_figure_CSB}). For detailed discussion see Appendix~\ref{Append:data_analysis} and Table~\ref{Table:Data-analysis}.
}~\label{Tab:CSB_exponents}
\end{table}

Here we address the stability of a disordered WSM  $(i)$ in the field-theoretical framework by using two different renormalization-group (RG) schemes: (a) an $\epsilon_m$-expansion about a critical disorder distribution, where $\epsilon_m=1-m$, with the Gaussian white noise distribution realized as $m \to 0$, and (b) $\epsilon_d=d-2$-expansion about $d_l=2$, the lower-critical spatial dimension for WSM-metal QPT, and $(ii)$ lattice-based numerical evaluation of average DOS by using the \emph{kernel polynomial method} (KPM) \cite{KPM-RMP} in the presence of generic chiral symmetric disorder [see Fig.~\ref{DOS_Fan_General} (upper panel)] as well as non-chiral disorder [see Fig.~\ref{DOS_Fan_General} (lower panel)]. Comparisons between the field theoretic predictions and numerical findings for all chiral disorders are given in Table~\ref{table-exponent}. Our central results can be summarized as follows.

1. From the scaling analysis we show in Sec.~\ref{dirtyWSM_intro} that all types of disorder (both CSP and CSB) are irrelevant perturbations in a WSM. This outcome is also supported numerically, see Fig.~\ref{DOS_Fan_General}, depicting that DOS scales as $\varrho(E) \sim |E|^2$ for small energy ($E$), when generic disorder is sufficiently weak.

2. We show in Sec.~\ref{CSP_disorder} that irrespectively of the details of two distinct $\epsilon$-expansions, in the presence of a CSP disorder, the WSM-metal QPT takes place through either a QCP (when either potential or axial potential disorder is present) or a line of QCPs (when both types of scalar disorder are present simultaneously), characterized by critical exponents
\begin{align}~\label{exponent_intro}
z=1+\frac{\epsilon}{2} + {\mathcal O} (\epsilon^2), \: \: \nu^{-1}=\epsilon + {\mathcal O} (\epsilon^2),
\end{align}
obtained from the leading order in $\epsilon$-expansions, where $\epsilon=\epsilon_m$ or $\epsilon_d$, and $\epsilon=1$ corresponds to the physical situation. Therefore, irrespective of the nature of elastic scatterers, the universality class of the WSM-metal QPT in the presence of a CSP disorder is unique, and we name such universality class \emph{chiral superuniversality}. Even though the exponent $\nu$ and $z$ can receive higher order corrections ${\mathcal O} (\epsilon^2)$, presently there is no controlled way to compute them beyond leading order in $\epsilon$~\cite{roy-dassarma-erratum, carpentier-1}.

3. In Sec.\ref{numerics_analysis} we carry out a thorough numerical analysis of DOS in the presence of all four CSP disorders,  obtained by using KPM from a lattice model [see Fig.~\ref{DOS_Fan_General} (a)-(d)]. Within the numerical accuracy we find that $z \approx 1.5$ and $\nu \approx 1$ across possible CSP disorder driven WSM-metal QPTs (see Fig.~\ref{numeric_analysis_figure} and Table~\ref{table-exponent}). Thus numerically extracted values of critical exponents are in excellent agreement with the field theoretic predictions from leading order $\epsilon$-expansions, and strongly support the proposed scenario of emergent chiral superuniversality.

4. In Sec.~\ref{CSB_disorder} we show that the CSB disorder can also drive a WSM-metal QPT through either an isolated QCP or a line of QCPs. Irrespective of the actual details of an $\epsilon$-expansion scheme, the values of the critical exponents at such QCP or line of QCPs are in a stark contrast to the ones reported in Eq.~(\ref{exponent_intro}), and typically $z>d$. In particular, the DSE varies continuously across the line of QCPs supported by a strong CSB disorder. On the other hand, $\nu^{-1}=\epsilon$ to the leading order in an $\epsilon$ expansion, irrespective of the RG scheme.

5. Since $z>d$ (always), the CSP disorder as well as the \emph{higher gradient terms} (inevitably present in a lattice model) become \emph{relevant} at the CSB disorder driven QCPs separating a WSM from a metallic phase. Consequently, in lattice-based simulations the WSM-metal QPT is expected to ultimately be controlled by the QCPs associated with CSP disorder. We anchor this outcome by numerically computing the DOS in the presence of all four internode scattering [see Fig.~\ref{DOS_Fan_General} (lower panel)] and find that across WSM-metal QPTs, driven by any CSB disorder $z \approx 1.5$ and $\nu \approx 1$ [see Table~\ref{Tab:CSB_exponents}]. \emph{Therefore, generic disorder driven WSM-metal QPT offers a rather sparse example of superuniversality, characterized by the critical exponents $z=3/2$ and $\nu=1$, to the leading order in $\epsilon$-expansions, which are in a reasonable good agreement with numerical findings (within error bars), see Eq.~(\ref{exponent_intro})}.

6. In Sec.~\ref{physicalobservable}, we show that various experimentally measurable quantities, such as average DOS, dynamic conductivity, specific heat and Gr\"uneisen ratio, exhibit distinct scaling behavior in terms of CLE and DSE in different phases of a dirty WSM. As such, they may be useful to distinguish types of disorder in a WSM. Most importantly, distinct scaling of observables can allow to pin the onset of various phases in real materials.

We point out that the notion of superuniversality is realized rather sparsely in condensed matter systems. Most prominent examples in this regard include the quantum Hall plateau transitions~\cite{KLZ1992,Lutken-Ross1993,Fradkin-Kivelson1996} and one-dimensional disordered superconducting wires~\cite{GRV2005}. Therefore, dirty Weyl semimetal represents, to the best of our knowledge, the only example of a three-dimensional system exhibiting superuniversality.

It is worth mentioning that for sufficiently strong disorder the metallic phase in a Weyl system undergoes a second continuous QPT into an Anderson insulating phase~\cite{fradkin, pixley-1, abrahams}, across the red dashed line shown in Fig.~\ref{Global_PD}. In Sec.~\ref{anderson}, we address the metal-insulator Anderson transition (AT), but only in the presence of random charge impurities. Our central achievements regarding the fate of the AT in strongly disordered Weyl metal are the followings:

1. We show that a Weyl metal undergoes a second transition at stronger disorder into an Anderson insulator (AI) phase. By numerically computing the \emph{typical density of states} (TDOS) at zero energy [$\varrho_{t}(0)$] we show that $\varrho_{t}(0)$ vanishes smoothly across the Weyl metal-AI QPT, while displaying critical and single-paramter scaling. In particular, $\varrho_{t}(0)$ is pinned at zero in the WSM and AI phases, while it is finite inside the entire metallic phase. By contrast, the average DOS at zero energy [$\varrho(0)$] remains finite in the metallic as well as AI phases, while being zero only in the weakly disordered WSM. Otherwise, $\varrho(0)$ decreases smoothly and monotonically across the Weyl metal-AI QCP.

2. We demonstrate that TDOS at zero energy displays single-parameter scaling across both (a) WSM-metal and (b) metal-AI QPTs. Specifically the order-parameter exponent for $\varrho_{t}(0)$, $\beta_t$, defined as $\varrho_{t}(0) \sim |\delta|^{\beta_t}$, where $\delta$ defines the reduced distance from transition point, is $\beta_t= 1.80 \pm 0.20$ across the WSM-metal QPT (which is different from the one for the average DOS at zero energy for which $\beta_a=1.50 \pm 0.05$).

3. We show that inside the metallic phase the mobility edge, separating the localized states from the extended ones reside at finite energy. With increasing strength of disorder the mobility edge slides down to smaller energy and across the AT the entire energy widow is occupied by localized states.

The rest of the paper is organized as follows. In Sec.~\ref{Weyl_Lattice}, we introduce a simple tight-binding model for a Weyl system and discuss possible phases and the phase transitions in the clean limit. In Sec.~\ref{WSM-Ins-QPT}, we demonstrate the effects of generic disorder near the clean WSM-insulator QCP, and perturbatively address the effects of strong disorder. In Sec.~\ref{dirtyWSM_intro} we set up the theoretical framework for addressing the role of randomness deep inside the WSM phase, and introduce the notion of $\epsilon_m$ and $\epsilon_d$ expansions for perturbative treatment of disorder. This section is rather technical and readers familiar with the formalism or interested in physical outcomes may wish to skip it. We devote Sec.~\ref{CSP_disorder} to the effects of CSP disorder and promote the notion of \emph{chiral superuniversality}. Detailed numerical analysis of the scaling of DOS is presented in Sec.~\ref{numerics_analysis}. Effects of CSB disorder are discussed in Sec~\ref{CSB_disorder} and scaling of various physical observables, such as DOS, specific heat, conductivity, etc., across the WSM-metal QPT is discussed in Sec.~\ref{physicalobservable}. We discuss the Anderson transition of the metallic phase at stronger disorder in Sec.~\ref{anderson}. Concluding remarks and a summary of our main findings are presented in Sec.~\ref{conclusions}. Some additional technical details have been relegated to the Appendices.


\section{Lattice model for Weyl system}~\label{Weyl_Lattice}

\begin{figure}
\includegraphics[scale=0.6]{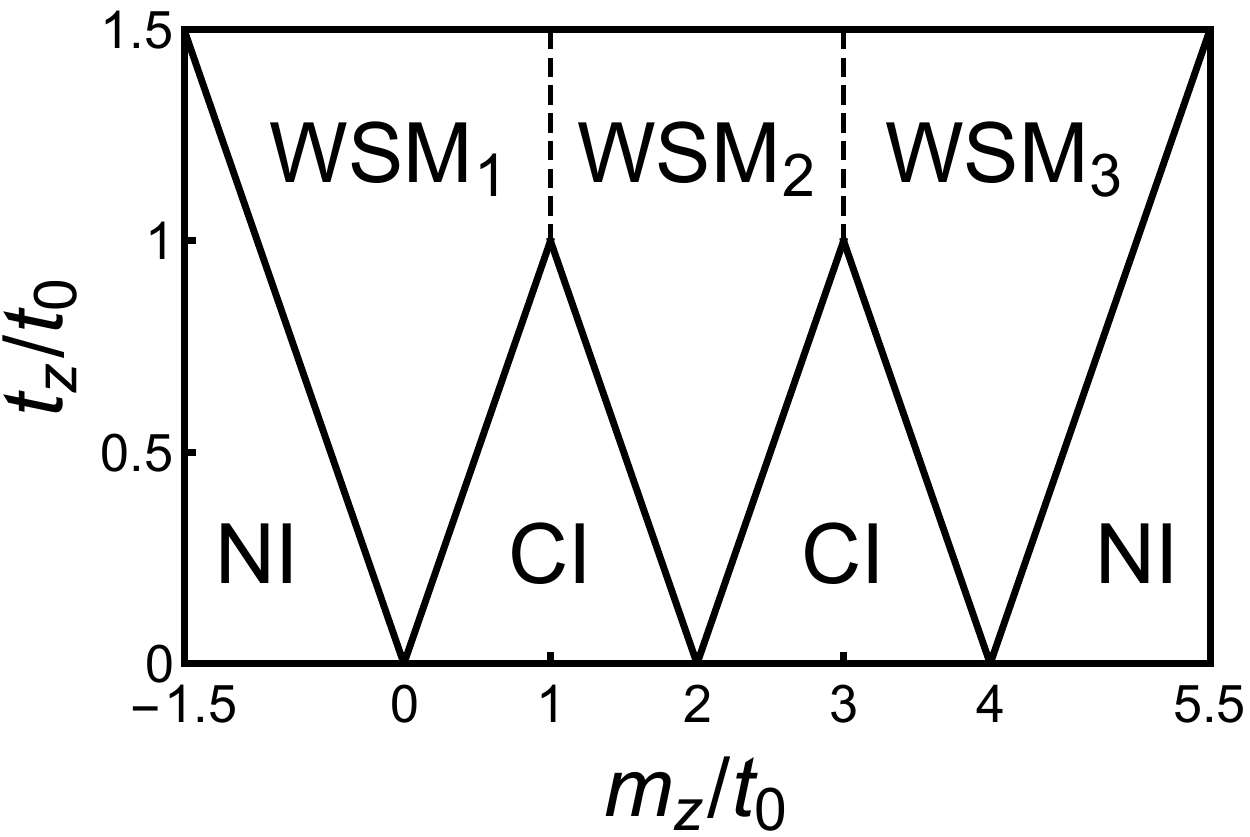}
\caption[]{The phase diagram of the clean noninteracting tight-binding model defined through Eqs.~(\ref{TB_Weyl}) and (\ref{wilson}). Here, NI and CI respectively represents trivial (normal) and Chern insulators. Weyl nodes in the WSM phase are always located along the $k_z$ direction. Respectively WSM$_{1,2,3}$ supports one, two and one pair of Weyl nodes. The projection of the Weyl nodes on the $xy$ plane in these phases are at the $(0,0)$ point, $(0,\pi)$ and $(\pi,0)$ points, and $(\pi, \pi)$ point. This model therefore supports \emph{translationally active} topological phases~~\cite{slager2013, slager2016}. The transitions between the WSM and insulating phases (solid lines) and the ones between two distinct WSM phases (dashed lines) are continuous. We emphasize that there is no symmetry distinction among these phases. }\label{TB_PD}
\end{figure}

Let us begin the discussion with a lattice realization of chiral Weyl fermions in a three-dimensional cubic lattice. Even though in most of the commonly known Weyl materials, such as the binary alloys TaAs and  NbP, Weyl fermions emerge from complex band structures in noncentrosymmetric lattices, their salient features can be captured from a simple tight-binding model
\begin{align}~\label{two-band-WSM}
H=\sum_{\boldsymbol k} \psi^\dagger_{\boldsymbol k} \left[ {\boldsymbol N}(\boldsymbol k) \cdot {\boldsymbol \sigma} \right] \psi_{\boldsymbol k}.
\end{align}
The two-component spinor is defined as $\psi^\top_{\boldsymbol k}=\left( c_{\boldsymbol k, \uparrow}, c_{\boldsymbol k, \downarrow} \right)$, where $c_{\boldsymbol k, s}$ is the fermionic annihilation operator with momentum $\boldsymbol k$ and spin/pseudospin projection $s=\uparrow, \downarrow$, and ${\boldsymbol \sigma}$s are standard Pauli matrices. We here choose
\begin{align}\label{wilson}
N_{3}({\boldsymbol k})= t_z \cos (k_z a)- m_z + t_0 \left[ 2-\cos (k_x a)-\cos (k_y a) \right],
\end{align}
where $a$ is the lattice spacing. The first term gives rise to two isolated Weyl nodes along the  $k_z$ axis at $k_z=\pm k^{0}_z$, where
\begin{align}\label{node_solution}
\cos (k^0_z a)= \frac{t_0}{t_z}\left[ \frac{m_z}{t_0} + \cos (k_x a)+\cos (k_y a)-2 \right],
\end{align}
with the following choice of pseudospin vectors
\begin{align}\label{TB_Weyl}
N_1({\boldsymbol k})= t \sin (k_x a), \: N_2({\boldsymbol k})= t \sin (k_y a).
\end{align}
The second term in Eq.~(\ref{wilson}), namely $N^M_3({\boldsymbol k})=t_0 \left[ 2-\cos (k_x a)-\cos (k_y a) \right]$, plays the role of a momentum dependent \emph{Wilson mass}~\cite{chen-song, roy-bera}. The resulting phase diagram of the above tight-binding model is displayed in Fig.~\ref{TB_PD}. We subscribe to this tight-binding model in Secs.~\ref{MCP_numerics_analysis}, ~\ref{numerics_analysis} and ~\ref{anderson} to numerically study the effects of randomness in various regimes of a dirty Weyl system.

For the sake of simplicity, we hereafter only consider the parameter regime $-t_0<m_z<t_0$ and $t_z \leq t_0$, so that only a single pair of Weyl fermions is realized at ${\bf k}^0=(0,0,\pm \cos^{-1}|m_z/t_z|)$. In the vicinity of these two points the Weyl quasiparticles can be identified as left and right chiral fermions, respectively. A WSM can be found when $|m_z/t|\leq 1$ and the system becomes an insulator for $|m_z/t|> 1$.  Even though we here restrict our analysis within the aforementioned parameter regime, this analysis can be generalized to study the semimetal-insulator QPTs in various other regimes shown in Fig.~\ref{TB_PD}.

Within this parameter regime, to capture the Weyl semimetal-insulator QPT which occurs along the line $t_z/m_z=1$, we expand the tight-binding model around the $\Gamma=(0,0,0)$ point of the Brillouin zone to arrive at the effective low energy Hamiltonian
\begin{align}~\label{Ins_WSM_QCP}
\hat{H}_{Q}(\Delta)= v \left( \sigma_1 k_x + \sigma_2 k_y \right) + \sigma_3 \left( b k^2_z -\Delta \right),
\end{align}
where $v=t a$ is the Fermi velocity in the $xy$ plane and  $b=t_z a^2/2$ bears the dimension of inverse mass. For $\Delta=t_z-m_z<0$ the system becomes an insulator (Chern or trivial). On the other hand, when $\Delta>0$, the lattice model describes a WSM. The QPT in this clean model between these two phases takes place at $\Delta=0$. Hence, $\Delta$ plays the role of a tuning parameter across the WSM-insulator QPT. The QCP separating these two phases is described by an \emph{anisotropic} semimetal, captured by the Hamiltonian $H_Q(0)$ in Eq.~(\ref{Ins_WSM_QCP}), that in turn also determines the universality class of the transition. Notice that the expansion of the lattice Hamiltonian [see Eq.~(\ref{TB_Weyl})] also yields terms $\sim k_x^2$ and $\sim k_y^2$ and higher order (from the Wilson mass), which are, however, irrelevant in the RG sense, and therefore do not affect the critical theory for the WSM-insulator QPT. Hence, we omit these higher gradient terms for now. We will discuss the paramount importance of such higher gradient terms close to the CSB disorder driven WSM-metal QPT in Sec.~\ref{CSB_disorder}. Next we address the stability of this quantum critical semimetal against disorder in the system using scaling theory and RG analysis.

\section{Effects of disorder on semimetal-insulator transition}~\label{WSM-Ins-QPT}

\begin{figure}
\includegraphics[width=8cm, height=2.0cm]{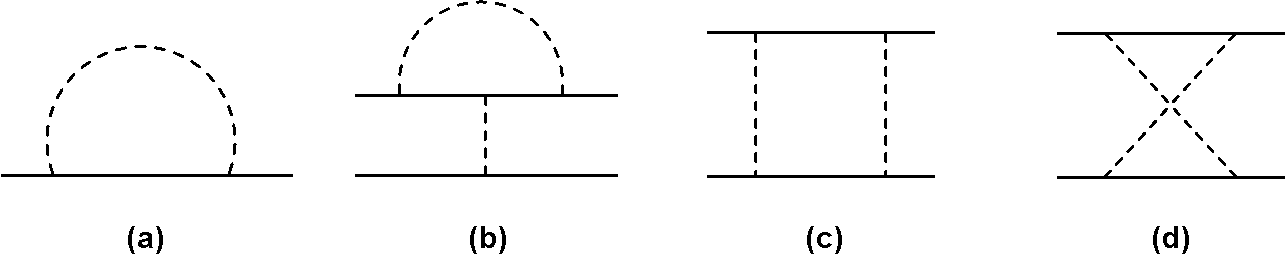}
\caption{ One-loop diagrams contributing to the self-energy correction [(a)], and renormalization of disorder coupling [(b)-(d)]. Notice that contributions from (c) ladder and (d) crossing diagram are ultraviolet divergent only in $\epsilon_n$ (Sec.~\ref{WSM-ins-RG}) and $\epsilon_d$ (Sec.~\ref{epsilond-formalism}) expansions, while they produce ultraviolet finite contribution in $\epsilon_m$ expansion (Sec.~\ref{epsilonm-formalism}). Here, solid (dashed) lines represent fermionic (disorder) field. }~\label{laddercrossing}
\end{figure}

The imaginary time ($\tau$) action associated with the low energy Hamiltonian [see Eq.~(\ref{Ins_WSM_QCP})] reads as
\begin{align}\label{eq:noninteracting-WSM-I}
S_0=\int d\tau d^2x_\perp dx_3 \:  \psi^\dagger \big[ \partial_\tau -i v \partial_j \sigma_j - \sigma_3 \left( b \partial^2_3 +\Delta \right) \big] \psi.
\end{align}
In the proximity to the Weyl semimetal-insulator QPT, the system can be susceptible to both random charge and random magnetic impurities, and their effect can be captured by the Euclidean action
\begin{align}~\label{generaldis_WSM_Ins}
S_D = \int d\tau d^2x_\perp dx_3 \: \psi^\dagger &\big[ V_0(\boldsymbol x) \sigma_0 + V_\perp(\boldsymbol x) \left( \sigma_1+\sigma_2 \right) \nonumber \\
&+ V_z(\boldsymbol x) \sigma_3 \big] \psi,
\end{align}
where $V_j(\boldsymbol x)$ are random variables. The effect of random charge impurities is captured by $V_0(\boldsymbol x)$, while $V_\perp(\boldsymbol x)$ and $V_z(\boldsymbol x)$ represents random magnetic impurities with the magnetic moment residing in the easy or $xy$ plane and in the $z$ direction (denoted here by $x_3$ for notational clarity), respectively, which we allow due to the anisotropy of the Hamiltonian [see Eq.~(\ref{Ins_WSM_QCP})]. All types of disorder are assumed to be characterized by Gaussian white noise distributions.

The scale invariance of the noninteracting action [see Eq.~(\ref{eq:noninteracting-WSM-I})] mandates the following scaling ansatz: $\tau \to e^l \tau$, $(x,y) \to e^l (x,y)$ and $x_3 \to e^{l/2} x_3$, followed by the rescaling of the field operator $\psi \to e^{-5 l/4} \psi$, where $l$ is the logarithm of running RG scale. The scaling dimension of the tuning parameter $\Delta$ is then given by $[\Delta]=1$, implying that $\Delta$ is a relevant perturbation at the WSM-insulator QCP, located at $\Delta=0$. The scaling dimension of the tuning parameter $\Delta$ plays the role of the \emph{correlation length exponent} ($\nu$) at this QCP, implying $\nu=1$. In the presence of disorder, as we show in Appendix~\ref{harris-generalized}, the Harris stability criterion~\cite{harris} can be  generalized for the WSM-insulator QCP with the quantum-critical theory of the form given by Eq.~(\ref{Ins_WSM_QCP}), but in a system with the topological or monopole charge $c$ [see Eq.~(\ref{general_WSM_INS_QCP})]. The generalized Harris criterion then suggests that WSM-insulator QCP in clean system remains stable against sufficiently weak disorder only if
\begin{equation}~\label{effective-dim}
\nu>\frac{2}{d_*}, \quad \mbox{with} \: \:\: \frac{2}{d_*}=\frac{4c}{(4+c)},
\end{equation}
and $d_*$ as the \emph{effective spatial dimensionality} of the system under the coarse graining procedure. At the WSM-insulator QCP $\nu=1$, and the critical excitations residing at $\Delta=0$ are therefore stable against weak disorder when $c=1$ [regular WSM, see Eq.~(\ref{Ins_WSM_QCP})]. We next analyze the effects of disorder on the WSM-insulator QCP using a RG approach. The same outcome can be arrived at from the computation of inverse scattering life-time ($1/\tau$) within the framework of self-consistent Born approximation [see Appendix~\ref{appendix-Born}].

\subsection{Perturbative RG analysis}~\label{WSM-ins-RG}

After performing the disorder averaging in the action [see Eq.~(\ref{generaldis_WSM_Ins})] within the replica formalism, we arrive at the replicated Euclidean action
\allowdisplaybreaks[4]
\begin{widetext}
\begin{align}~\label{action_replica}
\bar{S} &=\int d\tau d^2x_\perp dx_3 \;  \psi^\dagger_a \big[ \partial_\tau -i v\left( \partial_x \sigma_1 +\partial_y \sigma_2 \right)
+ \sigma_3 \left[ (-i)^n b_n \partial^n_3 -\Delta \right] \big] \psi_a - \int d\tau d\tau^\prime d^2x_\perp dx_3 \bigg[ \frac{\Delta_0}{2} \left( \psi^\dagger_a \psi_a \right)_{({\boldsymbol x}, \tau)}
\nonumber \\
&\times \left( \psi^\dagger_b \psi_b \right)_{({\boldsymbol x}, \tau^\prime)} + \frac{\Delta_\perp}{2} \sum_{j=1,2} \left( \psi^\dagger_a \sigma_j \psi_a \right)_{({\boldsymbol x}, \tau)} \left( \psi^\dagger_b \sigma_j \psi_b \right)_{({\boldsymbol x}, \tau^\prime)} + \frac{\Delta_z}{2} \left( \psi^\dagger_a \sigma_3 \psi_a \right)_{({\boldsymbol x}, \tau)} \left( \psi^\dagger_b \sigma_3 \psi_b \right)_{({\boldsymbol x}, \tau^\prime)} \bigg],
\end{align}
\end{widetext}
where $a, b$ are replica indices. Notice that here we have replaced $k^2_3 \to k^n_3$, with $n$ as an \emph{even integer} so that such deformation of spectrum does not change the symmetry of the system. We we will show that such deformation of the quasiparticle spectrum allows us to control the perturbative RG calculation in terms of disorder coupling. The above imaginary-time action ($\bar{S}$) remains invariant under the  space-time scaling $(x,y) \to e^{l} (x,y)$, $x_3 \to e^{l/n} x_3$ and $\tau \to e^{z l} \tau$. At the bare level the scale invariance of the free part of the action requires the field renormalization factor $Z_\psi=e^{-(2+1/n)l}$ and $\psi \to Z^{-1/2}_\psi \psi$. From this scaling analysis we immediately find that the scaling dimension of disorder couplings is $[\Delta_j]=-1/n$, for $j=0, \perp,z$. Therefore, at the WSM-insulator QCP, characterized by $n=2$, disorder is an irrelevant perturbation, in accordance with the prediction from the \emph{generalized Harris criterion}, implying the stability of this QCP against sufficiently weak randomness. Note that disorder couplings are  \emph{marginal} in a \emph{hypothetical} limit $n \to \infty$, for which the system effectively becomes a \emph{two-dimensional Weyl semimetal}. Therefore, perturbative analysis in the presence of generic disorder is controlled via an $\epsilon_n$-expansion, where $\epsilon_n=1/n$, about $n \to \infty$, following the spirit of $\epsilon$-expansions about upper or lower critical dimension~\cite{zinn-justin} and infinite monopole charge~\cite{roy-goswami-juricic, roy-foster}.

\begin{figure}
\subfigure[]{
\includegraphics[width=4cm, height=4cm]{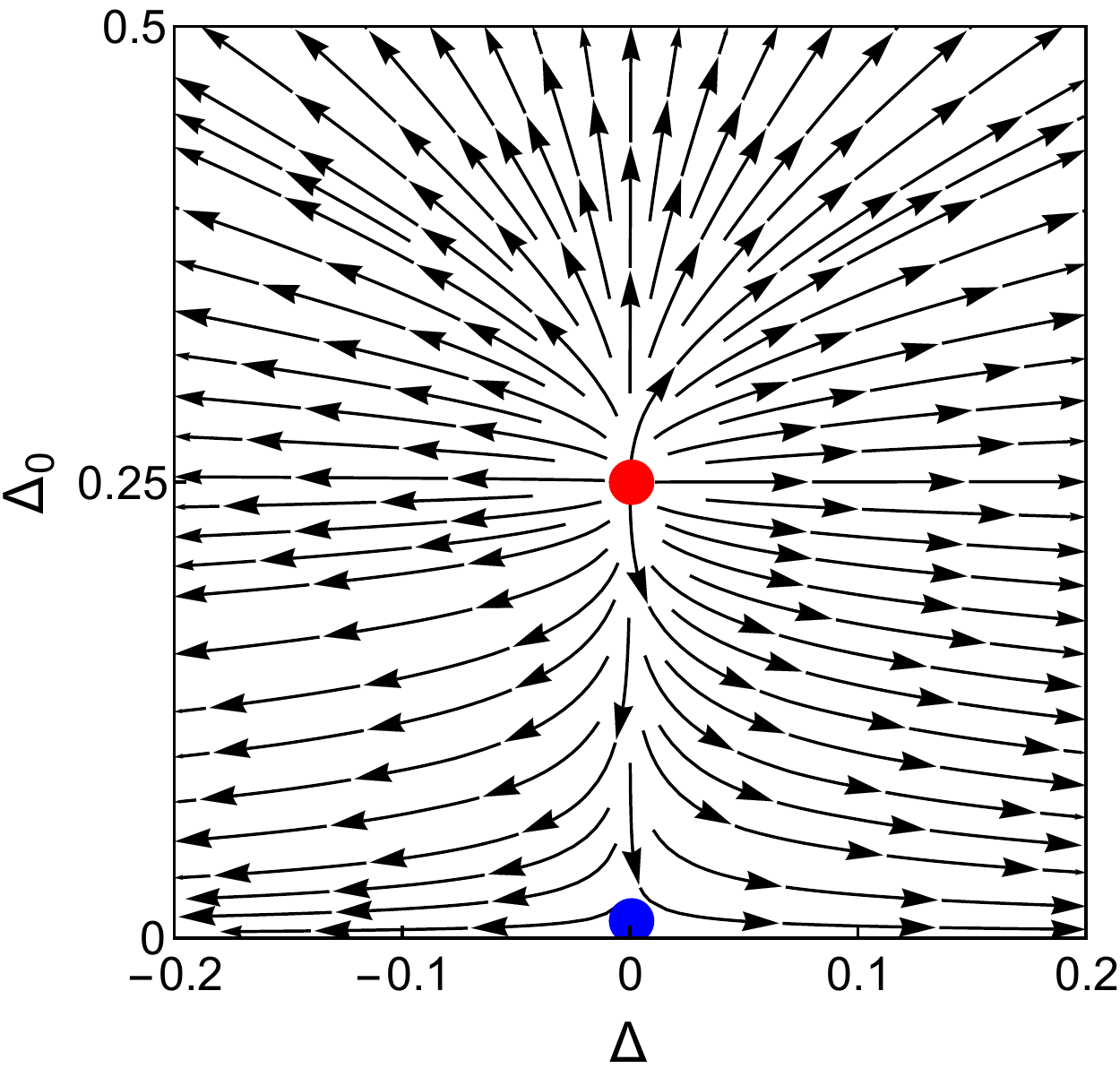}
\label{WSM_Ins_Flow}
}
\subfigure[]{
\includegraphics[width=4cm, height=4cm]{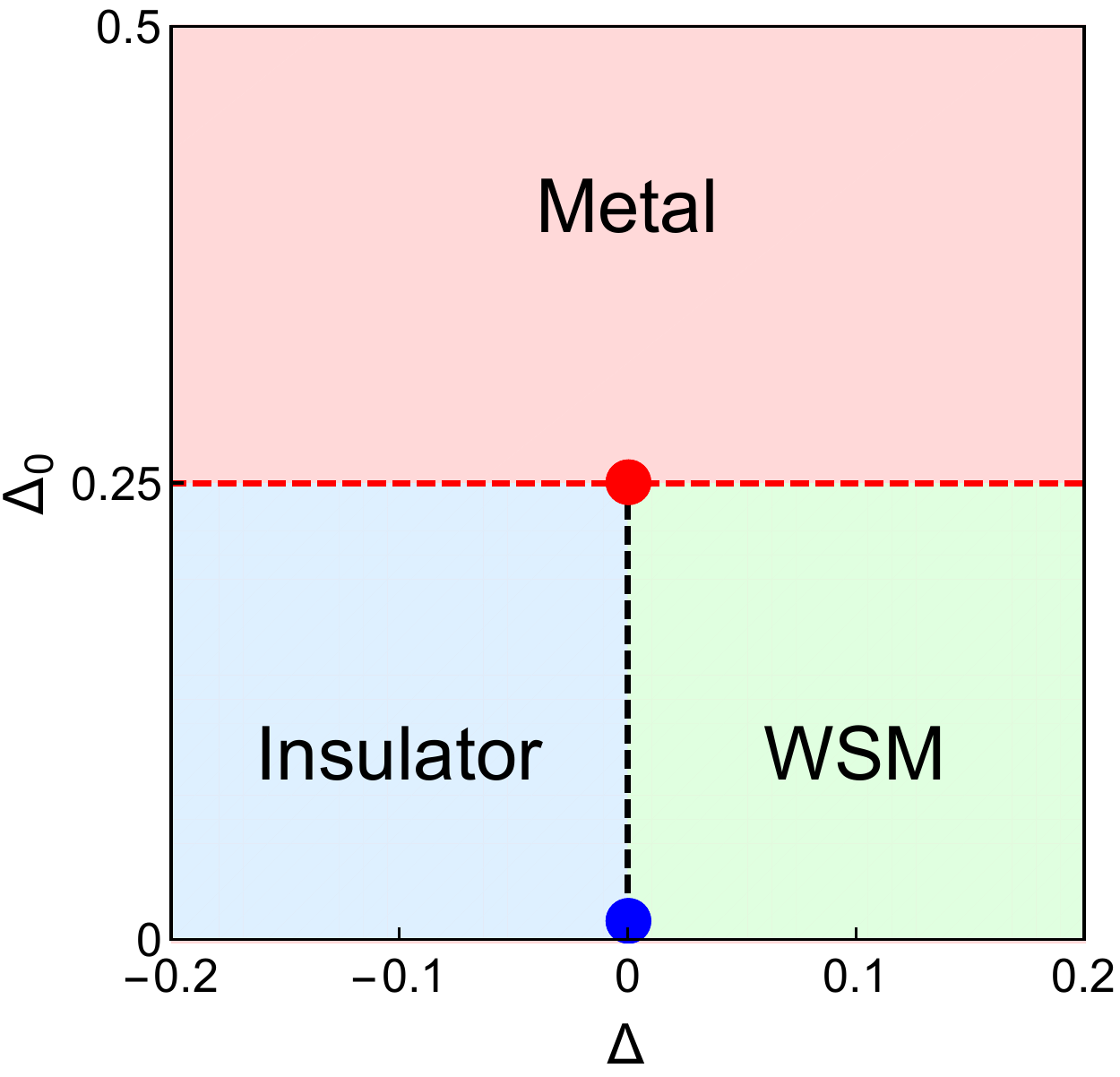}
\label{WSM_Ins_Phase}
}
\caption[]{(a) The RG flow diagram obtained from Eq.~(\ref{RG_WSM_INS}) and (b) the resulting phase diagram in the $\Delta-\Delta_0$ plane, for $\epsilon_n=1/2$. Here, $\Delta$ is the tuning parameter for WSM-insulator transition [see Eq.~(\ref{Ins_WSM_QCP})], and $\Delta_0$ is the strength of random charge impurities. Blue and red dot respectively represents a critical and a multicritical point. The metallicity sets in through the multicritical point.}\label{WSM_Ins_Flow_PD}
\end{figure}

Upon integrating out the fast Fourier modes within the momentum shell $\Lambda e^{-l} <k_\perp < \Lambda$, where $k_\perp=\sqrt{k^2_x+k^2_y}$, $0<k^2_3< \infty$ and accounting for pertubative corrections to one-loop order (see Fig.\ \ref{laddercrossing}), we arrive at the following flow equations
\allowdisplaybreaks[4]
\begin{align}~\label{RG_WSM_INS}
\beta_X&=-X\left( \Delta_0 + 2 \Delta_\perp +\Delta_z \right)=(1-z)X, \nonumber \\
\beta_\Delta&= \Delta \left[ 1 + \Delta_0 - 2 \Delta_\perp +\Delta_z \right], \nonumber \\
\beta_{\Delta_0}&=-\epsilon_n \Delta_0 + 2 \Delta_0 \left( \Delta_0 + 2 \Delta_\perp +\Delta_z \right), \\
\beta_{\Delta_\perp}&=-\epsilon_n \Delta_\perp + 2 \Delta_0 \Delta_z, \nonumber \\
\beta_{\Delta_z}&=-\epsilon_n \Delta_z + 2 \Delta_z \left(2 \Delta_\perp -\Delta_0 -\Delta_z\right) + 4 \Delta_0 \Delta_\perp, \notag
\end{align}
in terms of dimensionless parameters
\begin{align}
\hat{\Delta}=\frac{\Delta}{v \Lambda}, \: \hat{\Delta}_j=\Delta_j \left[ \frac{\Lambda^{\epsilon_n}}{(2 \pi)^2 b^{\epsilon_n}_n v^{2-\epsilon_n}} \right], \nonumber
\end{align}
for $X=v, b_n$, $j=0,\perp, z$, $\beta_Q \equiv dQ/dl$ is the $\beta$-function for the running parameter $Q$, and for brevity we omit the \emph{hat} notation in Eq.~(\ref{RG_WSM_INS}). In the above flow equations, we have kept only the leading divergent contribution that survives as $n \to \infty$. Inclusion of subleading divergences yields only nonuniversal corrections, as shown in Appendix~\ref{Append_WSM_INS}. The $\beta-$function for in-plane Fermi velocity ($v$) and $b_n$ leads to a scale dependent DSE
\begin{equation}~\label{DSE_MCP}
z(l)=1+ \left( \Delta_0 + 2 \Delta_\perp + \Delta_z \right) (l).
\end{equation}
Note that in this formalism the random charge-impurities do not generate any new disorder, allowing us to depict the RG flow in the ($\Delta, \Delta_0$) plane, as shown in Fig.~\ref{WSM_Ins_Flow}.

\begin{table}[t!]
\begin{tabular}{|l|l|l|l|l|l|l|}
\hline
{\bf Bilinear} & {\bf Physical quantity} & $\mathcal{T}$ & $\mathcal{P}$ & $U_c$ & $\mathcal{C}$ & {\bf Coupling} \\
\hline \hline
$\bar{\Psi} \gamma_0 \Psi$ & chemical potential   &  $\checkmark$ & $\checkmark$  & $\checkmark$ &  $\times$ & ${\Delta}_V$    \\
\hline
$\bar{\Psi} \gamma_0 \gamma_5 \Psi$ & axial potential  & $\checkmark$  & $\times$ & $\checkmark$ &  $\checkmark$ & $\Delta_A$  \\
\hline
$\bar{\Psi}\Psi$ & scalar mass & $\times$  & $\checkmark$  & $\times$ & $\checkmark$ & $\Delta_S$  \\
\hline
$\bar{\Psi}i \gamma_5\Psi$ & pseudo-scalar mass & $\checkmark$ & $\times$  & $\times$  & $\checkmark$ & $\Delta_{PS}$ \\
\hline
$\bar{\Psi}i \gamma_5 \gamma_j \Psi$ & axial current & $\times$  & $\checkmark$  & $\checkmark$ & $\checkmark$ & $\Delta_M$  \\
\hline
$\bar{\Psi}i \gamma_j \Psi$ & current & $\times$  & $\times$  & $\checkmark$ & $\times$ & $\Delta_C$  \\
\hline
$\bar{\Psi} i \Sigma_{0j} \Psi$ & temporal tensor & $\times$  & $\times$  & $\times$ & $\times$ & $\Delta_{SO}$ \\
\hline
$\bar{\Psi} \Sigma_{jk} \Psi$ & spatial tensor & $\checkmark$  & $\checkmark$  & $\times$ & $\times$ & $\Delta_{AM}$ \\
\hline \hline
\end{tabular}
\caption{ Various types of disorder represented by  fermionic bilinears ($j=1,2,3$), together with their symmetries under pseudo time-reversal ($\mathcal{T}$), parity ($\mathcal{P}$), continuous chiral rotation ($U_c$) and charge-conjugation ($\mathcal{C}$). The disorder couplings are represented by $\Delta_N$ and $\Sigma_{\mu \nu}=[\gamma_\mu, \gamma_\nu]/(2i)$. Note that true time-reversal symmetry in WSM in already broken. The pseudo time-reversal symmetry ${\mathcal T}$ is generated by an anti-unitary operator $\gamma_0 \gamma_2 K$, where $K$ is complex conjugation, such that ${\mathcal T}^2=-1$ (The true time-reversal operator is $\gamma_1 \gamma_3 K$). The parity operator is ${\mathcal P}=\gamma_0$, while the charge-conjugation operator is ${\mathcal C}=\gamma_2$. The continuous chiral symmetry ($U_c$) is generated by $\gamma_5$, the generator of translational symmetry in the continuum limit in a clean Weyl semimetal~\cite{roy-sau}. The Hermitain $\gamma$ matrices satisfy standard anti-commutation relation $\left\{ \gamma_\mu, \gamma_\nu \right\}=2\delta_{\mu \nu}$ for $\mu, \nu=0,1,2,3,5$, and for explicit representation of $\gamma$-matrices see Sec.~\ref{WSM:Hamiltonian-action}. Here $\checkmark$ and $\times$ signify even and odd under a symmetry operation, respectively. With a slightly different tight-binding model, where $N_j({\bf k})=t \cos(k_j a)$ and $N^M_3({\bf k})=[\sin(k_1 a)+\sin(k_2 a)-2 \sin(k_3 a)]$ [see Eq.~(\ref{two-band-WSM})], the axial current corresponds to magnetization, temporal and spatial tensors to spin-orbit and axial magnetization, respectively. However, such microscopic details do not alter any physical outcome. }~\label{table-disorder}
\end{table}

The coupled RG flow equations~(\ref{RG_WSM_INS}) support only two fixed points:

$\bullet$ \; $(\Delta,\Delta_0,\Delta_\perp, \Delta_z)=(0,0,0,0)$, which has only one unstable direction along the $\Delta$-direction that serves as the tuning parameter for WSM-insulator QPT. This fixed point stands as a QCP in the four dimensional coupling constant space. The correlation length exponent at this QCP is $\nu^{-1}=1$. All disorder couplings are irrelevant perturbations at this QCP [see the blue dot in Fig.~\ref{WSM_Ins_Flow}].

$\bullet$ \; $(\Delta,\Delta_0,\Delta_\perp, \Delta_z) \approx (0,\epsilon_n/2,0,0)$ stands as a multicritical point (MCP) with two unstable directions. At this MCP the WSM, an insulator and the metallic phase meet. Two correlation-length exponents are $\nu^{-1}_M =\epsilon_n$ determining the relevance of disorder coupling $\Delta_0$, which drives the anisotropic critical semimetal [described by $\hat{H}_Q(0)$] into a diffusive metallic phase, and  $\nu^{-1}=1$ that determines the relevance of the tuning parameter $\Delta$, controlling the WSM-insulator transition. The DSE for critical semimmetal-metal QPT is $z= 1+ \frac{\epsilon_n}{2}+{\mathcal O}(\epsilon^2_n)$. Therefore, for a three-dimensional anisotropic critical semimetal-metal QPT, setting $\epsilon_n=1/2$, the critical exponents are $\nu_M=2$ and $z=1.25$, to the leading order in $\epsilon_n$ expansion.

\begin{figure*}
\subfigure[]{
\includegraphics[width=4.5cm,height=4cm]{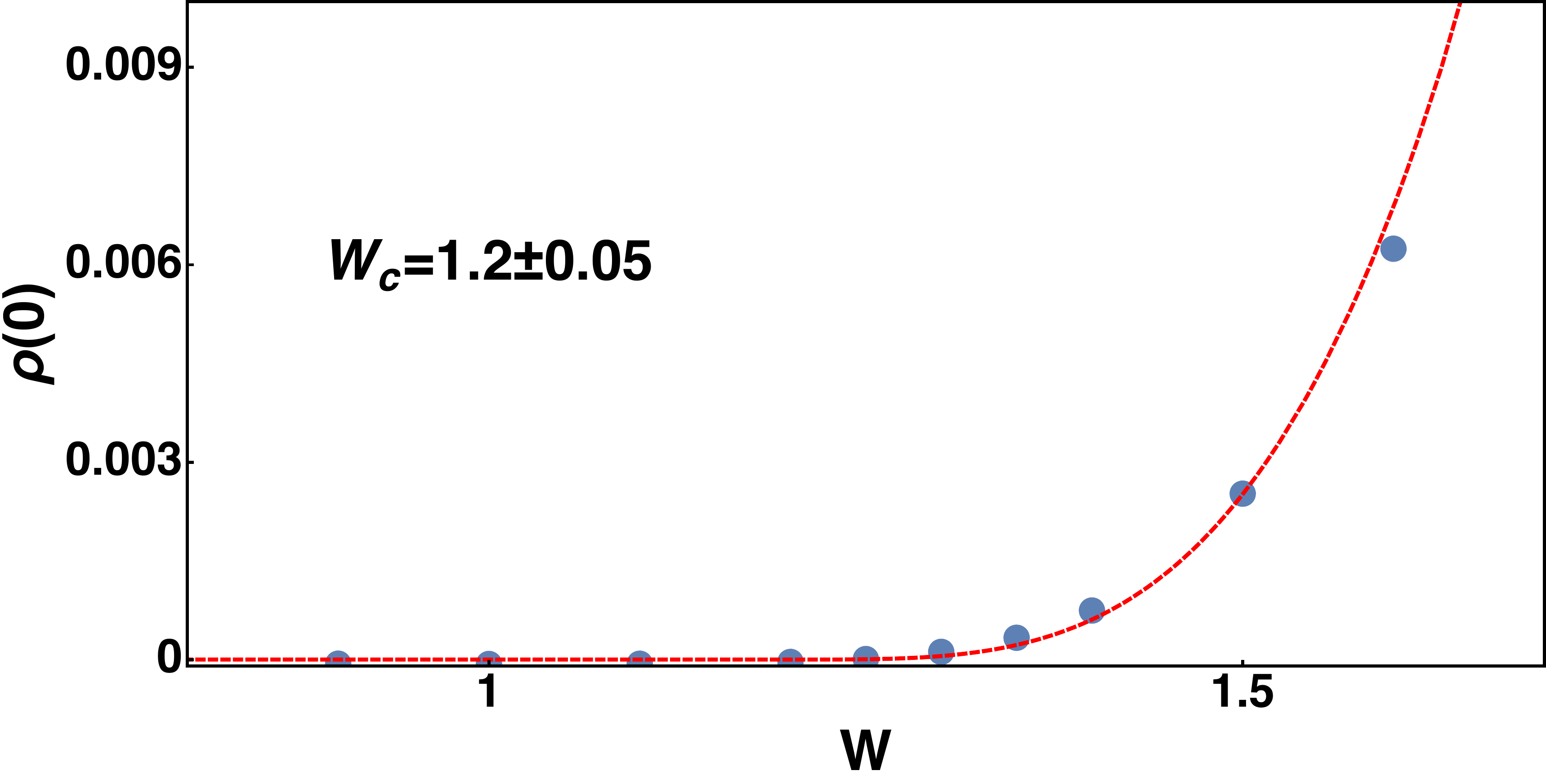}
\label{rhovsW_MCP}
}
\subfigure[]{
\includegraphics[width=4.5cm,height=4cm]{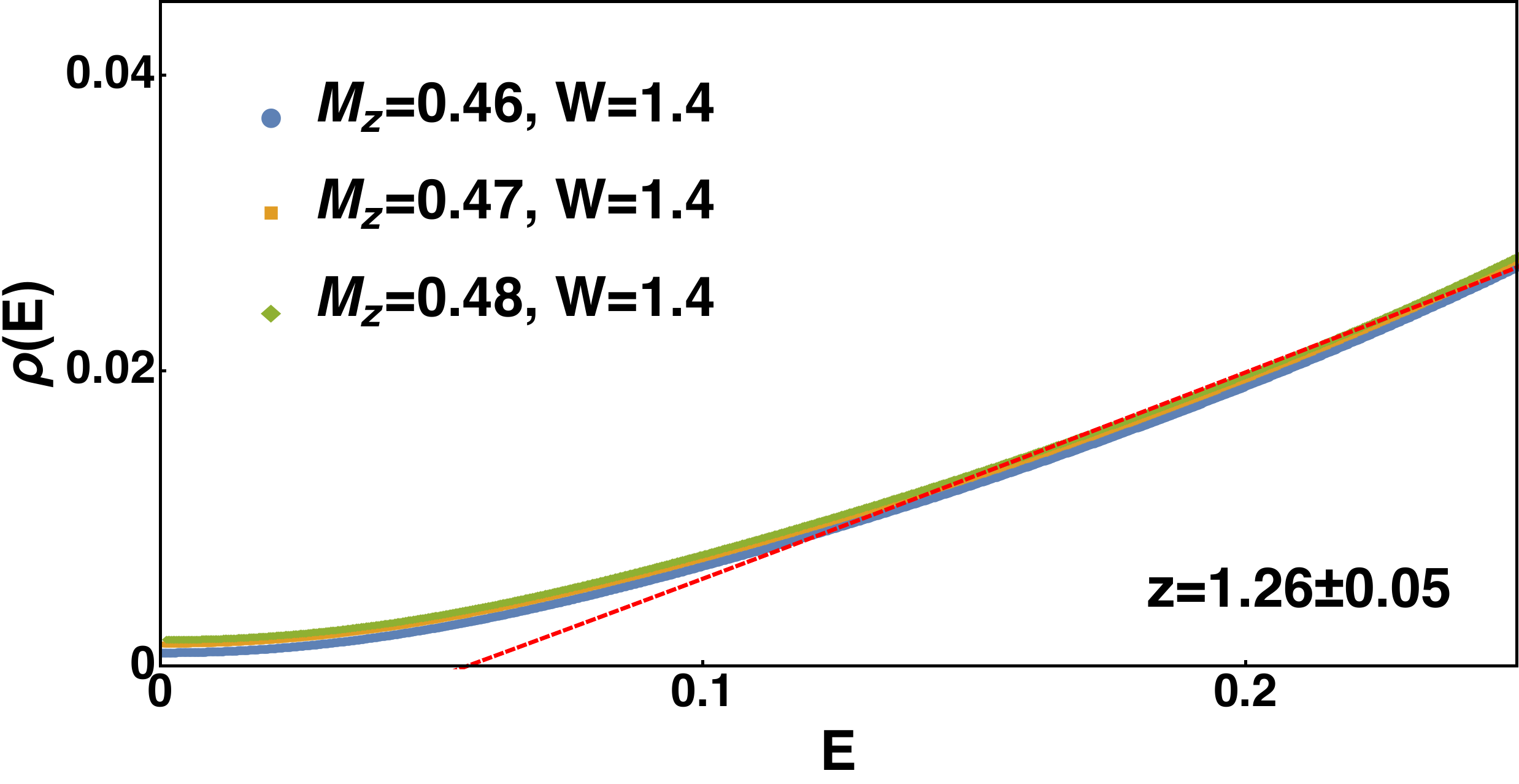}
~\label{MCP_DSE}
}
\subfigure[]{
\includegraphics[width=4.5cm,height=4cm]{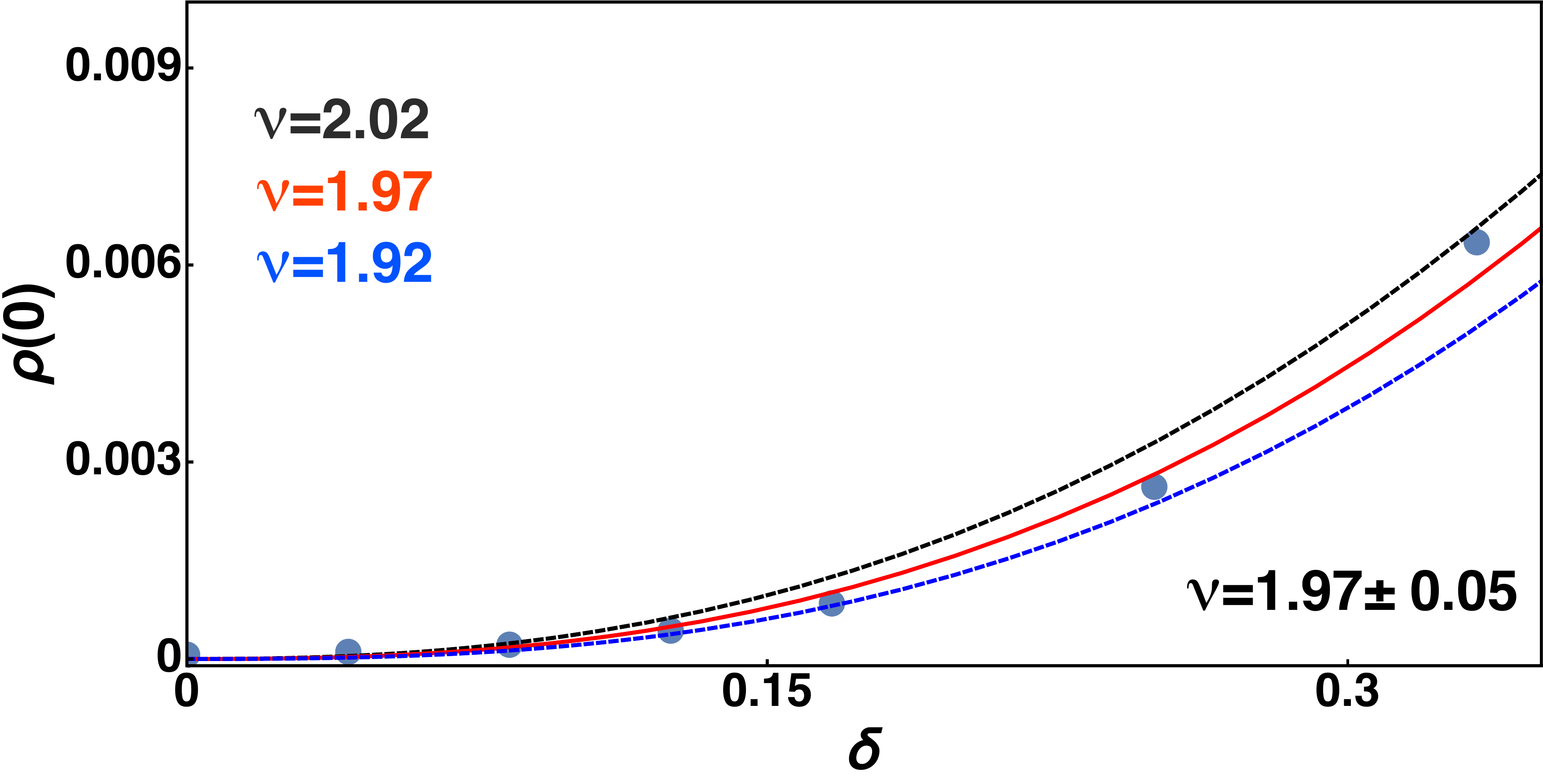}
~\label{MCP_CLE}
}
\caption{ Analysis of average density of states (DOS) in various regimes along the black dashed line shown in Fig.~\ref{MCP_numerics_PD} (left). Recall the black dashed line for weak disorder defines the phase boundary between the WSM and insulator, while when extended into the metallic phase [the red shaded regime in Fig.~\ref{MCP_numerics_PD} (left)] captures the instability of critical excitations residing at the WSM-insulator QCP toward the formation of a metallic phase. (a) Scaling of average DOS at zero energy [$\varrho(0)$] along the blacked dashed line as a function of increasing disorder ($W$), showing that $\varrho(0)$ remains pinned at zero up to a critical strength of disorder $W_c=1.20 \pm 0.05$. (b) Scaling of average DOS at finite energy [$\varrho(E)$] around the multi-critical point residing in the two dimensional coupling constant space $(m_z,W)$, indicating the dynamic scaling exponent for critical excitation-metal QPT is $z=1.26 \pm 0.05$. (c) Scaling of $\varrho(0)$ along the black dashed line inside the metallic phase indicating that correlation length exponent for critical excitation-metal QPT is $\nu=1.97 \pm 0.05$. Details of the data analysis are presented in Sec.~\ref{MCP_numerics_analysis}. The quoted error bars in $z$ and $\nu$ are \emph{fitting error bars}. See Appendix~\ref{Append:data_analysis} and Table~\ref{Table:Data-analysis} (last row) for further details of data analysis.
}~\label{MCP_analysis}
\end{figure*}

The RG flow and the resulting phase diagrams are shown in Fig.~\ref{WSM_Ins_Flow} and \ref{WSM_Ins_Phase}, respectively. At the multicritical point the average DOS scales as $\varrho(E) \sim |E|^{d_\ast/z-1} \approx |E|$ to one-loop order, since $d_*=5/2$ for $c=1$, as given by Eq.~(\ref{effective-dim}). Beyond the critical strength of disorder system becomes a metal where the average DOS at zero energy $[\varrho(0)]$ is finite and the order parameter exponent $\beta=(d_\ast-z)\nu= 2.5$ determines the scaling of $\varrho(0)$ according to $\varrho(0) \sim \delta^{\beta}=\delta^{2.5}$ in the metallic phase, where $\delta=\left( \Delta_0-\Delta^{\ast}_0 \right)/\Delta^{\ast}_0$ is the reduced disorder coupling from the critical one at $\Delta_0=\Delta^\ast_0$. Next we numerically demonstrate (a) stability of WSM-insulator QCP at weak disorder, (b) emergence of a metallic phase through a MCP at finite disorder coupling that masks the direct transition between WSM and insulator by numerically computing the average DOS using the kernel polynomial method. As a natural outcome of this exercise, we will also show that numerically extracted values of the exponents, $z$ and $\nu$, at the MCP, associated with the critical excitations-metal QPT agree with the predictions from the leading order $\epsilon_n$-expansion. We also note that the same spirit of RG analysis, controlled via ``band-flattening", can also be applied to address the effect of randomness deep inside the WSM phase. We, however, relegate that discussion to Appendix~\ref{epsilon_n:WSM-metal_Append}.

For the sake of simplicity, we here neglect quantum corrections to RG flow equations due to non-trivial dispersion along $k_z$. Nonetheless, our formal approach allows to systematically account for such quantum corrections, controlled via another small parameter $1/n$ (in the spirit of an $1/N$-expansion, where $N$ counts the number of fermion flavors~\cite{zinn-justin}). Therefore, our RG analysis is ultimately controlled by two small parameters $\epsilon_n$ (measuring the deviation from the marginality condition for disorder, i.e. two spatial dimensions, leading to non-trivial bare scaling dimension $[\Delta_j]=-\epsilon_n$ for all disorder couplings with $j=0,\perp, z$) and $1/n$ (measuring the strength of the band dispersion in $k_z$ direction and thus controlling the quantum (loop) corrections arising from finite band curvature in this direction). In this regard the RG analysis follows the spirit of simultaneous $\epsilon$- and $1/N$- expansions~\cite{zinn-justin}. Only at the very end of the calculation we set $\epsilon_n = 1/2$ and $n = 2$ (physically relevant situation). This analysis is presented in details in Appendix~\ref{subleading}. The resulting exponents (after accounting for $1/n$ quantum corrections), namely $z = 1.245$ and $\nu = 2$ are sufficiently close to the ones we report here by taking $n \to  \infty$ in the perturbative loop corrections.

\subsection{Scaling of density of states near WSM-insulator QCP: Numerical demonstration of the MCP}~\label{MCP_numerics_analysis}

Before we discuss the scaling behavior of the average DOS along the WSM-insulator phase boundary and inside the metallic phase, setting in through the instability of critical semimetallic phase, let us point out some crucial subtle issues associated with such analysis. Note that the average DOS of the critical semimetal [described by $\hat{H}_Q(0)$ in Eq.~(\ref{Ins_WSM_QCP})] vanishes as $\varrho(E) \sim |E|^{3/2}$, while that in the WSM phase vanishes as $\varrho(E) \sim |E|^2$. But, in the insulating phase average DOS displays \emph{hard gap}. Based on scaling analysis we expect WSM, insulator and the critical semimetal to be stable against sufficiently weak disorder. We exploit these characteristic features to pin the WSM-insulator phase boundary for weak disorder. On the other hand, for stronger disorder onset of a metallic phase can be identified from the existence of finite average DOS at zero energy. Following these diagnostic tools we arrive at the phase diagram of a Weyl materials residing in the close proximity to the WSM-insulator QPT; see Fig.~\ref{MCP_numerics_PD} (left). We are ultimately interested in exposing the existence of a MCP in the ($m_z, t_z$) plane [the red dot in Fig.~\ref{WSM_Ins_Flow}] which has \emph{two} relevant directions. One of them controls critical semimetal-metal QPT, while the other one drives WSM-insulator QPT. Since we consider the former transition, our focus will be restricted on the black dashed line shown in Fig.~\ref{MCP_numerics_PD}.

More specifically, we here compute the average DOS by employing the KPM~\cite{KPM-RMP} starting with the tight-binding model, introduced in Eqs.~(\ref{two-band-WSM}), ~(\ref{wilson}) and (\ref{TB_Weyl}), and staying in the close vicinity of $m_z/t_0=0.5$ and $t_z/t_0=0.5$ (see the phase diargam in Fig.~\ref{TB_PD}). The tight-binding model is implemented on a cubic lattice with periodic boundary conditions in all three directions and the linear dimensionality of the system in each direction is $L=140$. Even though average DOS is a self-averaged quantity, we perform average over 20 random disorder realization to minimize the residual statistical fluctuations, compute 4096 Chebyshev moments and take trace over 12 random vector to obtain average DOS. For the sake of simplicity we here account for only random charge impurities. Potential disorder is distributed uniformly and randomly within the range $[-W, W]$. The scaling of average DOS can be derived in the following way.

Since we are following only one relevant direction associated with the MCP, effectively it can be treated as a simple QCP across which various physical observables (such as average DOS) display single parameter scaling. Note that total number of states $N(E,L)$ in a $d$-dimensional system of linear dimension $L$, below the energy $E$ is proportional to $L^d$, and in general is a function of two dimensionless parameters $L/\xi$ and $E/E_0$. Here, $\xi \sim \delta^{-\nu}$ is the correlation length that diverges at the QCP, located at $\delta=0$, where $\delta=\frac{W-W_c}{W_c}$ is the reduced distance from the QCP, located at $W=W_c$. Consequently, the correlation energy, defined as $E_0 \sim \delta^{\nu z}$ vanishes as the QCP is approached from either side of the transition~\cite{sachdev-book}. Following the standard formalism of scaling theory we then can write
\begin{equation}
N(E,L)= \left( L/\xi\right)^d \; G \left( E/\delta^{\nu z}, L/\delta^{-\nu} \right),
\end{equation}
where $G$ is an universal but unknown scaling function. Therefore, from the definition of average DOS $\varrho(E,L)=L^{-d} dN(E,L)/dE$ we arrive at the following scaling form
\begin{equation}~\label{DOS_Scaling_numerics}
\varrho(E,L)=\delta^{\nu (d-z)} \; F\left( |E|\delta^{-\nu z}, \delta L^{1/\nu} \right),
\end{equation}
where $F$ is yet another universal, but typically unknown scaling function. However, we can access the behavior of the scaling function in different regimes along the black dashed line shown in Figs.~\ref{MCP_numerics_PD} (left), which we exploit to compute critical exponents characterizing the critical semimetal-metal QPT across the MCP. In the final step we have used the fact that average DOS remains particle-hole symmetric, but on average. Note we will use exactly the same scaling function deep inside the WSM phase in the presence of generic disorder, discussed in Sec.~\ref{numerics_analysis}.
We must stress here that in the above expression $d=d_\ast$, the effective dimensionality of the system, defined in Eq.~(\ref{effective-dim}), when we address the scaling of ADOS along the phase boundary between the WSM and an insulator, and across the QPT to a metallic phase through the MCP, shown in Fig.~\ref{MCP_numerics_PD}(left). On the other hand, we set $d=3$ (physical dimensionality) while addressing the WSM-metal transition since the electronic dispersion is linear and isotropic in a WSM.

First of all, notice that average DOS $\varrho(0)$ is pinned to zero along the phase boundary between the WSM and insulator for weak enough disorder, as shown in Fig.~\ref{rhovsW_MCP}. Therefore, critical semimetal separating these two phases remains stable against weak disorder and the nature of the WSM-insulator direct transition remains unchanged for weak enough randomness. However, beyond a critical strength of disorder, $W_c=1.20 \pm 0.05$, $\varrho(0)$ becomes finite and metallicity sets in through the MCP, see Figs.~\ref{MCP_numerics_PD} (left) and ~\ref{rhovsW_MCP}. Beyond this point there exists no direct transition between the WSM and an insulator. Also note for $W \ll W_c$, $\varrho(E) \sim |E|^{1.5}$ as shown in Fig.~\ref{MCP_numerics_PD} (right), as expected, since in the clean system $z=1$ and $d_\ast=5/2$.

\begin{figure}[t!]
\includegraphics[width=8cm,height=3.75cm]{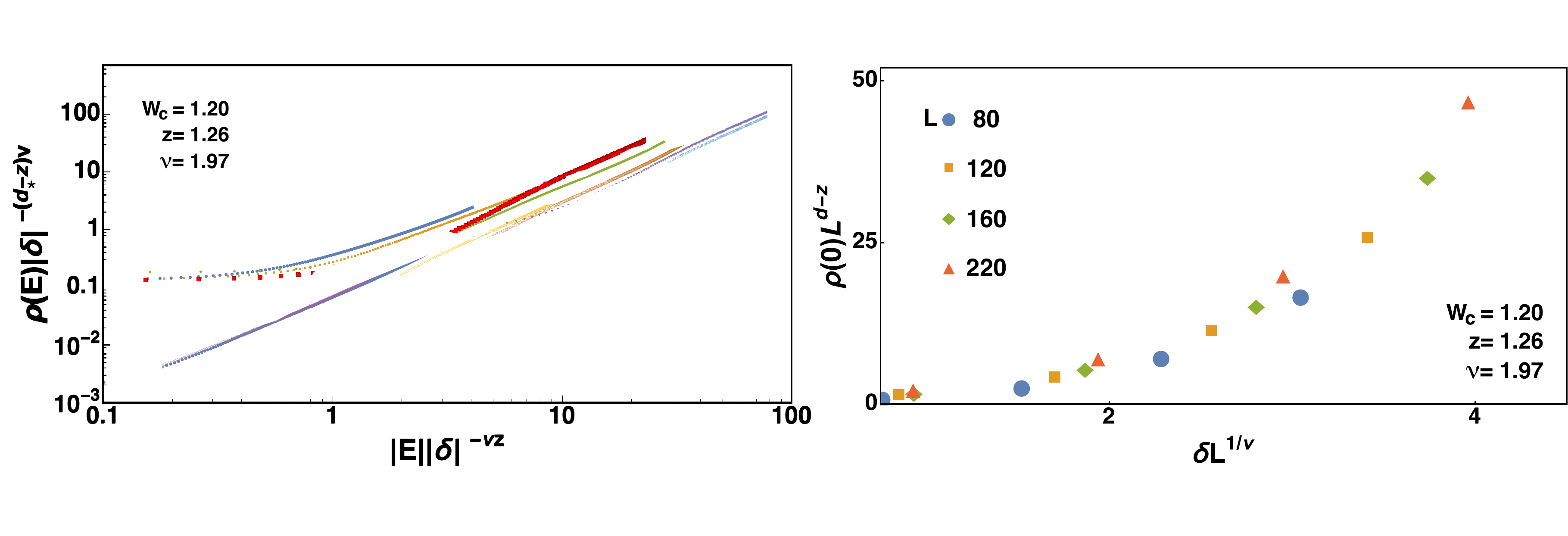}
\caption{ (a) Collapse of average DOS at finite energy (obtained in system with $L=220$) across the multi-critical point (MCP) shown in Fig.~\ref{MCP_numerics_PD} (left). All data collapse reasonably well onto two branches corresponding to anisotropic semimetal (upper branch) and metallic phase (lower branch), which tend to meet in the critical regime. (b) Data collapse of average DOS at zero energy for different system sizes inside the metallic phase, appearing across the MCP. These two data collapses are obtained with numerically extracted critical exponents $z=1.26$ and $\nu=1.97$ [see Fig.~\ref{MCP_analysis}]; with $d_\ast=5/2$.
}~\label{collapse_MCP}
\end{figure}

Now we consider very close proximity to the MCP, located at $W=W_c$ along the disorder axis. At this MCP average DOS becomes independent of $\delta$, yielding $F(x) \sim x^{\frac{d_\ast}{z}-1}$. By comparing $\varrho(E)$ with $E$, we obtain the DSE associated with critical semimetal-metal QPT to be $z=1.26 \pm 0.05$, see Fig.~\ref{MCP_DSE}.

Next we move into the metallic phase, but continue to follow the black dashed line from Fig.~\ref{MCP_numerics_PD} (left). In the metallic phase $\varrho(0)$ becomes finite [see Fig.~\ref{rhovsW_MCP}]. Thus to the leading order $F(x) \sim x^0$ and consequently $\varrho(0) \sim \delta^{(d_\ast-z) \nu}$. With the prior notion of $z=1.26 \pm 0.05$, now by comparing $\varrho(0)$ vs. $\delta$ we obtain the CLE at the MCP associated with the critical semimetal-metal QPT to be $\nu=1.97 \pm 0.05$, as shown in Fig.~\ref{MCP_CLE}~\footnote{ 
After accounting for the variation in the location of $W_c$ and determination of $z$, we finally obtain $\nu= 1.98 \pm 0.10$, see Appendix~\ref{Append:data_analysis} for discussion and Table~\ref{Table:Data-analysis} (last row) for analysis.}.

Therefore, numerically extracted values of two critical exponents, namely $\nu=1.97$ and $z=1.26$, at the MCP associated with the critical semimetal-metal QPT match quite satisfactorily with the field theoretic prediction obtained from an $\epsilon_n$-expansion introduced in this work, which allows to control the RG calculation by tuning the flatness of the quasiparticle dispersion along $k_z$ direction: \emph{a controlled ascent from two spatial dimension}.

We now discuss two different types of data collapses across the disorder-driven MCP. The results are shown in Fig.~\ref{collapse_MCP}. First we focus on the largest system with $L=220$. From Eq.~(\ref{DOS_Scaling_numerics}), upon neglecting the finite size effects, we compare $\varrho(E) |\delta|^{-(d_\ast-z)\nu}$ vs. $|E| |\delta|^{-\nu z}$ along the black line from Fig.~\ref{MCP_numerics_PD}(left). With numerically obtained values of $\nu$ and $z$ we find that all data nicely collapse onto two branches (corresponding to the anisotropic semimetal and metallic sides of the QPT), which meet in the critical regime, as shown in Fig.~\ref{collapse_MCP}(left). Next we compare the average DOS at zero energy in the metallic phase, namely $\varrho(0) L^{d_\ast-z}$ vs. $L^{1/\nu} \delta$, in systems of different sizes ($L$), as shown in Fig.~\ref{collapse_MCP}(right). We also obtain excellent finite-size data collapse for a wide range of system sizes using already numerically extracted values of $\nu$ and $z$. Therefore, field-theoretic predictions and numerical findings across the disorder-driven MCP  are in good agreement with each other. Next we address the effects of disorder inside the WSM phase by pursing complementary field theoretic and numeric approaches.

Note that the MCP, where WSM, an insulator, a metal and the critical anisotropic semimetal meet, possesses two relevant directions, see Fig.~\ref{WSM_Ins_Flow}. Hence, at finite energies two quantum critical fans associated with (1) critical anisotropic semimetal-metal and (2) WSM-metal QPTs (characterized by distinct sets of critical exponents) \emph{interwine}. Thus, obtaining a high quality data collapse at finite energies [see Fig.~\ref{collapse_MCP}(left)] across this MCP is quite challenging, and qualitatively it is slightly worse than that across the WSM-metal QPT (sufficiently far from the MCP), shown in Figs.~\ref{numeric_analysis_figure} and \ref{numeric_analysis_figure_CSB} (third column). Still, roughly 300 data points effectively fall on two branches [top (bottom) one representing metallic (anisotropic semimetallic) phase] with numerically extracted mean values of the exponents, $z = 1.26$ and $\nu = 1.97$, in good agreement with analytical predictions from leading order in $\epsilon_n$-expansion ($z=1.25$ and $\nu=2$). The quality of finite-size data collapse obtained from the scaling of $\varrho(0)$ in different systems [see Fig.~\ref{collapse_MCP}(right)] is yet quite comparable to the ones shown in Figs.~\ref{numeric_analysis_figure} and \ref{numeric_analysis_figure_CSB}(forth column) across the WSM-metal QPT.

\section{Dirty Weyl semimetal: Model and scaling analysis}~\label{dirtyWSM_intro}

In this section, we set up the field theoretical framework to analyze the role of disorder when the system is deep inside the WSM phase. We will introduce the notion of two different $\epsilon$-expansions: (a) an $\epsilon_m-$expansion about a critical disorder distribution, where $\epsilon_m=1-m$ with Gaussian white noise distribution recovered as $m \to 0$; (b) an $\epsilon_d-$expansion, with Gaussian white noise distribution from outset, about the lower critical dimension $d_c=2$ for WSM-metal QPT, where $\epsilon_d=d-2$, and therefore for three spatial dimensions $\epsilon_d=1$.

\subsection{Hamiltonian and action}~\label{WSM:Hamiltonian-action}

The effective low energy description of WSM can be obtained by expanding the lattice Hamiltonian [see Eq.~(\ref{TB_Weyl})] around the Weyl nodes located at ${\bf k}^0=(0,0,\pm k_z^0)$, with $k_z^0=\cos^{-1}(\frac{m_z}{t_z})$. The resulting low energy Hamiltonian reads
\begin{align}\label{eq:ham-WSM1}
H_W=\tau_0 \otimes v \left( k_x \sigma_1 + k_y \sigma_2 \right)+ \tau_3 \otimes \sigma_3 v_z  k_z,
\end{align}
where $v=ta$, $v_z=a\sqrt{t_z^2-m_z^2}$, and the momentum is measured from the Weyl nodes. For simplicity we hereafter take the Fermi velocity to be isotropic, $v=v_z$, so that the low energy Hamiltonian becomes rotationally symmetric. Upon performing a unitary rotation with $U=\sigma_0 \oplus \sigma_3$, the above Hamiltonian assumes a quasirelativistic form $H_W=i \gamma_0 \gamma_j v k_j$, where $\gamma_0=\tau_1 \otimes \sigma_0$, $\gamma_j=\tau_2 \otimes \sigma_j$ for $j=1,2,3$ are mutually anti-commuting $4\times4$ Hermitian matrices, and summation over repeated spatial indices is assumed. To close the Clifford algebra of five mutually anticommuting matrices we define $\gamma_5=\tau_3 \otimes \sigma_0$. Two sets of Pauli matrices $\sigma_\mu$ and $\tau_\mu$ respectively operate on spin/pseudospin and valley or chiral (left and right) indices. The low energy effective Hamiltonian enjoys variety of \emph{emergent} discrete and continuous symmetries. The above Hamiltonian is invariant under a pseudo-time-reversal symmetry, generated by anti-unitary operator ${\mathcal T}=\gamma_0 \gamma_2K$, where $K$ is the complex conjugation, a charge conjugation symmetry, generated by ${\mathcal C}=\gamma_2$, and parity or inversion symmetry generated by ${\mathcal P}=\gamma_0$. Furthermore, the Hamiltonian [see Eq.~(\ref{eq:ham-WSM1})] also possesses a global chiral $U(1)$ symmetry, generated by $\gamma_5$, which in the low energy limit corresponds to the generator of translational symmetry~\cite{roy-sau}.

To incorporate the effects of disorder we consider the following minimal continuum action for a dirty WSM
\be\label{eq:action}
S=\int d^d {\bf x} d\tau \left[{\bar \Psi}(\gamma_0\partial_\tau+v\gamma_j\partial_j)\Psi-\varphi_N({\bar \Psi}N{ \Psi})\right],
\ee
with ${\bf x}$ as $d-$dimensional spatial coordinates, the four-component spinor $\Psi^\dagger=(u_{\uparrow,+}^\dagger,u_{\downarrow,+}^\dagger,u_{\uparrow,-}^\dagger,u_{\downarrow,-}^\dagger)$, where $u^\dagger_{\sigma,\tau}$ is the fermionic creation operator near the Weyl point at $\tau {\bf k}^0$ for $\tau=\pm$ (left/right) and with spin $\sigma=\uparrow,\downarrow$, while ${\bar\Psi}=\Psi^\dagger\gamma_0$, as usual. Various disorder fields $\varphi_N$, coupled to the fermion bilinears, are realized with different choices of $4 \times 4$ matrices, $N$, as shown in Table~\ref{table-disorder}. Notice that the matrices associated with four types of disorder anticommute with $\gamma_5$ and represent chiral symmetric disorder, while for the other four types of disorder $[N, \gamma_5]=0$ and the corresponding disorder vertex breaks the $U(1)$ chiral symmetry. \emph{As we demonstrate in this paper, such a global chiral symmetry plays a fundamental role in classifying the disorder-driven WSM-metal QPTs.}
\\

\subsection{$\epsilon_m$ expansion in three dimensions}~\label{epsilonm-formalism}

We assume that the disorder field obeys the distribution~\cite{kim-moon, halperin-disorder}
\be\label{eq:disorder-distribution-realspace}
\langle \varphi_N({\bf x}) \varphi_N({\bf y})\rangle={\Delta}_N\frac{1}{|{\bf x}-{\bf y}|^{d-m}} ,
\ee
or in the momentum space
\be\label{eq:disorder-distribution-momentumspace}
\langle \varphi_N({\bf q}) \varphi_N({\bf 0})\rangle={\tilde \Delta}_N\frac{1}{|{\bf q}|^{m}} ,
\ee
and the limit $m\rightarrow0$ corresponds to the Gaussian white noise distribution, which we are ultimately interested in. This form of the white noise distribution stems from the following representation of the $d-$dimensional $\delta-$function~\cite{pallab-sudip2016}
\be
\delta^{(d)}({\bf x}-{\bf y})=\lim_{m\rightarrow0}\frac{\Gamma\left(\frac{d-m}{2}\right)}{2^m\pi^{d/2}\Gamma(m/2)}\frac{1}{|{\bf x}-{\bf y}|^{d-m}}.
\ee

We now carry out the scaling analysis of the continuum action for a WSM given by Eq.~(\ref{eq:action}). The scaling dimensions of the momentum and frequency are $[q]=1$, and $[\omega]=z$. The form of the Euclidean action [see Eq.~(\ref{eq:action})] then implies that the engineering scaling dimension of the fermionic field  $[\Psi]=d/2$ and $[v]=z-1$, while the scaling dimension of the disorder field is  $[\varphi_N]=z+\eta_{\varphi_N}$, since the engineering dimension of the disorder field is equal to the DSE $z$ for any choice of $N$, and $\eta_{\varphi_N}$ is its anomalous dimension. Eq.~(\ref{eq:disorder-distribution-realspace}) then yields
\be\label{eq:scaling-dimension-disorderstrength}
[\Delta_N]=2(z+\eta_{\varphi_N})-d+m.
\ee
Due to linearly dispersing low-energy quasiparticles, a WSM corresponds to $z=1$ fixed point, and in $d=3$ the engineering dimension of the disorder strength is $[\Delta_N]=m-1$. A first implication of this result is that the white noise disorder, $m=0$, is \emph{irrelevant} close to the WSM ground state in $d=3$. Second, for $m=1$, the disorder is marginal and we use that to introduce the deviation from this value as an expansion parameter  $\epsilon_m=1-m$.

\begin{figure}
\includegraphics[width=8cm, height=3.75cm]{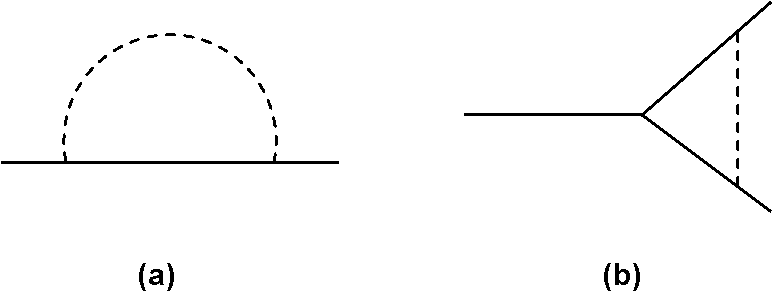}
\caption{ One-loop (a) self-energy and (b) vertex diagram. Contributions from only these two diagrams are ultraviolet divergent in $\epsilon_m=m-1$ expansion. Evaluations of these two diagrams are shown in Appendix~\ref{detailsfieldtheory}. Here, solid (dashed) lines represent fermion (disorder) fields. }~\label{selfvertex}
\end{figure}

The $\beta-$function (infrared) for the disorder coupling $\Delta_N$ in the $\epsilon_m$ expansion is given in terms of its scaling dimension in Eq.~(\ref{eq:scaling-dimension-disorderstrength}), yielding
\be\label{eq:beta-function-general}
\beta_{\Delta_N}=\Delta_N[-\epsilon_m+2(z-1)+2\eta_{\varphi_N}],
\ee
in $d=3$. Therefore, to obtain the explicit form of this $\beta-$function in terms of the disorder couplings, we have to compute the DSE and the anomalous dimension of the disorder field. The former is obtained from the fermion self-energy with the diagram shown in Fig.~\ref{selfvertex}(a), while the latter is found from the vertex diagram in Fig.~\ref{selfvertex}(b). Evaluation of these two diagrams has been carried out using field-theoretic method (see Appendix~\ref{detailsfieldtheory}). Alternatively, one may choose to integrate out the fast modes within the momentum shell $\Lambda e^{-l}<k<\Lambda$, with  $\Lambda$ as an ultraviolet cutoff in the momentum, to arrive at the RG flow equations for $\Delta_N$. We note that in the $\epsilon_m$-expansion two ladder diagrams shown in Fig.~\ref{laddercrossing} [(c), (d)] are \emph{ultraviolet convergent} (see Appendix~\ref{Append:Ladder_Crossing_epsilonm}) irrespective of the choice of disorder vertices. Therefore, during the coarse graining no new or short-range disorder gets generated (see also Appendix~\ref{Sec:epsilonm_8coupling}). This conclusion remains operative even beyond the leading order in $\epsilon_m$-expansion.

\subsubsection{Self-energy and dynamic scaling exponent}

We first show the computation of the self-energy diagram, shown in Fig.\ \ref{selfvertex}(a), yielding the dynamical exponent and the anomalous dimension for the fermion field within the regularization scheme defined by the parameter $\epsilon_m=1-m$, the deviation from the critical disorder distribution. All the integrals are therefore performed in $d=3$. The divergent part of the integral appears as a pole $\sim1/\epsilon_m$, analogously to the case of the dimensional regularization where the deviation from the upper or lower critical space-time dimension plays the role of an expansion parameter. To find renormalization constants, we use minimal subtraction, i.e. we keep only divergent part appearing in the corresponding diagrams.

\begin{figure}[t!]
\subfigure[]{
\includegraphics[width=4cm, height=4cm]{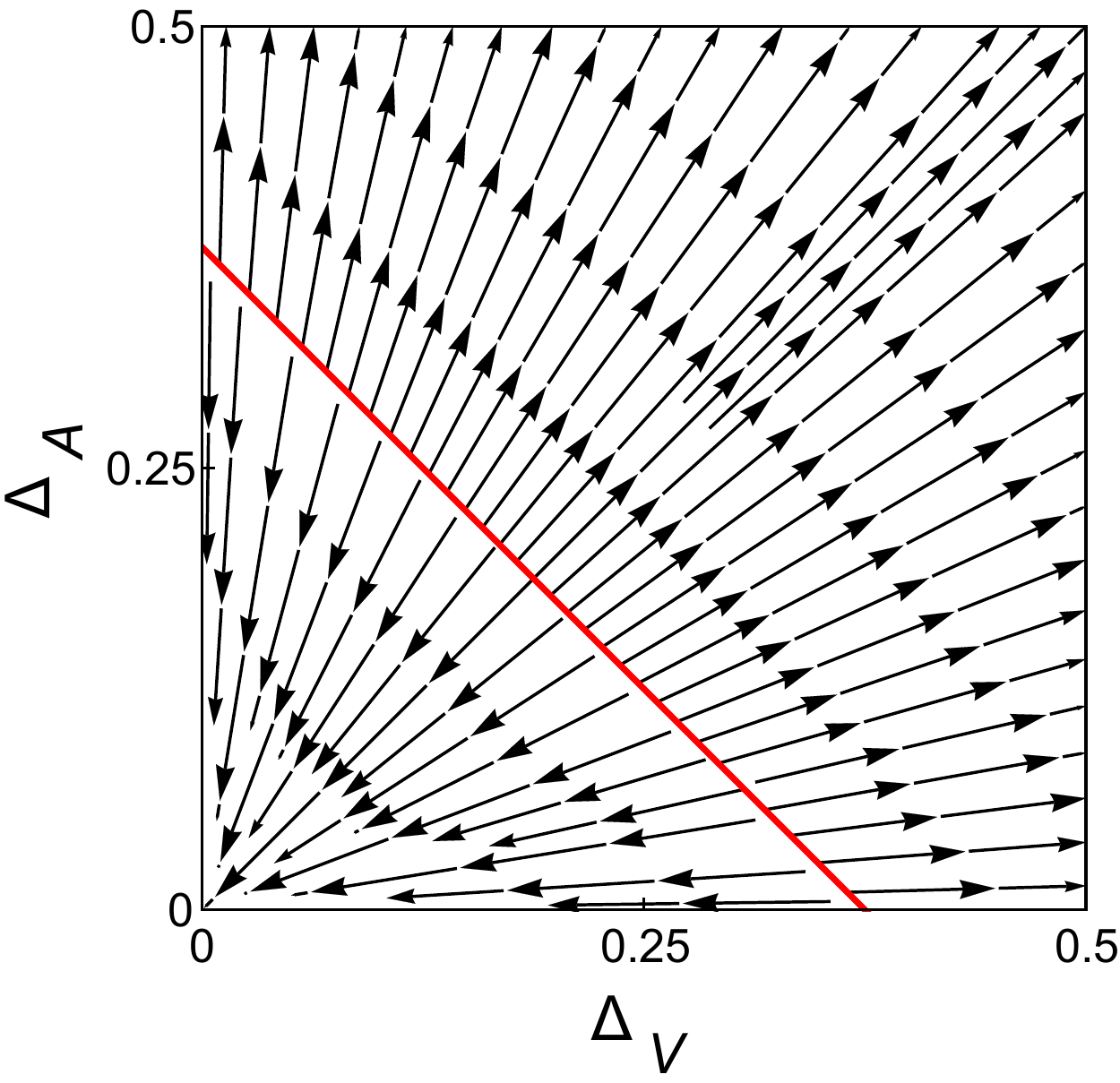}
\label{potaxialflow}
}
\subfigure[]{
\includegraphics[width=4cm, height=4cm]{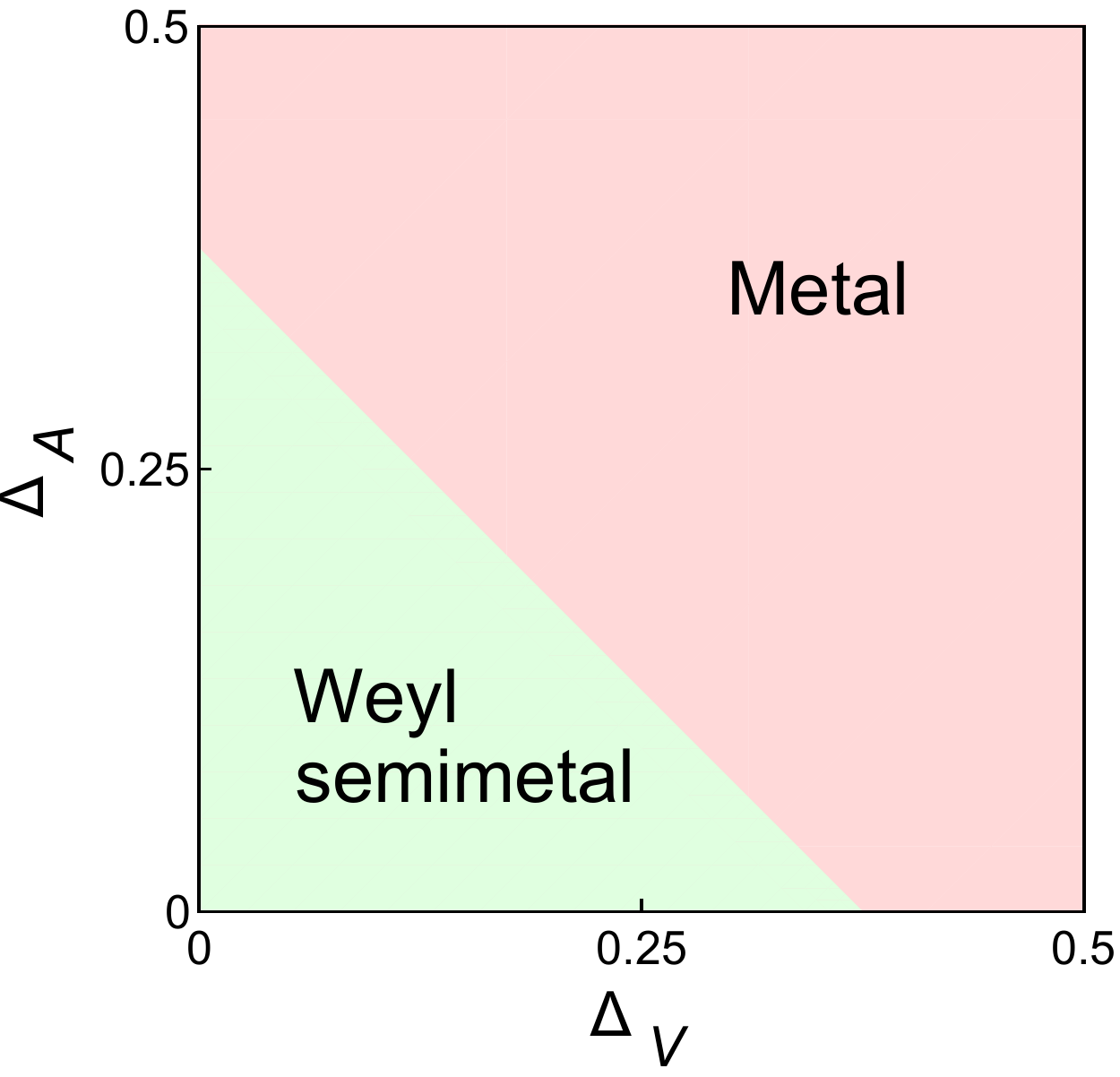}
\label{potaxialphase}
}
\caption[]{(a) The RG flow diagram and (b) the phase diagram in the $\Delta_V-\Delta_A$ plane, for $\epsilon_m=1$, obtained from Eq.~(\ref{ChiralRG_epsilon}). Here $\Delta_V$ and $\Delta_A$ are respectively the strength of potential and axial potential disorder. The red line in (a) corresponds to the line of quantum critical points [see Eq.~(\ref{LineQCP_chiral})] that in turn defines the phase boundary between the Weyl semimetal and metallic phases, as shown in panel (b). A similar flow and phase diagram is obtained from the RG calculation performed within the framework of an $\epsilon_d$ expansion [see Eq.~(\ref{LineQCP_chiral_hard})]~\cite{chakravarty, roy-dassarma-intdis, pallab-sudip2016}. }\label{Potential_Axial}
\end{figure}

The action [see Eq.~(\ref{eq:action})] without the disorder yields the inverse free fermion propagator $G_0^{-1}(i\omega,{\bf k})=i(\gamma_0 \omega+v_0\gamma_j k_j)$, with $v_0$ as the bare Fermi velocity. Taking into account the self-energy correction, the inverse dressed fermion propagator is
\be  \label{eq:fermion-dressed}
G^{-1}(i\omega,{\bf k})=G_0^{-1}(i\omega,{\bf k})+\Sigma(i\omega,{\bf k}),
\ee
with $\Sigma(i\omega,{\bf k})$ as the self-energy. After accounting for all possible disorders, we arrive at the following compact expression for the self-energy (see Appendix~\ref{detailsfieldtheory} for details)
\begin{align}\label{self-energy-final}
\Sigma(i\omega, \boldsymbol k)= i \gamma_0\omega \left(\frac{ f_1 (\Delta_j)}{\epsilon_m}\right) +
i v_0 \gamma_j k_j \left(\frac{f_2 (\Delta_j)}{3\epsilon_m}\right),
\end{align}
where
\begin{align}
f_1(\Delta_j)&= \Delta_V+\Delta_A+3\Delta_M+3 \Delta_C+3\Delta_{SO}+3 \Delta_{AM} \nonumber \\
&+\Delta_S+\Delta_{PS}, \label{f1}\\
f_2(\Delta_j)&=-\Delta_V-\Delta_A+\Delta_M+\Delta_C-\Delta_{SO}- \Delta_{AM} \nonumber \\
&+\Delta_S + \Delta_{PS},
\end{align}
with $\hat{\Delta}_j=\Delta_j k^{\epsilon_m}/(2\pi^2 v^2)$ as the dimensionless disorder strength, and for brevity we here drop the hat symbol in the final expression. From the above expression of the self-energy, together with the renormalization condition $G^{-1}(\omega,{\bf k})=Z_\Psi(i\gamma_0\omega+Z_v v i\gamma_j k_j)$, with $v$ as the renormalized Fermi velocity, we arrive at the expression for  the fermion-field renormalization $(Z_\Psi)$ and velocity renormalization $(Z_v)$
\begin{align}\label{eq:Zv}
Z_\Psi=1+\frac{f_1(\Delta_j)}{\epsilon_m}, \;
Z_v=1-\frac{1}{\epsilon_m} \left[f_1(\Delta_j) - \frac{f_2(\Delta_j)}{3} \right].
\end{align}
This equation then yields the anomalous dimension for the fermion field
\be\label{eq:eta-psi}
\eta_\Psi=-\sum_j\frac{d\ln Z_\Psi}{d\Delta_j} \beta_{\Delta_j}.
\ee
Furthermore, the renormalization factor $Z_v$ enters the renormalization condition for the Fermi velocity $Z_v v=v_0$. Using Eq.~(\ref{eq:Zv}), together with $\beta_{\Delta_N}=-\epsilon_m\Delta_N+\mathcal{O}(\Delta_j^2)$, we find
\be \label{eq:beta-v}
\beta_v=-\frac{1}{3}v\left[3f_1(\Delta_j)-f_2(\Delta_j)\right].
\ee
Finally, the $\beta-$function of the Fermi velocity is $\beta_v=(1-z)v$, which together with Eq.~(\ref{eq:beta-v}) determines the DSE
\be\label{eq:z}
z=1+\frac{1}{3} \left[ 3 f_1(\Delta_j)-f_2(\Delta_j) \right].
\ee

\subsubsection{Vertex correction: Anomalous dimension of disorder field}

We now turn to the vertex correction due to the disorder, shown in Fig.~\ref{selfvertex}(b), which yields the anomalous dimension of the disorder field. As shown in Appendix~\ref{detailsfieldtheory}, the vertex represented
by the matrix $N$ receives the correction of the form
\be \label{eq:vertex-general}
V_N({\bf k})=\sum_M \left[ M\gamma_j N \gamma_j M \right] \: \frac{\Delta_M}{3\epsilon_m}.
\ee
The corresponding renormalization condition that determines the renormalization constant $Z_{\varphi_N}$ for the disorder field  reads
\be\label{eq:Zphi}
Z_\Psi Z_{\varphi_{N}} N+V_N = N,
\ee
with $Z_\Psi$ given by Eq.~(\ref{eq:Zv}). The above condition in turn yields the anomalous dimension of the disorder field as
\be\label{eq:etaphi}
\eta_{\varphi_N}= -\sum_j\frac{d\ln Z_{\varphi_N}}{d\Delta_j} \beta_{\Delta_j},
\ee
which we then use to write the explicit form of the $\beta-$function, given by Eq.~(\ref{eq:beta-function-general}) in terms of the disorder couplings.

\subsection{$\epsilon_d$-expansion about $d=2$}~\label{epsilond-formalism}

Alternatively, one may take the Gaussian white noise distribution in Eq.~(\ref{eq:disorder-distribution-realspace}) with $m\rightarrow 0$ from the outset. In that case, the engineering dimension of the disorder coupling is equal to $2-d$, since $z=1$ in a clean WSM. Therefore, $d=2$ is the \emph{lower} critical dimension in the problem and we can use $\epsilon_d=d-2$ as an expansion parameter, following the spirit of $\epsilon$-expansion~\cite{chakravarty, roy-dassarma, radzihovsky, roy-dassarma-erratum, roy-dassarma-intdis, carpentier-1, pixley-2, zinn-justin, mirlin-2}. In this scheme, after performing the disorder averaging using the replica method, the imaginary time action assumes a similar form of Eq~(\ref{action_replica}).

\begin{figure}[t!]
\subfigure[]{
\includegraphics[width=4cm, height=4cm]{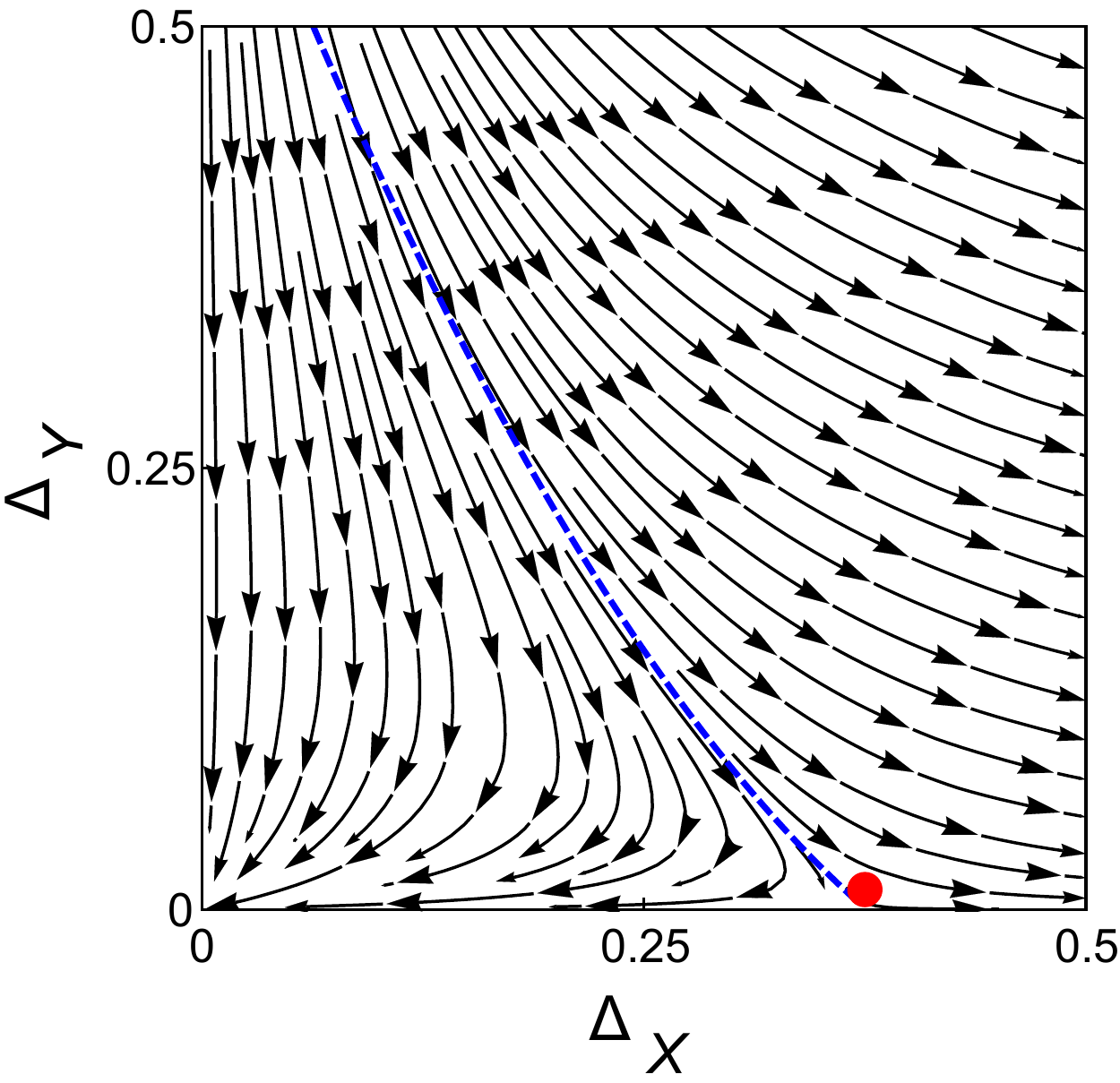}
\label{potaxmagcurflow}
}
\subfigure[]{
\includegraphics[width=4cm, height=4cm]{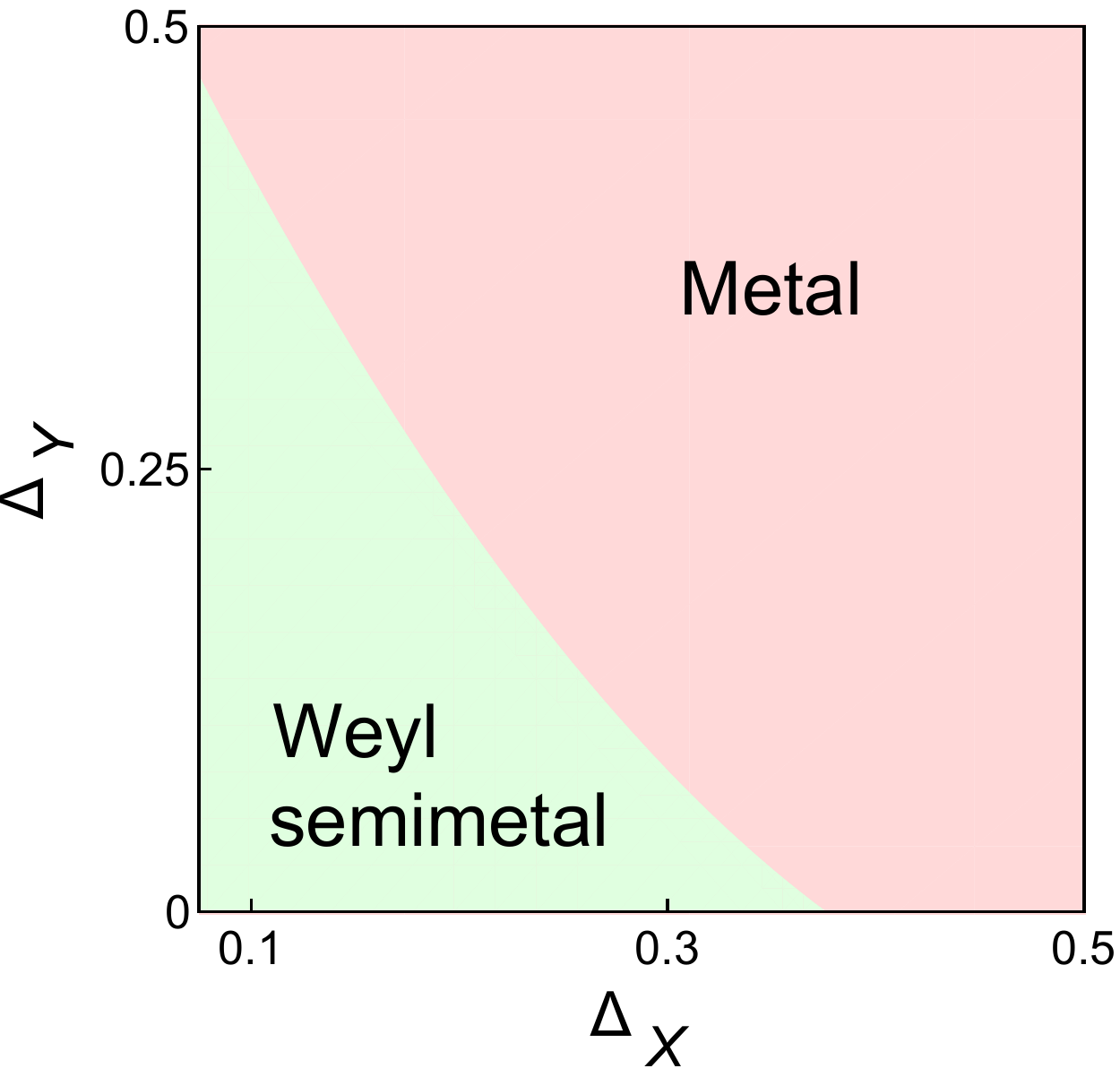}
\label{potaxmagcurphase}
}
\caption[]{(a) The renormalization group flow diagram and (b) corresponding phase diagram in the $\Delta_X-\Delta_Y$ plane, where $X=V,A$ and $Y=M,C$ obtained from Eq.~(\ref{ChiralRG_epsilon}). In these planes there is only one QCP at $\Delta_X=3\epsilon_m/8, \Delta_Y=0$ (the red dot). The phase boundary between the Weyl semimetal and metal in panel (b) is determined by the irrelevant direction, shown by blue dotted line in panel (a).  }\label{PotAxialMagCur}
\end{figure}

Within the framework of the $\epsilon_d$ expansion only the temporal (frequency-dependent) component of self energy acquires a disorder-dependent correction to the leading order. The self-energy correction due to disorder reads as
\begin{align}
\Sigma(i \omega, \boldsymbol k)=  i\gamma_0 \omega  \left(\frac{f_1(\Delta_j)}{\epsilon_d}\right),
\end{align}
with the function $f_1(\Delta_j)$ given by Eq.~(\ref{f1}), and $\Delta_j \Lambda^{\epsilon_d}/(2\pi v^2)\rightarrow\Delta_j$. This result is obtained from Eq.\ (\ref{eq:self-energy}) with $d=2+\epsilon_d$ and $m=0$. As a result, the field renormalization factor $Z_\Psi=1+ f_1(\Delta_j)/\epsilon_d$ and the velocity renormalization factor is $Z_v=1-f_1(\Delta_j)/\epsilon_d$. Using the renormalization condition $Z_v v=v_0$, together with $\beta_{\Delta_N}=-\epsilon_d\Delta_N+\mathcal{O}(\Delta_j^2)$, we obtain the leading order RG flow equation for the Fermi velocity
\begin{align}
\beta_v= v(1-z)=-v f_1 \left(\Delta_j \right),
\end{align}
which yields a scale dependent dynamic exponent $z=1+ f_1(\Delta_j)$. The seemingly different expressions for the flow equation and DSE in these two schemes stems from underlying different methodology of capturing the ultraviolet divergences of various diagrams. However, such details do not alter any physical outcome.  While extracting the RG flow of all disorder couplings, we first complete the $\gamma$ matrix algebra in $d=3$ and subsequently perform the momentum integral in $d=2+\epsilon$. Such procedure is safe at least to the leading order in $\epsilon_d$-expansion as the relevant Feynman diagrams [see Fig.~\ref{laddercrossing}] do not contain any \emph{overlapping divergence}. For next to the leading order calculation one also needs to perform the $\gamma$-matrix algebra in $d=2+\epsilon$. However, in the $\epsilon_m$-expansion scheme we do not need to continue the $\gamma$ matrix algebra in general dimension, as the entire analysis is performed in $d=3$.

\section{Chiral symmetric or intra-node disorder}~\label{CSP_disorder}

We first focus on chiral-symmetric disorders. For a single pair of Weyl fermions there are four such disorders, namely chemical potential, axial potential, current and axial current disorders, as shown in Table \ref{table-disorder}. With appropriate lattice model axial current disorder corresponds to magnetic impurities and from here onward we use this terminology. We will address the effect of weak and strong chiral symmetric disorder using both $\epsilon_m$ and $\epsilon_d$ expansions.

\subsection{$\epsilon_m$ expansion}

Let us first analyze this problem pursuing the $\epsilon_m$ expansion. Using Eqs.~(\ref{eq:beta-function-general}), (\ref{eq:z}), (\ref{eq:Zphi}) and (\ref{eq:etaphi}), we obtain the following RG flow equations for the coupling constants to the leading order in $\epsilon_m$
\begin{align}\label{ChiralRG_epsilon}
\beta_{\Delta_V}&=\Delta_V \left[-\epsilon_m + \frac{8}{3} \left( \Delta_V + \Delta_A \right) + \frac{16}{3} \left( \Delta_C + \Delta_M \right) \right], \nonumber \\
\beta_{\Delta_A}&=\Delta_A \left[-\epsilon_m + \frac{8}{3} \left( \Delta_V + \Delta_A \right) + \frac{16}{3} \left( \Delta_C + \Delta_M \right) \right], \nonumber \\
\beta_{\Delta_M}&= -\epsilon_m \; \Delta_M, \: \: \beta_{\Delta_C}= -\epsilon_m \; \Delta_C.
\end{align}
The above set of flow equations supports a \emph{line of quantum critical points} in the $\Delta_V-\Delta_A$ plane, determined by
\begin{equation}\label{LineQCP_chiral}
\Delta_{V,\ast}+\Delta_{A,\ast}=\frac{3}{8} \epsilon_m,
\end{equation}
where the quantities with subscript ``$\ast$" represent their critical values for WSM-metal QPT. The RG flow in this plane is shown in Fig.~\ref{potaxialflow}. The line of QCPs also determines the WSM-metal phase boundary, and the corresponding phase diagram in the $\Delta_V-\Delta_A$ plane is shown in Fig.~\ref{potaxialphase}. At each point of this line of QCPs the DSE and CLE are respectively given by
\begin{equation}\label{eq:exponents-chiral}
z=1+\frac{\epsilon_m}{2} + {\mathcal O}(\epsilon^2_m), \: \nu^{-1}= \epsilon_m + {\mathcal O}(\epsilon^2_m).
\end{equation}
Therefore, for the Gaussian white noise distribution, realized for $\epsilon_m=1$, we obtain $z=3/2$ and $\nu=1$ from the leading order $\epsilon_m$ expansion. If the bare value of either the chemical potential or axial potential disorder strength is zero, the quantum-critical behavior is governed by the QCP corresponding to the disorder of a nonvanishing bare value~\cite{pallab-sudip2016}. This QCP features the critical exponents of the same value to the one-loop order as in the case of the quantum-critical line, given by Eq.~(\ref{eq:exponents-chiral}).

\begin{figure}[t!]
\subfigure[]{
\includegraphics[width=4cm, height=4cm]{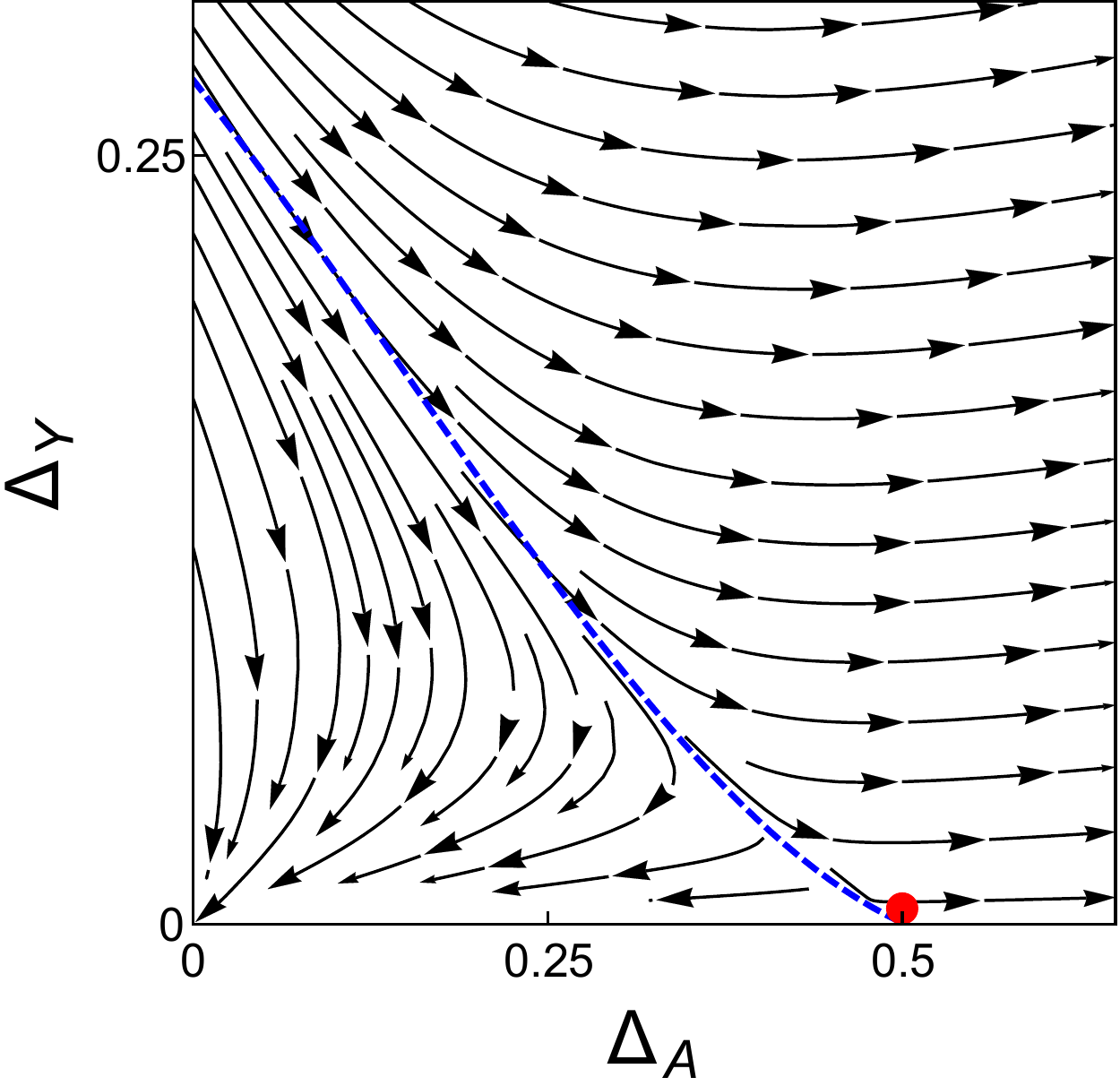}
\label{axialmagcurflow}
}
\subfigure[]{
\includegraphics[width=4cm, height=4cm]{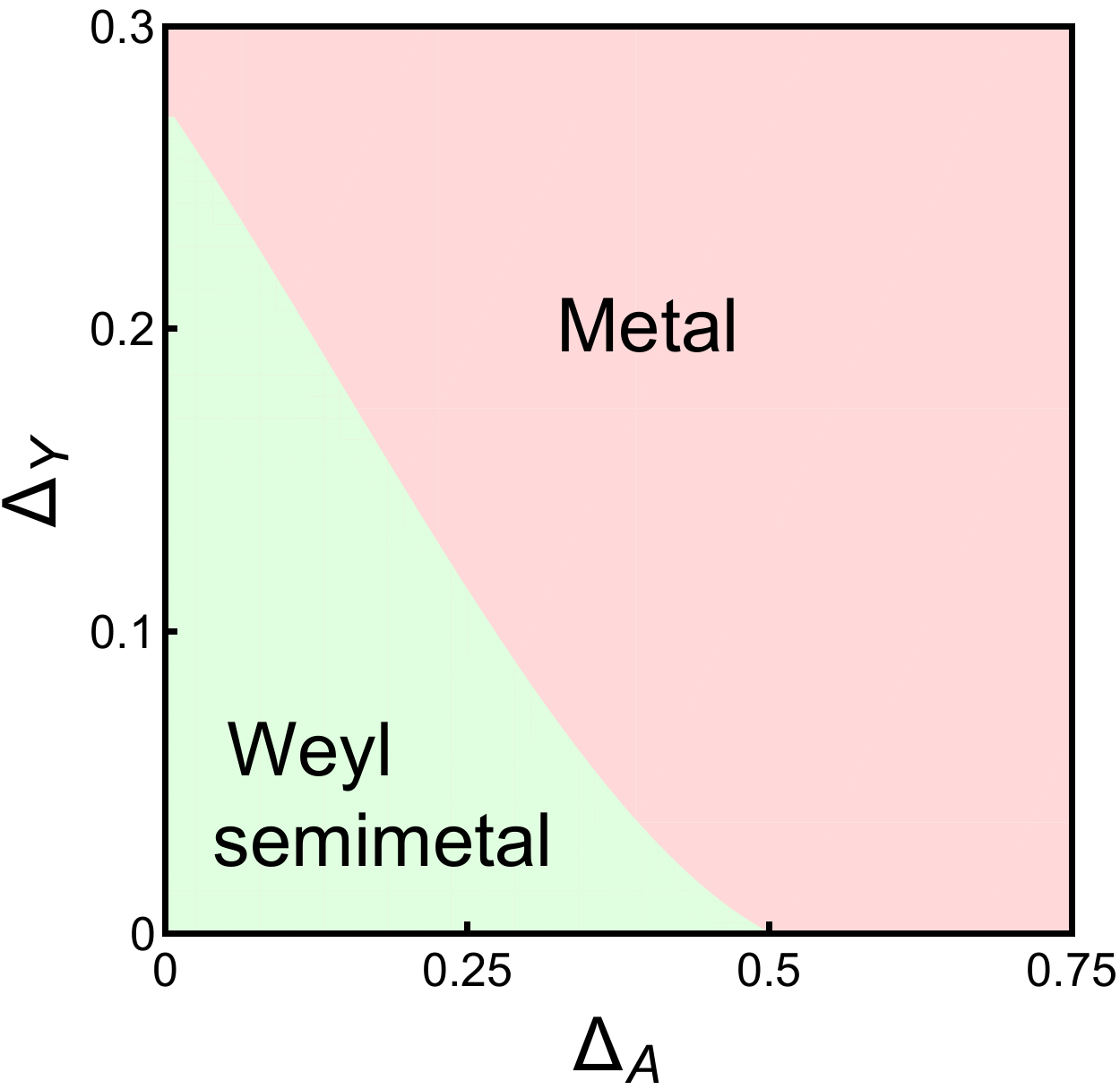}
\label{axialmagcurphase}
}
\caption[]{(a) The renormalization group flow diagram and (b) corresponding phase diagram in the $\Delta_A-\Delta_Y$ plane, where $Y=M,C$ obtained from $\epsilon_d$ expansion for $\epsilon_d=1$. There is only one quantum critical point at $\Delta_A=\epsilon_d/2, \Delta_Y=0$ (the red dot). The phase boundary between the Weyl semimetal and metal in panel (b) is determined by the irrelevant direction, shown by the blue dotted line in panel (a). These figures are qualitatively similar to the ones shown in Fig.~\ref{PotAxialMagCur}, apart from the nonuniversal shift in the phase boundary. }\label{AxialMagCur}
\end{figure}

From the RG flow equations [see Eq.~(\ref{ChiralRG_epsilon})], we find that both magnetic and current disorder are always \emph{irrelevant} perturbations, at least to the leading order in the $\epsilon_m$-expansion. In the $\Delta_X-\Delta_Y$ plane, where $X=V,A$ and $Y=M,C$ the RG flow diagram is shown in Fig.~\ref{potaxmagcurflow} and the corresponding phase diagram is shown in Fig.~\ref{potaxmagcurphase}. Importantly, the QPT separating the metallic and the semimetallic phase in any $\Delta_X-\Delta_Y$ plane is governed by the QCP located at $\Delta_{X,\ast}=3 \epsilon_m/8$. The phase boundary between these two phases is determined by the \emph{irrelevant} direction at this QCP. Therefore, across the entire WSM-metal phase boundary in these planes the universality class of the QPT is identical and characterized by $z=1+\epsilon_m/2 + {\mathcal O}(\epsilon^2_m)$ and $\nu^{-1}=\epsilon_m+ {\mathcal O}(\epsilon^2_m)$ to the leading order in $\epsilon_m$-expansion.

\subsection{$\epsilon_d$ expansion}

The RG flow equations for the chiral symmetric disorder coupling constants within the framework of the leading order $\epsilon_d$-expansion are
\begin{align}\label{chiralRG_hard}
\beta_{\Delta_V}&= \Delta_V \left[ -\epsilon_d + 2 F_+ (\Delta_j) \right] + 8 \Delta_M \Delta_C \nonumber \\
\beta_{\Delta_A}&= \Delta_A \left[ -\epsilon_d + 2 F_+ (\Delta_j) \right] +  4 \left( \Delta^2_M+\Delta^2_C \right) \\
\beta_{\Delta_M}&= \Delta_M \left[ -\epsilon_d + \frac{2}{3} F_- (\Delta_j) \right] + \frac{8}{3} \left( \Delta_C\Delta_V +\Delta_A \Delta_M \right) \nonumber \\
\beta_{\Delta_C}&= \Delta_C \left[ -\epsilon_d + \frac{2}{3} F_- (\Delta_j) \right] + \frac{8}{3} \left( \Delta_C\Delta_V +\Delta_A \Delta_M \right), \nonumber
\end{align}
where $F_\pm (\Delta_j)=\left( \Delta_V+\Delta_A \right) \pm \left( \Delta_C+\Delta_M \right)$. These coupled flow equations also support only a line of QCPs in the $\Delta_V-\Delta_A$ plane, as we previously found from Eq.~(\ref{ChiralRG_epsilon}) using $\epsilon_m$-expansion, now determined by
\begin{align}~\label{LineQCP_chiral_hard}
\Delta_{V,\ast} + \Delta_{A,\ast} =\frac{\epsilon_d}{2},
\end{align}
similar to the one  in Eq.~(\ref{LineQCP_chiral}). The critical exponents at each point of such line of QCPs are $z=1+\epsilon_d/2 + {\mathcal O}(\epsilon^2_d)$ and $\nu^{-1}=\epsilon_d + {\mathcal O}(\epsilon^2_d)$. We here stress that presently there is no known method to compute these two exponents beyond leading order in $\epsilon_d$ in a controlled fashion~\cite{roy-dassarma-erratum, carpentier-1}. Therefore, in three spatial dimensions $\epsilon_d=1$ and we find $z=3/2$ and $\nu=1$~\cite{chakravarty,roy-dassarma-intdis}. The RG flow diagram and the corresponding phase diagram are similar to the ones shown in Figs.~\ref{potaxialflow} and \ref{potaxialphase}. Only the location of the line of QCPs and the phase boundary shift in a nonuniversal fashion. The differences in the flow equations [~(\ref{LineQCP_chiral}) and ~(\ref{chiralRG_hard})], arise from two diagrams shown in Fig.~\ref{laddercrossing} (c) and (d), which produce ultraviolet divergent contributions, but only within the $\epsilon_d$ expansion scheme. In the presence of only potential disorder we find $z=3/2$ and $\nu=1$~\cite{chakravarty, roy-dassarma, radzihovsky, roy-dassarma-erratum, roy-dassarma-intdis, carpentier-1}.

\begin{figure*}
\includegraphics[width=17cm,height=3.25cm]{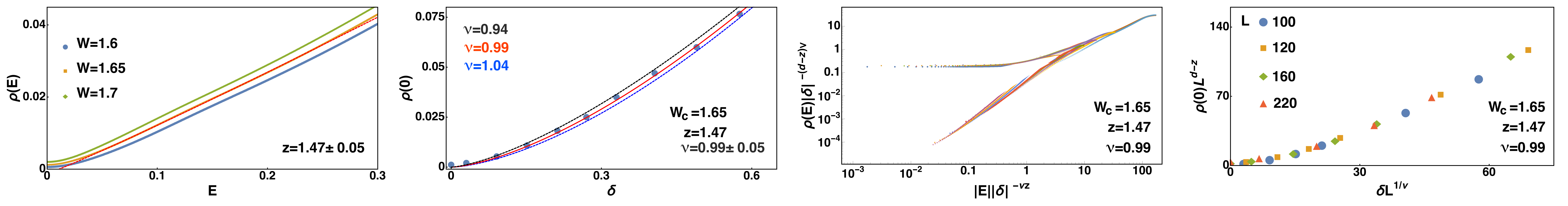}
\includegraphics[width=17cm,height=3.25cm]{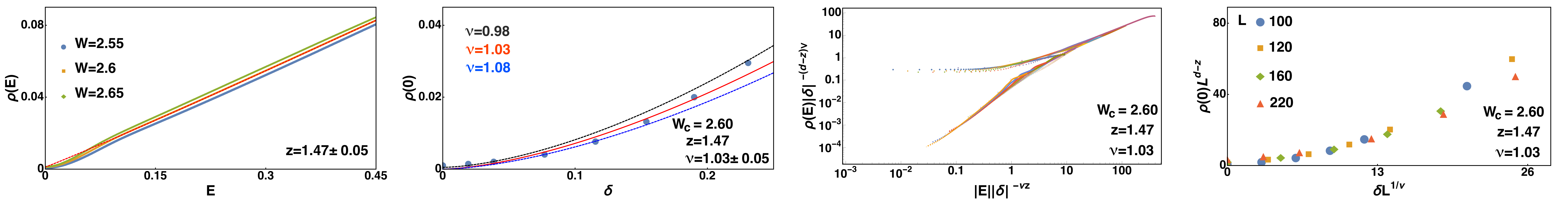}
\includegraphics[width=17cm,height=3.25cm]{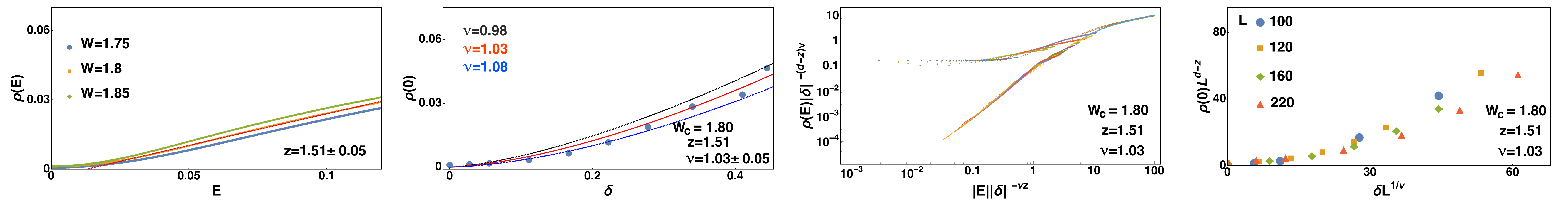}
\includegraphics[width=17cm,height=3.25cm]{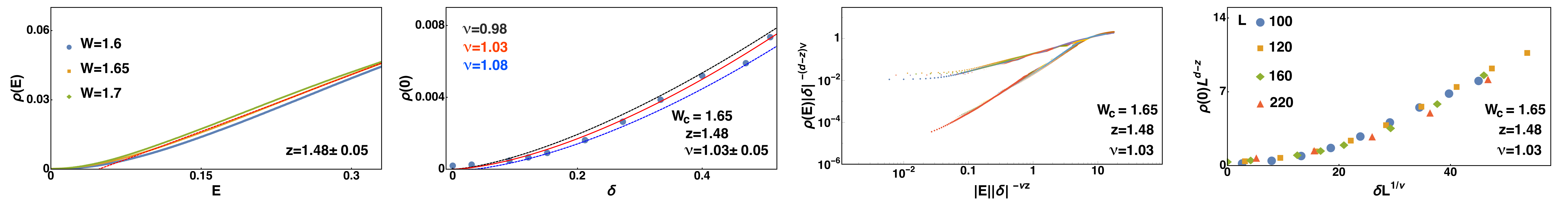}
\caption{Scaling analysis of average density of states (ADOS) in various regimes of the phase diagram of a dirty WSM for all four possible intranode scatterings; plots from top to bottom rows correspond to potential ($\Delta_V$), axial potential ($\Delta_A$), axial current ($\Delta_M$) and current ($\Delta_C$) disorder. 
First column shows the scaling of ADOS $\varrho(E)$ vs. $E$ around the critical strength of disorder ($W=W_c$). The second column depicts the scaling of ADOS at zero energy $\varrho(0)$ vs. $\delta$, the reduced distance from the critical disorder defined as $\delta=\frac{W-W_c}{W_c}$. In the third column we display $\varrho(E) \delta^{-(d-z)\nu}$ vs. $|E||\delta|^{-\nu z}$ for weak ($W<W_c$) and strong ($W>W_c$) disorder and $|E| \ll t(=1)$. All data collapse onto two branches. The top branch represents the metallic phase, while the lower branch represents WSM. Note that these two branches meet at large values of $|E||\delta|^{-\nu z}$, corresponding to the quantum critical regime. All data in first three columns are obtained from a system of linear dimension $L=220$. The finite size data collapse inside the metallic phase is shown in the forth column, where we compare $\varrho(0)L^{d-z}$ vs. $\delta L^{1/\nu}$ for $100 \leq L \leq 220$. Notice that all data collapse onto one branch for small to moderate values of $\delta L^{1/\nu}$, with the numerically extracted values of the critical exponents $z$ and $\nu$, quoted in the figure and summarized in Table~\ref{table-exponent}. The quality of the data collapse progressively worsens for larger values of $\delta L^{1/\nu}$ due to the existence of a second QPT of a three-dimensional dirty Weyl metal into the Anderson insulating phase, discussed in Sec.~\ref{anderson}. Scaling of ADOS and data analysis are discussed in details in Sec.~\ref{numerics_analysis}. The quoted error bars in $z$ and $\nu$ are \emph{fitting error bars}. See Appendix~\ref{Append:data_analysis} and Table~\ref{Table:Data-analysis} (first four rows) for details of data analysis.
}~\label{numeric_analysis_figure}
\end{figure*}

\begin{figure*}
\includegraphics[width=17cm,height=3.25cm]{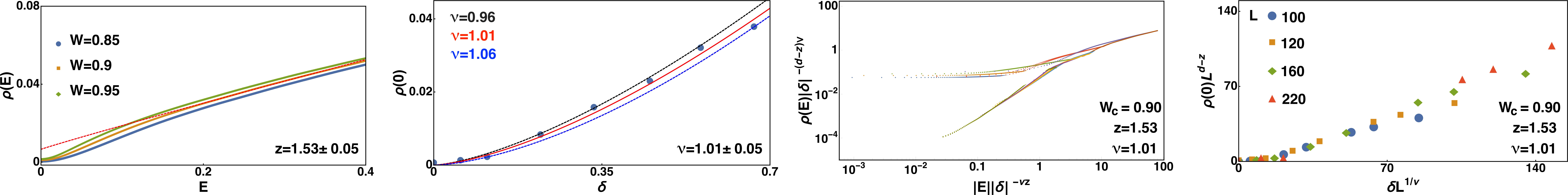}
\includegraphics[width=17cm,height=3.25cm]{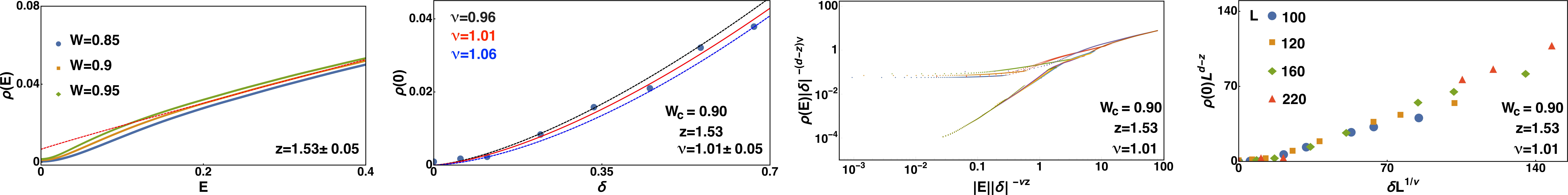}
\includegraphics[width=17cm,height=3.25cm]{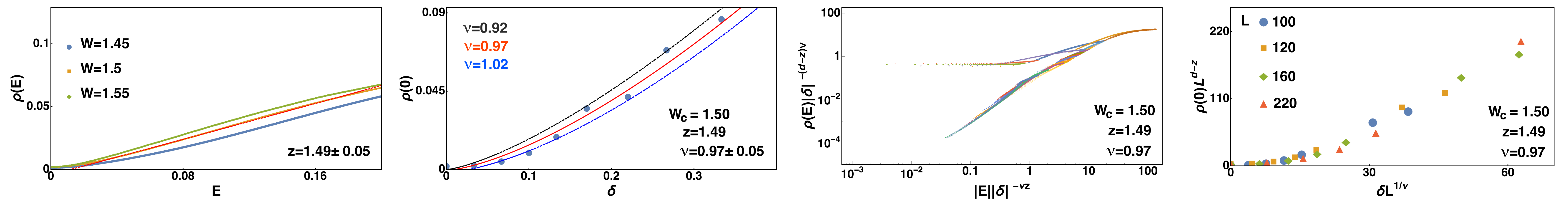}
\includegraphics[width=17cm,height=3.25cm]{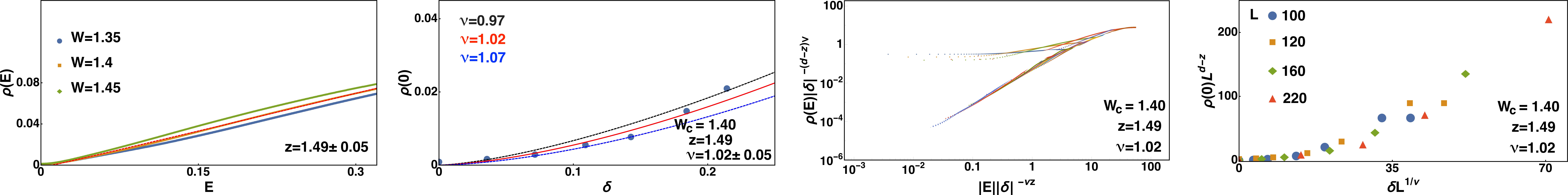}
\caption{ Scaling analysis of numerically extracted ADOS in various regimes of the phase diagram of a dirty WSM in the presence of inter-node scattering. Each column is identical to the corresponding one in Fig.~\ref{numeric_analysis_figure} [including methods of analysis and system size]. The plots from top to bottom rows correspond to temporal ($\Delta_{SO}$) and spatial ($\Delta_{AM}$) component of tensor, scalar ($\Delta_S$) and pseudo scalar ($\Delta_{PS}$) mass disorder [see Table~\ref{table-disorder}]. 
Final results of our analysis are quoted in Table~\ref{Tab:CSB_exponents}. The quoted error bars in $z$ and $\nu$ are \emph{fitting error bars}. See also Appendix~\ref{Append:data_analysis} and Table~\ref{Table:Data-analysis} (from 4$^{\rm th}$-8$^{\rm th}$ row) for additional details.
 }~\label{numeric_analysis_figure_CSB}
\end{figure*}

Notice that if we start with only magnetic or current disorder, the axial disorder gets generated from Feynman diagrams (c) and (d) in Fig.~\ref{laddercrossing}. Thus, to close the RG flow equations, we need to account for $\Delta_A$ coupling from the outset, and the resulting RG flow equations read
\begin{align}
\beta_{\Delta_A}&= \Delta_A \left[ -\epsilon_d + 2 \left(\Delta_A +3 \Delta_Y \right) \right] + 4 \Delta^2_Y \nonumber \\
\beta_{\Delta_Y}&= \Delta_Y \left[ -\epsilon_d + \frac{2}{3} \left( \Delta_Y-\Delta_A \right) \right] + \frac{8}{3} \Delta_A \Delta_Y,
\end{align}
for $Y=M, C$. The above set of coupled RG flow equations supports only one QCP, located at $\Delta_{A,\ast}=\epsilon_d/2, \Delta_Y=0$. The RG flow and the resulting phase diagrams are shown in Figs.~\ref{axialmagcurflow} and ~\ref{axialmagcurphase}, respectively. Hence, in the presence of magnetic and current disorder the transition to the metallic phase is controlled by the QCP due to axial disorder. If we also take into account the presence of potential disorder, then such a semimetal-metal QPT takes place through one of the points residing on the line of QCPs in the $\Delta_V-\Delta_A$ plane, depending on the bare relative strength of these two disorder couplings.

\subsection{Chiral superuniversality}

From the discussion in previous two subsections, we can conclude that in the presence of chiral-symmetric disorder in a WSM, the semimetal-metal QPT takes place either through a QCP or a line of QCPs. The location of the line of QCPs and the resulting phase boundaries are nonuniversal and thus dependent on the RG scheme. However, the universal quantum critical behavior with chiral symmetric disorder couplings is insensitive of these details, at least to the leading order in the expansion parameter, and all QPTs in the four-dimensional hyperplane of disorder coupling constants, are characterized by an identical set of critical exponents, namely $z=1+\epsilon/2 +{\mathcal O} (\epsilon^2)$ and $\nu^{-1}=\epsilon+{\mathcal O} (\epsilon^2)$, with $\epsilon=1$. The importance of the higher order corrections is presently unknown. Therefore, emergent quantum critical behavior for strong chiral-symmetric disorder stands as a rare example of superuniversality, and we name it \emph{chiral superuniversality}. Next we demonstrate emergence of such superuniversality across WSM-metal QPT by \emph{numerically} analyzing the scaling of average DOS in the presence of generic chiral symmetric disorder.

\section{Numerical demonstration of chiral superuniversality}~\label{numerics_analysis}

Motivated by the field-theoretic prediction of emergent chiral superuniversality across the WSM-metal QPTs driven by CSP disorder, next we numerically investigate the scaling of average DOS across such QPTs. Since $\varrho(0)$ vanishes and is finite in the WSM and metallic phases, respectively, it can be promoted as a bonafide order-parameter across the WSM-metal QPT~\cite{herbut-disorder, pixley-1, pixley-2, roy-bera, ohtsuki, pixley-4}. In addition, such analysis endows an opportunity to extract the critical exponents for the transition non-perturbatively and, at the same time, test the validity of the proposed scenario for chiral superuniversality. The WSM phase is realized from the tight-binding model, defined through Eqs.~(\ref{wilson}) and (\ref{TB_Weyl}), which we implement on a cubic lattice of linear dimension $L$. For numerical analysis we always set $m_z=0$, and for current disorder take $t=t_z=1=t_0$, while $t=1=t_0, t_z=\frac{1}{2}$ for remaining seven types of elastic scatterers [see Table~\ref{table-disorder}], in the clean model, given by Eqs.~(\ref{two-band-WSM})-(\ref{TB_Weyl}). We use lattice realizations of disorder introduced in Appendix~\ref{disorder_lattice}. We impose periodic boundary condition in all three directions. The average DOS is computed by using the \emph{kernel polynomial method}~\cite{KPM-RMP}. The average is taken over 20 random realization of disorder that minimizes the residual statistical error in average DOS, which is a self-averaged quantity. We typically compute 4096 Chebyshev moments and take trace over $\sim$ 12 random vectors to compute average DOS. All types of disorder are distributed uniformly and randomly within the range $[-W, W]$. The scaling theory for average DOS has already been discussed in Sec.~\ref{MCP_numerics_analysis}. Thus, we can readily start from the final expression of the general scaling form of the average DOS, presented in Eq.~(\ref{DOS_Scaling_numerics}) and continue with our numerical analysis.

\subsection{Numerical analysis with random intra-node scatterers or chiral-symmetric disorder}~\label{chiral-numerics}

We begin the discussion on the effects of randomness on WSM by focusing on the intra-node or chiral symmetric disorder. Let us first focus on the quantum critical regime and for now we assume that the system size is sufficiently large so that we can neglect the $L$-dependence in Eq.~(\ref{DOS_Scaling_numerics}). In this regime the scaling function must be independent of $\delta$, dictating $F(x) \sim x^{\frac{d}{z}-1}$. Therefore, when $W = W_c$ we compare $\varrho(E)$ vs. $E^{\frac{d}{z}-1}$ and extract the DSE $z$. Such analysis for all four possible CSP disorders is shown in the first column of Fig.~\ref{numeric_analysis_figure} and numerically extracted values of $z$ are quoted in Table~\ref{table-exponent}. Within the numerical accuracy, we always find $z \approx 1.5$ in excellent agreement with the field-theoretic result, obtained from the leading order $\epsilon$ expansions.

Next we proceed to the metallic side of the transition, where average DOS at zero energy becomes finite. From the scaling function in Eq.~(\ref{DOS_Scaling_numerics}), we obtain $\varrho(0) \sim \delta^{(d-z) \nu}$. Thus by comparing $\varrho(0)$ vs. $\delta$, we extract the CLE $\nu$, using already obtained value of the DSE $z$, as shown in the second column of Fig.~\ref{numeric_analysis_figure}. The numerically found CLE is also quoted in Table~\ref{table-exponent}, and within numerical accuracy $\nu \approx 1$ always, irrespective of the nature of CSP disorder. Once again we find an excellent agreement of numerically extracted values of $\nu$ with the one obtained from the leading order $\epsilon$-expansions. These two results strongly support the picture of chiral superuniversality.

To test the quality of our numerical analysis we search for two types of data collapse. First, we compare $\varrho(E) |\delta|^{-\nu(d-z)}$ vs. $|\delta|^{-\nu z} |E|$, motivated by the scaling form of average DOS, displayed in Eq.~(\ref{DOS_Scaling_numerics}). Using numerically obtained values of $\nu$ and $z$, we find that for energies much smaller than the bandwidth ($|E| \ll 1$), all data collapse onto two separate branches for all four disorders, as shown in the third column of Fig.~\ref{numeric_analysis_figure}. While the top branch corresponds to the metallic phase, the lower one stems from the WSM phase and eventually these two branches meet in the quantum critical regime.

Finally, we demonstrate a finite size data collapse for $\varrho(0)$ for different system sizes ($L$) by focusing on the metallic side of the transition. Setting $E=0$ in Eq.~(\ref{DOS_Scaling_numerics}), we obtain $\varrho(0)=L^{z-d} F(0,\delta L^{1/\nu})$. Hence, we compare $\varrho(0) L^{d-z}$ vs. $\delta L^{1/\nu}$ and find an excellent data collapse for $100<L<220$, using numerically obtained values of $\nu$ and $z$ for all four disorders, as shown in the fourth column of Fig.~\ref{numeric_analysis_figure}. The data collapse becomes systematically worse for large values of $\delta$ or stronger disorder due to the existence of a second transition that takes the system from a metallic phase to an Anderson insulator. Therefore, our thorough numerical analysis provides a valuable and unprecedented insight into the nature of the WSM-metal QPTs driven by generic chiral symmetric disorder, and staunchly supports the proposal of an emergent chiral superuniversality across such QPTs.

Finally, we note that one can attempt to extract the CLE ($\nu$) from the scaling of ADOS at finite energy in the semimetallic side of the transition in the following way. In the WSM phase the universal scaling function (after neglecting the $L$-dependence) $F(x) \sim x^{d-1}$, yielding $\varrho(E) \sim \delta^{(1-z)d \nu} |E|^{d-1}$, see Eq.~(\ref{DOS_Scaling_numerics}) for sufficiently small energy. By contrast, for moderately high-energy (still $|E| \ll 1$) $\varrho(E) \sim |E|$ inside the critical regime. Therefore, by tracking the scaling of the \emph{crossover} boundary between the WSM (displaying $\varrho(E) \sim |E|^2$) and critical regime (displaying $\varrho(E) \sim |E|^2$) at finite energy for subcritical disorder one can extract the CLE $\nu$. However, determination of such crossover boundary does not rest on any strict criterion and is often (if not always) associated with a large \emph{error}, which in turn produces a large error bar in the determination of CLE~\cite{herbut-disorder, pixley-2, ohtsuki, roy-bera}. Therefore, this methodology of determining $\nu$ and corresponding error bar is questionable.

\begin{figure}[t!]
\subfigure[]{
\includegraphics[width=4cm, height=4cm]{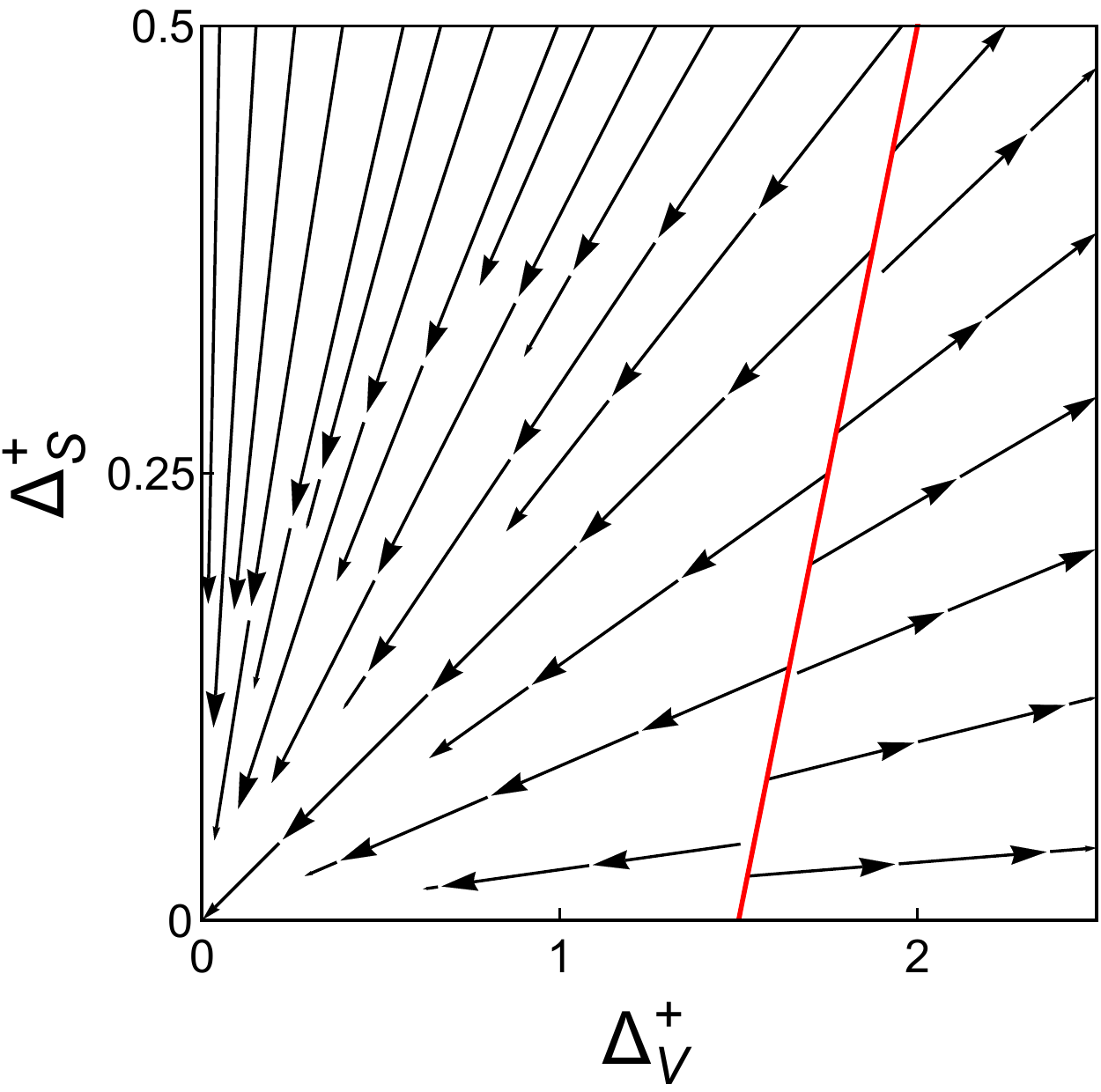}
\label{CSBDisorder_flow}
}
\subfigure[]{
\includegraphics[width=4cm, height=4cm]{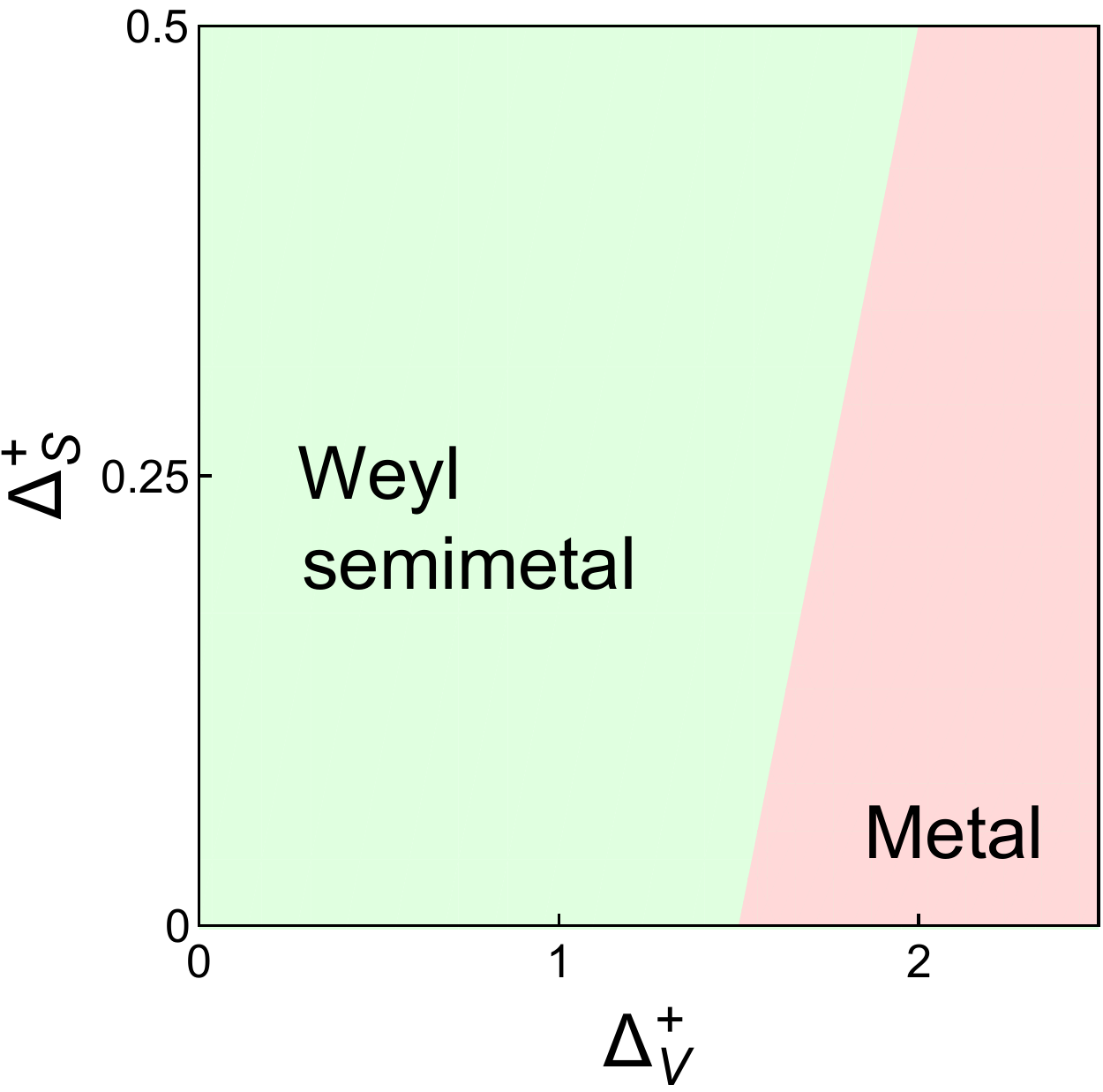}
\label{CSBDisorder_Phase}
}
\caption[]{(a) The renormalization group flow diagram and (b) corresponding phase diagram in the $\Delta^{+}_{V}-\Delta^{+}_{S}$ plane obtained from $\epsilon_m$ expansion for $\epsilon_m=1$. The WSM-metal QPT in this coupling constant space is controlled by the line of QCPs [see Eq.~(\ref{LQCP_CSB})], shown by the red line in panel (a) that in turn also determines the phase boundary between these two phases, as shown in panel (b). }\label{CSBDisorder}
\end{figure}

\subsection{Numerical analysis with random inter-node scatterers or non-chiral disorder}~\label{non-chiral-numerics}

Motivated by the intriguing possibility of realizing an \emph{emergent superuniversality} we further seek to examine its robustness in the presence of inter-node scattering (also referred as non-chiral disorder). In the simplest version of a Weyl semimetal comprised of only two Weyl nodes there are four sources of internode scattering, highlighted in Table~\ref{table-disorder}, and their lattice realization is shown in Appendix~\ref{disorder_lattice}. We rely on the scaling of average DOS in the presence of non-chiral disorder as well, and all the parameters and numerical strategies are identical to the ones pursued for chiral symmetric (intranode) disorder. The analyses of average DOS in various regimes of the phase diagram of disordered WSM are performed in the same fashion. The locations of WSM-metal QPT are shown in Fig.~\ref{rho0vsW} (lower row), and numerically extracted values of two critical exponents $\nu$ and $z$ are reported in Table~\ref{Tab:CSB_exponents}. The details of the data analysis are displayed in Fig.~\ref{numeric_analysis_figure_CSB}.

Within the numerical accuracy we find that the WSM-metal QPT driven by CSB disorder is also characterized by $\nu \approx 1$ and $z \approx 1.5$. \emph{Therefore, the chiral superuniversality appears to be generic in a dirty WSM, and the WSM-metal QPTs belong to the same universality class, irrespective of the nature of impurities.} Such an intriguing outcome further motivates us to understand the effect of internode scattering in a WSM from a field theoretic point of view, which we present in the following section by carrying out two different $\epsilon$-expansions, described in Secs.~\ref{epsilonm-formalism} and ~\ref{epsilond-formalism}.

\section{Chiral symmetry breaking or inter-node disorder}~\label{CSB_disorder}

In a WSM constituted by a single pair of Weyl nodes, there are four CSB disorders, namely temporal and spatial components of a tensor disorder, which in a suitable lattice model respectively represents spin-orbit and  axial magnetic disorder, as well as scalar and pseudoscalar mass disorder, see Table~\ref{table-disorder}. We will address the effects of weak and strong CSB disorder by using both $\epsilon_m$ and $\epsilon_d$ expansions.

\subsection{$\epsilon_m$ expansion}

\begin{figure}
\subfigure[]{
\includegraphics[width=4cm, height=4cm]{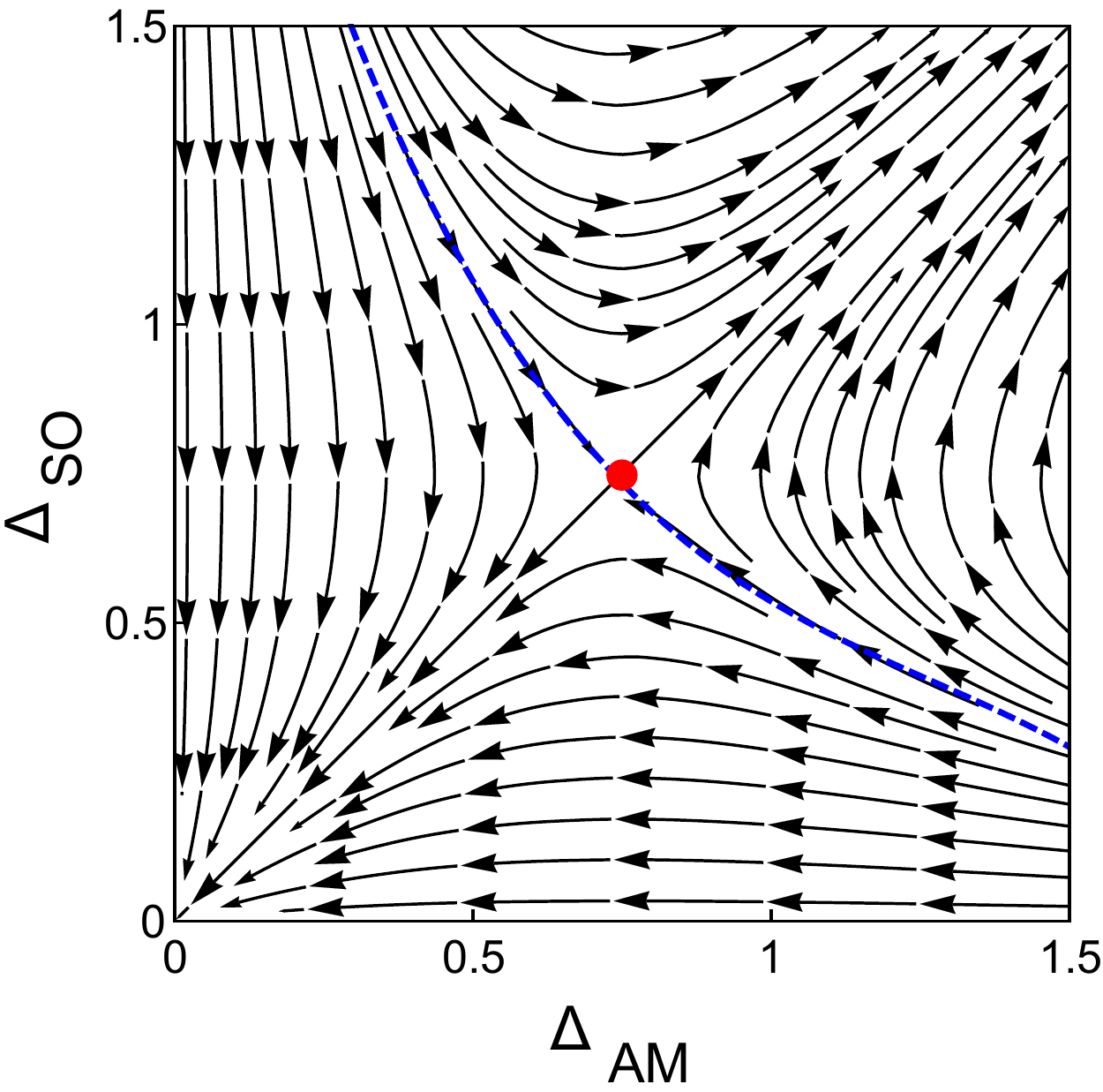}
\label{SpOrAxMg_flow}
}
\subfigure[]{
\includegraphics[width=4cm, height=4cm]{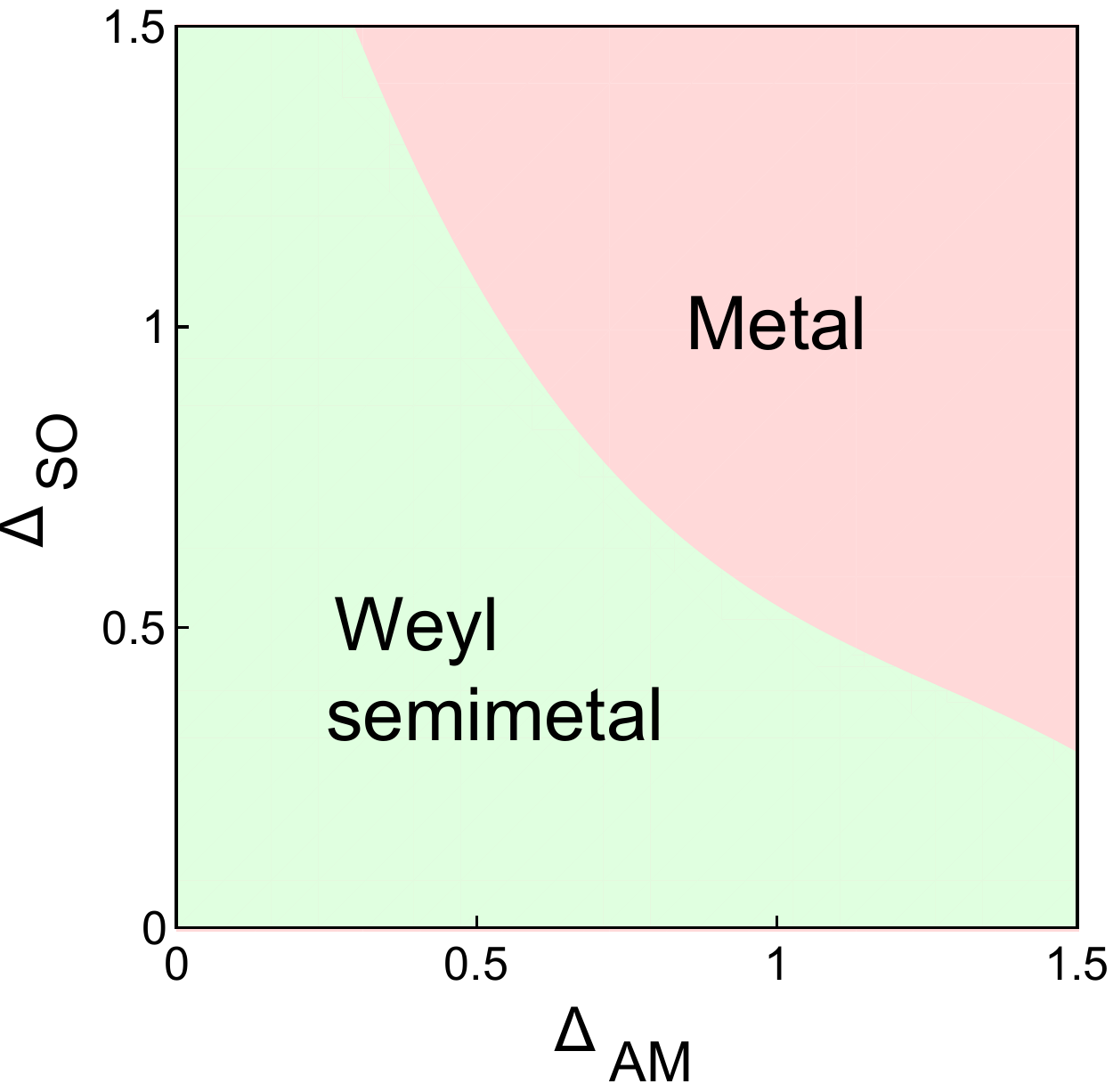}
\label{SpOrAxMg_Phase}
}
\caption[]{(a) The renormalization group flow diagram and (b) corresponding phase diagram in the $\Delta_{AM}-\Delta_{SO}$ plane obtained from $\epsilon_m$ expansion for $\epsilon_m=1$. There is only one quantum critical point at $\Delta_{AM}=\Delta_{SO}=3\epsilon_m/4$ (the red dot). The phase boundary between the Weyl semimetal and metal in panel (b) is determined by the irrelevant direction, shown by blue dotted line in panel (a). }\label{SpOrAxMg}
\end{figure}

 Within the framework of an $\epsilon_m$ expansion the RG flow equations to one-loop order read as
\allowdisplaybreaks[4]
\begin{align}
\beta_{\Delta_{SO}}&=\Delta_{SO} \left[ -\epsilon_m + \frac{4}{3} \left( \Delta_{AM}-\Delta_S \right) \right], \nonumber \\
\beta_{\Delta_{AM}}&=\Delta_{AM} \left[ -\epsilon_m + \frac{4}{3} \left( \Delta_{SO}-\Delta_{P} \right) \right], \\
\beta_{\Delta_{S}} &=\Delta_{S} \left[ -\epsilon_m + \frac{4}{3} \left(5 \Delta_{SO}-4 \Delta_{AM}-2 \Delta_S+\Delta_{PS} \right) \right], \nonumber \\
\beta_{\Delta_{PS}}&=\Delta_{PS} \left[ -\epsilon_m + \frac{4}{3} \left(5 \Delta_{AM}-4 \Delta_{SO}-2 \Delta_{PS}+\Delta_{S} \right) \right]. \nonumber
\end{align}
Therefore, individually each CSB disorder is always an irrelevant perturbation, at least to the leading order in the $\epsilon_m$ expansion, and as such does not lead to any QPTs. However, in the absence of chiral symmetry all four disorder couplings are present and to address the critical properties in this situation we recast the above flow equations in terms of newly defined coupling constants as
\begin{align}
\beta_{\Delta^{+}_V}&=-\epsilon_m \Delta^{+}_V +\frac{2}{3} \left[ \left( g^{+}_V \right)^2-\left( \Delta^{-}_V \right)^2 - \Delta^{+}_V \Delta^{+}_S - \Delta^{-}_V \Delta^{-}_S \right] \nonumber \\
\beta_{\Delta^{-}_V}&=-\epsilon_m \Delta^{-}_V -\frac{2}{3} \left[ \Delta^{-}_V \Delta^{+}_S + \Delta^{+}_V \Delta^{-}_S \right]  \\
\beta_{\Delta^{+}_S}&=-\epsilon_m \Delta^{+}_S -\frac{2}{3} \left[ \left( \Delta^{+}_S \right)^2 + 3 \left( \Delta^{-}_S \right)^2 - \Delta^{+}_V \Delta^{+}_S -9 \Delta^{-}_V \Delta^{-}_S  \right] \nonumber \\
\beta_{\Delta^{-}_S}&=-\epsilon_m \Delta^{-}_S-\frac{2}{3} \left[ \Delta^{-}_S \Delta^{+}_V-4 \Delta^{+}_S \Delta^{-}_S -9 \Delta^{+}_S \Delta^{-}_V  \right], \nonumber
\end{align}
where $\Delta^{\pm}_V=\Delta_{SO} \pm \Delta_{AM}, \: \Delta^{\pm}_S=\Delta_{S} \pm \Delta_{PS}$. The above set of RG flow equations supports a line of QCPs determined by the equation
\begin{equation}~\label{LQCP_CSB}
\Delta^{+}_{V,\ast}= \Delta^{+}_{S,\ast} + \frac{3 \epsilon_m}{2}, \; \Delta^{-}_{V,\ast}=0, \; \Delta^{-}_{S,\ast}=0.
\end{equation}
Notice that if we tune the CSB disorders, so that $\Delta^{-}_V=\Delta^{-}_S=0$, these two coupling constants do not get generated through quantum corrections, and the plane with $\Delta^{-}_V=\Delta^{-}_S=0$, shown in Fig.~\ref{CSBDisorder}, remains invariant under the RG. The RG flow in this plane is shown in Fig.~\ref{CSBDisorder_flow}, and the corresponding phase diagram is presented in Fig.~\ref{CSBDisorder_Phase}. The WSM-metal phase boundary in the $\Delta^{+}_V-\Delta^{+}_S$ plane is determined by the line of QCPs, given by Eq.~(\ref{LQCP_CSB}), qualitatively similar to the situation in the presence of potential and axial disorders, as shown in Fig.~\ref{Potential_Axial}. However, these two scenarios are fundamentally different in the sense that while the DSE $z=1+\epsilon/2$, with $\epsilon=\epsilon_m$ or $\epsilon_d$, is fixed along the entire line of QCPs in the $\Delta_V-\Delta_A$ plane, it varies continuously along the line of QCPs in the $\Delta^{+}_V-\Delta^{-}_S$ plane according to
\begin{align}\label{DCE-CSBdisorder}
z=1+ \frac{2}{3}\left[ 5 \Delta^{+}_{V,\ast} + \Delta^{+}_{S,\ast} \right]=1+ 5 \epsilon_m + 4  \Delta^{+}_{S,\ast},
\end{align}
where the quantity with subscript ``$\ast$" denote the critical value for WSM-metal transition. Such continuously varying DSE leaves its signature in critical scaling of various physical observables, as we discuss below, and qualitatively mimics the picture of Kosterlitz-Thouless transition. Notice that the end point of such line of QCPs on the $\Delta^{+}_V$ axis reside in the $\Delta_{SO}-\Delta_{AM}$ plane at $\Delta_{SO}=\Delta_{AM}=3 \epsilon_m/4$, and the RG flow in this plane is shown in Fig.~\ref{SpOrAxMg_flow}. The phase diagram of a dirty WSM containing only spin-orbit and axial magnetic disorder in this plane is shown in Fig.~\ref{SpOrAxMg_Phase}, with $z=1+5\epsilon_m$, which is directly obtained from Eq.~(\ref{DCE-CSBdisorder}) by setting $\Delta^{+}_S=0$. It is worth pointing out that in the $\Delta_{SO}-\Delta_{AM}$ plane the phase boundary between the WSM and metallic phase is set by the irrelevant parameter associated with the QCP, while when such QCP percolates through $\Delta^{+}_V-\Delta^{+}_S$ plane in the form of a line of QCPs, it is determined by the relevant direction at each point on the line of QCPs.

\begin{figure}
\subfigure[]{
\includegraphics[width=4cm, height=4cm]{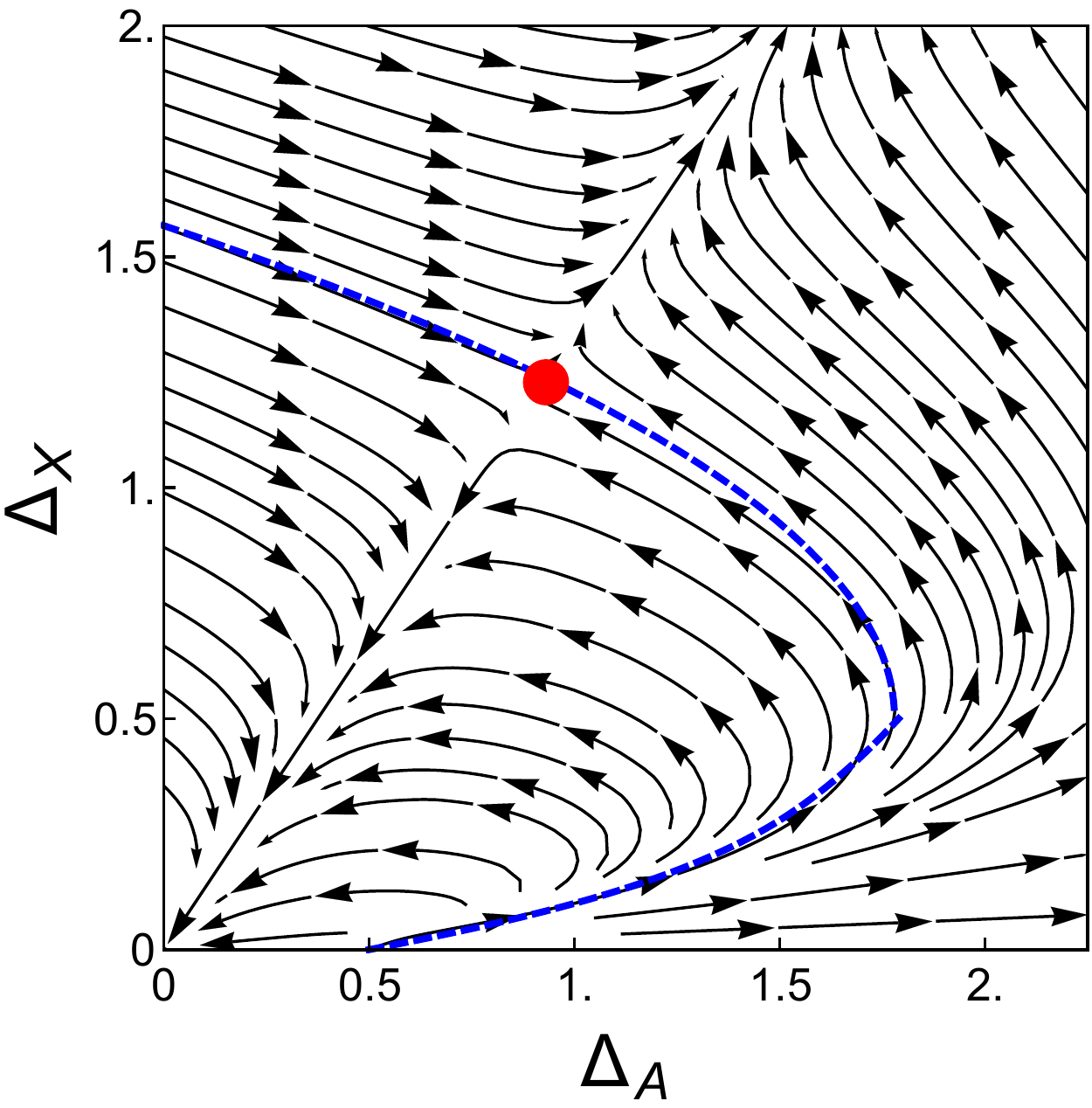}
\label{SoAxHard_flow}
}
\subfigure[]{
\includegraphics[width=4cm, height=4cm]{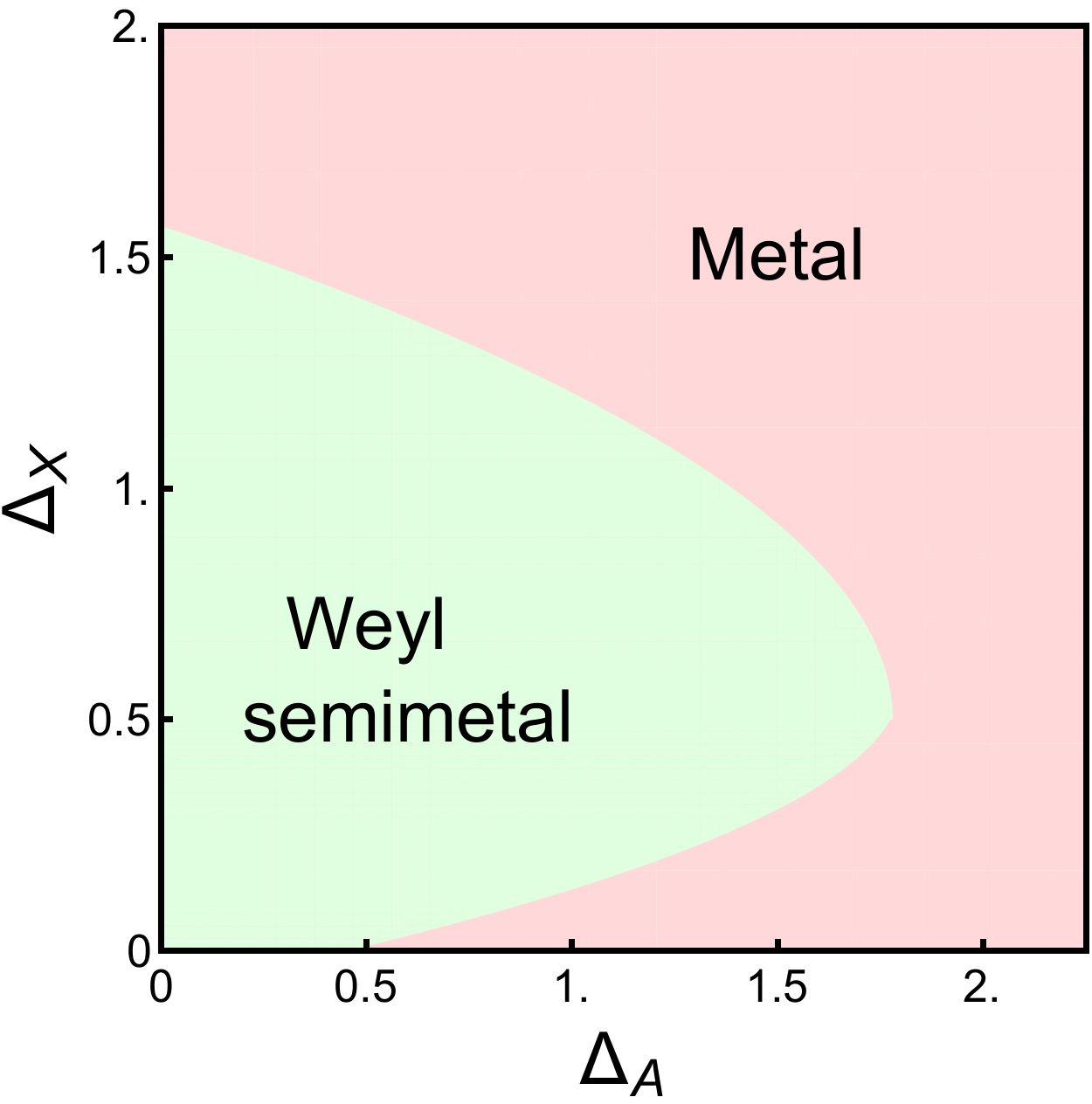}
\label{SoAxHard_phase}
}
\caption[]{(a) The renormalization group flow diagram and (b) corresponding phase diagram in the $\Delta_{A}-\Delta_{X}$ plane obtained from $\epsilon_d$ expansion for $\epsilon_d=1$, where $X=SO, AM$. There exists only one QCP at $\Delta_{A}=6 \epsilon_d/5$, $\Delta_{X}=9 \epsilon_d/10$. The QCP at $\Delta_A=\epsilon_d/2$ in the absence of a CSB disorder now possesses two unstable directions. Note that a new critical point emerges from the competition between the chiral and non-chiral disorder~\cite{chakravarty, roy-dassarma-intdis}. }\label{SoAx_Hard}
\end{figure}

\subsection{$\epsilon_d$ expansion}

Next let us address the effects of CSB disorder within the framework of an $\epsilon_d$ expansion. In this method the RG flow equations become very complicated due to the ultraviolet divergent contribution arising from the class of the Feynman diagrams shown in Fig.~\ref{laddercrossing} (c) and (d), and it is challenging to decode the emergent quantum-critical phenomena. Thus we attempt to unearth critical properties by focusing on various coupling constant subspaces that remain closed under the RG, at least to the leading order. Let us first focus on spin-orbit or axial magnetic disorder. The RG flow equations read
\begin{align}
\beta_{\Delta_{X}}&=-\epsilon_d \Delta_{X} -\frac{2}{3} \Delta^2_X + 2 \Delta_X \Delta_A \nonumber \\
\beta_{\Delta_{A}}&=-\epsilon_d \Delta_{A} + 2 \Delta^2_A -6 \Delta_A \Delta_X + 4 \Delta^2_X,
\end{align}
where $X=SO, AM$. Notice that even though the bare theory contains only spin-orbit or axial magnetic disorders, the CSP axial disorder gets generated and in order to keep the RG flow equations closed we need to include the latter from the outset. The coupled flow equations support one QCP, located at $\Delta_{X,\ast}=9 \epsilon_d/10, \Delta_{A,\ast}=6 \epsilon_d/5$~\cite{chakravarty, roy-dassarma-intdis}. The RG flow diagram is shown in Fig.~\ref{SoAxHard_flow}, and the resulting phase diagram is displayed in Fig.~\ref{SoAxHard_phase}. Note that QCP obtained in the absence of the CSB disorders, located at $\Delta_{A,\ast}=\epsilon_d/2$ now becomes unstable in the presence of either spin-orbit or axial magnetic disorder, and a new QCP results from the competition between these two disorders, as mentioned above. This outcome although is in contrast with our previously reported results obtained from $\epsilon_m$ expansion, still shows some qualitative similarities, as we argue below. Notice that the DSE and CLE at the new QCP, shown in Fig.~\ref{SoAxHard_flow}, are respectively given by
\begin{align}
z=1 + \frac{9}{2} \epsilon_d + {\mathcal O} (\epsilon^2_d), \:\:\: \nu^{-1}=\epsilon_d + {\mathcal O} (\epsilon^2_d).
\end{align}
As a result the mean DOS at the QCP diverges as $\varrho(E) \sim |E|^{-5/11}$ for $\epsilon_d=1$ or $d=3$, since $z>d$. Hence, both $\epsilon$-expansions give rise to diverging DOS at the QCP controlled via spin orbit and axial magnetic disorder. Although the calculated values of DSE depend on RG scheme, to the leading order they do not differ significantly, $z=6$ for $\epsilon_m=1$, and $z=11/2$ for $\epsilon_d=1$,  while $\nu=1$, is independent of the RG scheme.

\subsection{Mass disorder}

We now discuss the role of mass disorder in WSMs. It should be noted that a WSM can be susceptible to two different types of mass disorder (a) scalar mass disorder and (b) pseudo-scalar mass disorder. Both of them break the chiral symmetry, but can be rotated into each other by the generator of the chiral symmetry $\gamma_5$. The flow equation for mass disorder within the framework of an $\epsilon$ expansion reads as
\begin{align}
\beta_{\Delta_X}=-\epsilon_j \Delta_X - \alpha_j \Delta^2_X,
\end{align}
for $X=S, PS$, where $\alpha_m=8/3$ and $\alpha_d=2$, $j=m,d$ corresponds to $\epsilon_m$ and $\epsilon_d$ expansions, respectively. Hence, by itself scalar or pseudoscalar mass disorder does not drive any WSM-metal QCP, at least within the leading order in $\epsilon$-expansions. In this regard both $\epsilon_m$ and $\epsilon_d$ expansions yield an identical result.

Finally, we discuss yet another interesting aspect of mass disorder, when it coexists with the axial one. The flow equations in the presence of these two disorders are
\begin{eqnarray}
\beta_{\Delta_A} = -\Delta_A \left[\epsilon_j  - \tilde{\alpha}_j \Delta_{-} \right], \;
\beta_{\Delta_X} = -\Delta_X \left[ \epsilon_j + \tilde{\alpha}_j \Delta_{-} \right], \nonumber \\
\end{eqnarray}
for $X=S, PS$, where $\Delta_{-}=\Delta_A-\Delta_X$, $\tilde{\alpha}_m=8/3$, $\tilde{\alpha}_d=2$, and respectively $j=m,d$ corresponds to $\epsilon_m$ and $\epsilon_d$ expansions. These two flow equations support a line of QCPs, determined by
\begin{equation}
\Delta_{A,\ast}=\frac{\epsilon_j}{\tilde{\alpha}_j} + 2 \Delta_{X,\ast}.
\end{equation}
The location of such line of QCPs is regularization dependent (through $\tilde{\alpha}_j$), along which the DSE and CLE, given by
\begin{equation}
z=1+\frac{\epsilon_j}{2}+2 \Delta_{S,\ast}, \; \nu^{-1}=\epsilon_j
\end{equation}
are identical in both $\epsilon$-expansion schemes. Therefore, in a WSM with these two disorders the DSE continuously increases from $z=3/2$ in an unbounded fashion, while the CLE remains fixed. The numerical investigation of such interesting possibility is left for a future work.

\subsection{Why is the chiral superuniversality so robust?}~\label{chiralsuperuniversality_explanation}

Leaving aside the interesting possibilities of realizing such as line of QCPs with continuously varying critical exponents, perhaps the most urgent issue to be addressed is the following: \emph{Why does the disorder-driven WSM-metal QPT always display same universality class, characterized by $\nu \approx 1$ and $z \approx 1.5$?}

The answer to this question in presence of intra-node or chiral-symmetric disorders has already been provided in Sec.~\ref{CSP_disorder}. Note that scaling dimension of any disorder coupling in a $d$-dimensional WSM is $[\Delta_a]=2 z-d$. But at all CSB disorder driven QCPs,  controlling the WSM-metal QPT, $z>d$ irrespective of the RG methodology. Therefore, even though the bare values of CSP disorders in lattice-based simulations are set to be zero, discussed in Sec~\ref{non-chiral-numerics}, they do get generated as we approach the Weyl points through the coarse graining procedure. Ultimately the CSP disorder becomes relevant at CSB disorder driven WSM-metal QCPs. As a result, the dirty system even though initially tends to flow toward the QCPs with $z>d$, described in this section, it flows back toward the chiral symmetric QCP or line of QCPs shown in Fig.~\ref{potaxialflow}. This is the reason why the WSM-metal QPTs are always characterized by CLE $\nu \approx 1$ and DSE $z \approx 1.5$ (within numerical accuracy), the characteristics of the proposed \emph{chiral superuniversality}. The above argument is very generic and does not depend on the number of Weyl nodes. Therefore, in any lattice system, we expect WSM-metal QPT to always belong to the chiral superuniversality. This outcome can be anchored from the RG calculation in the presence of all eight possible disorder couplings (since in strong disorder regime all disorders get generated even if the bare coupling for some specific channel is set to be zero), as shown in Appendix~\ref{RG_8couplings} within the framework of both $\epsilon_m$ and $\epsilon_d$ expansions. Such analysis confirms that only the line of QCPs, defined through Eq.~(\ref{LineQCP_chiral}) or Eq.~(\ref{LineQCP_chiral_hard}), and shown in Fig.~\ref{potaxialflow}, ultimately controls the quantum-critical behavior.
Among all possible WSM-metal QCPs, we note that along the entire line of QCPs in the plane of regular and axial potential disorders, shown in Fig.~\ref{Potential_Axial}, the DSE possesses the least (and constant) value. As a consequence, ADOS is smallest along this line of QCPs, which is thus expected to be robust against any perturbation. Therefore, we believe that the proposed notion of emergent superuniverslaity across such a line of QCPs in the chiral-symmetric hyperplane is non-perturbative in nature, which is further substantiated by our complementary numerical analysis, always yielding $z \approx 1.5$ and $\nu \approx 1$ (within numerical error bars), see Table~\ref{table-exponent} and Table~\ref{Tab:CSB_exponents}. 
This strongly supports the above argument in favor of chiral superuniversality under generic circumstances~\footnote{We note that the quality of data collapses for CSB disorders, shown in Fig.~\ref{numeric_analysis_figure_CSB}, is slightly less pronouced than those for CSP disorder, displayed in Fig.~\ref{numeric_analysis_figure}, which can qualitatively be understood in the following way. In the presence of only inter-node scatterers system first tends to flow toward the line of QCPs set by purely CSB disorder, discussed early in this section. Only when disorder gets sufficiently strong the intra-node disorder becomes relevant and the system starts flowing toward the line of QCPs discussed in Sec.~\ref{CSP_disorder}. The system then gets stuck in the  crossover regime dominated by CSB disorder, and consequently the data collapse (involving finite energy states) becomes slightly less prominent. To achieve equally good quality data collapse even in the presence of CSB disorder we therefore need to subscribe to larger systems, which can be numerically challenging.  }.

The specific tight-binding model we subscribe in this work [see Sec.~\ref{Weyl_Lattice}] also contains Wilson mass that bears higher gradient terms, such that $\tau_3 b_\perp (k^2_x+k^2_y)$, with $b_\perp=t_0 a^2/2$. The scaling dimension of such operator is $[b_\perp]=z-2$. Hence, the higher gradient terms are \emph{irrelevant} at clean WSM fixed point ($[b_\perp]=-1$) as well as at the chiral symmetric line of QCPs ($[b_\perp]=-1/2$), but becomes relevant at pure CSB disorder-driven QCPs (since $z>d>2$). This is also the reason why chiral superuniversality is such a generic and utmost stable situation.

Furthermore, we also show that the chiral superuniversality does not depend on the choice of disorder distribution. For example, in Appendix~\ref{Append:Correlated_Disorder} we perform similar analysis of average DOS in the presence of \emph{correlated} potential disorder that by construction significantly suppresses the inter-valley scattering (at least when disorder is sufficiently weak). However, the universality class of the WSM-metal QPT (characterized by $z$ and $\nu$) remains unchanged (within numerical accuracy) by the profile of the distribution function. This observation should further strengthen the proposed scenario of emergent superuniversality (insensitive to the nature of disorder and its distribution) across the WSM-metal QPT.

Nevertheless, we believe pure CSB disorder driven QCPs (with $z>d$) can in principle be realized in a numerical simulation performed in momentum space, where forward/  or intranode or CSP scattering processes can be suppressed deliberately and higher gradient terms can be avoided completely. Such an analysis is an interesting exercise of a pure academic interest, and we leave it for a future investigation.

\section{Quantum critical scaling of physical observables}~\label{physicalobservable}

As demonstrated in the previous two sections that QPT from a WSM to  a diffusive metal can be driven by different types of elastic scatters, and the critical exponents are remarkably independent of the actual nature of randomness. We here highlight how these exponents can affect the scaling behavior of measurable quantities as the Weyl material undergoes this QPT~\footnote{In spite of the emergent superuniversality, the putative line of QCPs driven by CSB disorders with continuously varying DSE $z>d$ may leave its imprint on the physical observables in the crossover regime before the CSP disorders take over and ultimately the system flows toward the chiral symmetric quantum-critical line with $z=3/2$ and  $\nu=1$. In that sense the physical observables we discuss in this section can also distinguish between different types of disorder (inter-node vs intra-node). }.

\subsection{Residue of quasiparticle pole}

As the WSM-metal QCP is approached from the semimetallic phase, the residue of quasiparticle pole vanishes and beyond the critical strength of disorder Weyl fermions cease to exist as sharp quasiparticle excitations, similar to the situation for two-dimensional Dirac fermion-Mott insulator QPT in the presence of a strong Hubbard interaction \cite{HJR, sorella}. The residue of quasiparticle pole ($Z$) vanishes as
\begin{align}~\label{residue}
Z \sim \left( \frac{\Delta_\ast-\Delta}{\Delta_\ast}\right)^{\nu \eta_\Psi} \equiv \delta^{\nu \eta_\Psi},
\end{align}
where $\eta_\Psi$ is the fermionic anomalous dimension at the critical point located at the disorder strength $\Delta=\Delta_\ast$. Within the framework of an $\epsilon_d$ expansion $\eta_\Psi=0$ to the leading order in $\epsilon_d$, and one needs to account for two-loop diagrams to obtain finite $\eta_\Psi$. In contrast, in the $\epsilon_m$ expansion we obtain nontrivial fermionic anomalous dimension even to the one-loop order, and $\eta_\psi \sim \epsilon$, as shown in Eq.~(\ref{eq:eta-psi}). Therefore, at the WSM-metal QCP, the quasiparticle spectrum displays \emph{a branch-cut} and the critical point represents a strongly coupled \emph{non-Fermi liquid}. Alternatively, the residue of quasiparticle pole plays the role of a \emph{bonafide} order parameter on the semimetallic side. It is worth mentioning that the disappearance of residue of quasiparticle pole has recently been tracked in quantum Monte Carlo simulations for Hubbard model in two-dimensional honeycomb lattice~\cite{sorella}, and we can expect that future numerical work can verify our proposed scaling form in Eq.~(\ref{residue}) across the disorder driven WSM-metal QPTs. The Fermi velocity scales as $v\sim |\delta|^{\nu(z-1)}$, and since $z>1$ at the QCP or the quantum-critical line, the Fermi velocity vanishes at the transition to the metallic phase. A subsequent numerical work has demonstrated the suppression of residue of quasiparticle pole~\cite{pixley-residue}.

\subsection{Average density of states} 

The most widely studied physical quantity in numerical simulations across the WSM-metal QPT is the average DOS~\cite{herbut-disorder, pixley-1, pixley-2, roy-bera, ohtsuki, pixley-4}. Since throughout the paper we have already extensively used the average DOS to characterize phases, for the sake of completeness we here review only its salient features. We can infer the scaling form of the average DOS in the thermodynamic limit $L\to\infty$ in different phases by using its scaling function [see Eq.~(\ref{DOS_Scaling_numerics})]. In the quantum critical regime $\varrho(E)$ should be independent of $\delta$, yielding $\varrho_Q(E) \sim E^{d/z-1}$. Inside the WSM phase, the average DOS scales as $\varrho_W(E) \sim \delta^{(1-z)d \nu} |E|^{2}$. In the metallic phase average DOS at zero energy is finite and scales as $\varrho(0)\sim \delta^{(d-z)\nu}$. From the quoted values of DSE  and CLE, it is straightforward to find the scaling of average DOS in these three regimes of the phase diagram in a dirty WSM, which we have used in the numerical analysis of this observable in the previous sections.

\subsection{Conductivity} 

The optical  conductivity ($\sigma$) at $T=0$ can as well serve as an order parameter across the WSM-metal QPT, and assumes the following scaling ansatz for frequency ($\Omega$) much smaller than the bandwidth~\cite{juricic-disorder}
\begin{align}~\label{OC_Scaling}
\sigma(\Omega)= \delta^{\nu(d-2)} {\mathcal G} \left( \Omega \delta^{-\nu z} \right),
\end{align}
where ${\mathcal G}$ is yet another unknown universal scaling function. This scaling form remains operative even at finite temperature as long as $\Omega \gg T$, i.e., in the \emph{collisionless} regime. In the \emph{collision dominated} regime at $T \gg \Omega$, the dc conductivity also assumes a similar scaling form as in Eq.~(\ref{OC_Scaling}), upon replacing the frequency ($\Omega$) by temperature ($T$)~\cite{wegner, radzihovsky, brouwer-1}. In the WSM side of the transition, the optical conductivity vanishes linearly with $\Omega$ and scales as $\sigma_W(\Omega) \sim \delta^{\nu(1-z)(d-2)} \Omega^{d-2}$. Inside the critical regime the optical conductivity scales as $\sigma_Q(\Omega) \sim \Omega^{(d-2)/z}$. In the presence of strong CSP disorder $z \approx 3/2$, and the optical conductivity inside the quantum critical regime thus vanishes as $\sigma_Q(\Omega) \sim \Omega^{2/3}$. Since for non-chiral disorder the DSE is typically much bigger than in the presence of chiral symmetric one, the optical conductivity vanishes with a \emph{weaker} power as $\Omega \to 0$ when the system is still dominated by CSB disorder before CSP disorder takes over. Hence, in this regime the system becomes \emph{more metallic} in the presence of CSB disorder than with only CSP disorder. Inside the metallic phase, the optical conductivity becomes finite and scales as $\sigma_M(0) \sim \delta^{\nu(d-2)}$ as $\Omega \to 0$. Within the leading order $\epsilon_m$ or $\epsilon_d$ expansions, the conductivity of the metal is therefore always independent of the actual nature of elastic scatterers, since $\nu^{-1}=\epsilon_m$ or $\epsilon_d$, and $\epsilon_m=1$, $\epsilon_d=1$. Otherwise, weak disorder (such as potential) causes enhancement of optical conductivity without altering $\sigma \sim \Omega$ scaling~\cite{juricic-disorder} [see also Appendix~\ref{OC:alternative} for a simple derivation].

\subsection{Specific heat} 

The specific heat ($C_v$) also displays distinct scaling behavior in three regimes of the phase diagram of a dirty WSM. The scaling of specific heat at temperature much smaller than bandwidth follows the ansatz~\cite{roy-dassarma}
\begin{align}~\label{CV_Scaling}
C_V(T)=\frac{T^{d/z}}{v^d} {\mathcal H} \left( T \delta^{-\nu z} \right),
\end{align}
where ${\mathcal H}$ is also an unknown universal scaling function. In the WSM phase, ${\mathcal H} (x) \sim x^{d(z-1)/z}$ and the specific heat scales as $C_V \sim \delta^{d\nu(1-z)} T^d$, so that we recover $T^3$ dependence for three dimensional Weyl fermion. Inside the metallic phase, ${\mathcal H} (x) \sim x^{1-d/z}$, yielding $C_V \sim \delta^{\nu(d-z)} T$ and we obtain $T$-linear specific heat, similar to the situation in Fermi liquids. By contrast, inside the critical regime $H(x) \sim x^{0}$, yielding $C_V \sim T^{3/z}$. Therefore, the specific heat, analogous to the conductivity, displays distinct power-law dependence on temperature inside the quantum critical regime depending on the dominant source of disorder, while its scaling inside the WSM and metallic phases is insensitive to the nature of random impurities. Hence, the scaling of specific heat can be used to extract the extent of the critical regime and crossover boundaries among different phases of a dirty Weyl system at finite temperature~\cite{pixley-2}.

\subsection{Mean-free path}

The quasiparticle mean-free path ($\mathscr{L}$) also follows the critical scaling
\begin{align}~\label{MFP_Scaling}
\left[ \mathscr{L}(E) \right]^{-1}= \delta^{\nu} {\mathcal J} \left( E\delta^{-\nu z}\right),
\end{align}
where ${\mathcal J}$ is a universal, but unknown scaling function, with energy much smaller than bandwidth. At the QCP ($\delta=0$) the mean-free path should be independent of $\delta$, implying ${\mathcal J}(x)\sim x^{-1/z}$. Therefore, inside the quantum critical fan, the mean-free path at zero energy diverges as $\mathscr{L}(E) \sim E^{-1/z}$. In the metallic phase, ${\mathcal J}(x) \sim x^0$ as $x \to 0$, leading to finite mean-free path at zero energy, and $\mathscr{L}(0) \sim \delta^{-\nu}$. On the other hand, in the WSM phase, the mean-free path  $\mathscr{L}_W(E) \sim \delta^{\nu (z-1)}E^{-1}$, as $E \to 0$. Since at all disorder driven QCPs $z>1$, $\mathscr{L}_W(E)$ decreases with increasing disorder, indicating propensity toward the onset of a metallicity in the system.

\subsection{Gr$\ddot{\mbox{u}}$neisen parameter}

Yet another directly measurable quantity is the Gr$\ddot{\mbox{u}}$neisen parameter, defined as $\gamma=\alpha/C_P$, where $\alpha$ is the thermal expansion parameter, and $C_P$ is the specific heat measured at constant pressure. The Gr$\ddot{\mbox{u}}$neisen ratio in the WSM phase  $\gamma_W \sim T^{-4}$, while inside the critical regime $\gamma_Q \sim T^{-(1+d/z)}$. Inside the metallic phase $\gamma_M \sim T^{-2}$. Therefore, the Gr$\ddot{\mbox{u}}$neisen parameter displays distinct power law behavior in three different phases of a dirty WSM.

Fascinating scaling behavior can also be observed for the magnetic Gr$\ddot{\mbox{u}}$neisen ratio, defined as $\Gamma_H = \left( \partial M/\partial T\right)_H/C_H$, where $M\propto H$ is magnetization, $C_H$ is the molar specific heat, and $H$ is the magnetic field strength. In the presence of sufficiently weak randomness when Landau quantization is sharp ($\omega_c \tau \gg 1$, where $\omega_c$ is cyclotron frequency and $\tau$ is scattering lifetime) and it dominates over the Zeeman coupling, leading to $\Gamma_H \sim T^{-4/z}$. On the other hand, in the presence of strong elastic scattering when $\omega_c \tau \ll 1$ the Landau levels are sufficiently broadened and the dominant energy scale is set by Zeeman coupling, yielding $\Gamma_H \sim T^{-2}$, which is independent of dimensionality ($d$) or DSE ($z$). Therefore, for a fixed weak magnetic field, as the strength of impurities is gradually increased, the magnetic Gr$\ddot{\mbox{u}}$neisen ratio should display a smooth crossover from $T^{-4}$ to $T^{-2}$ dependence. Note that such a crossover will take place even before the system enters  the quantum critical regime and will persist in the metallic regime as well, since elastic scattering is strong in these two phases.

\section{Anderson Transition}~\label{anderson}

\begin{figure}[t!]
\subfigure[]{
\includegraphics[width=4cm,height=3.75cm]{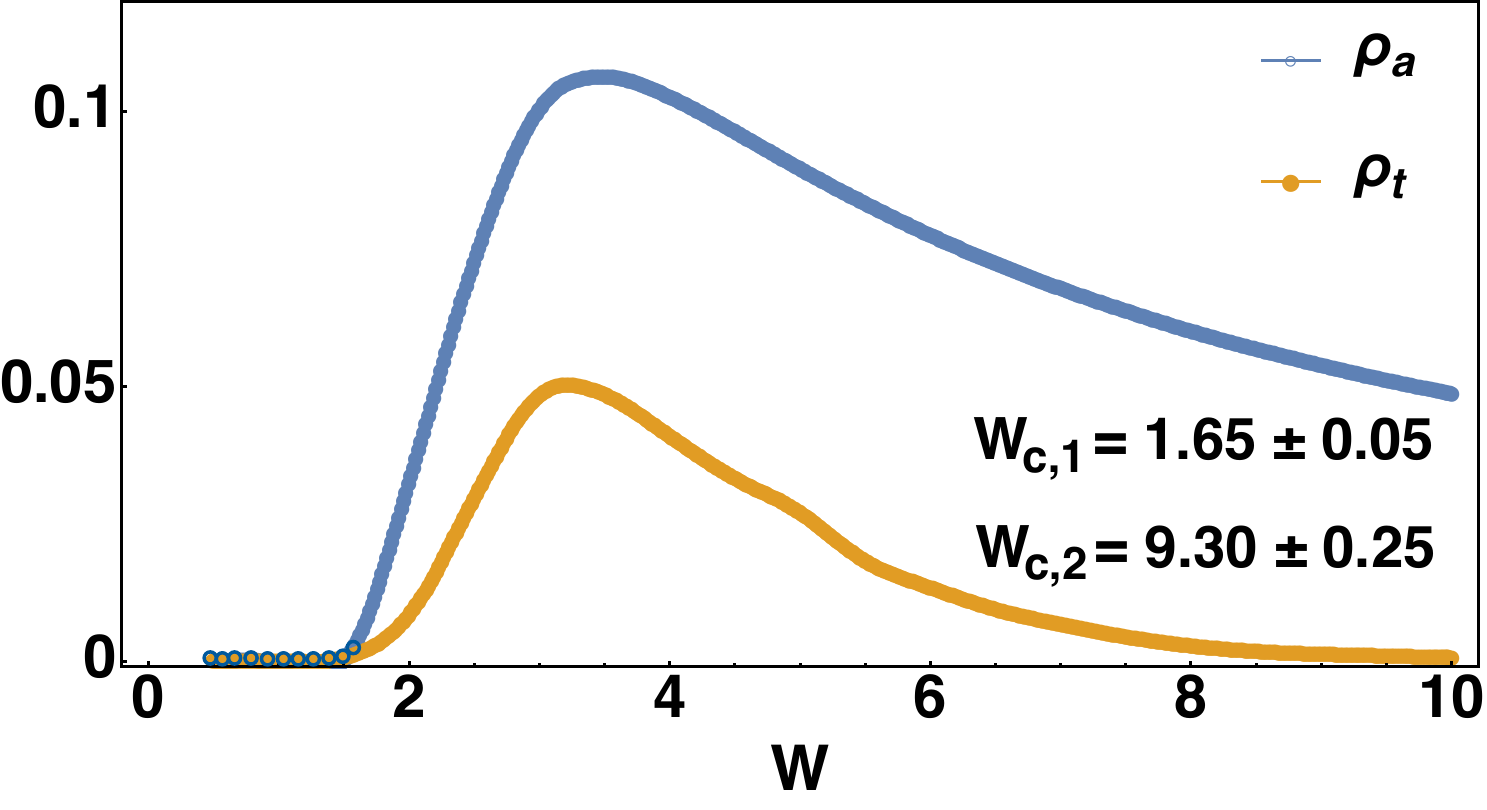}~\label{anderson:1}
}
\subfigure[]{
\includegraphics[width=4cm,height=3.75cm]{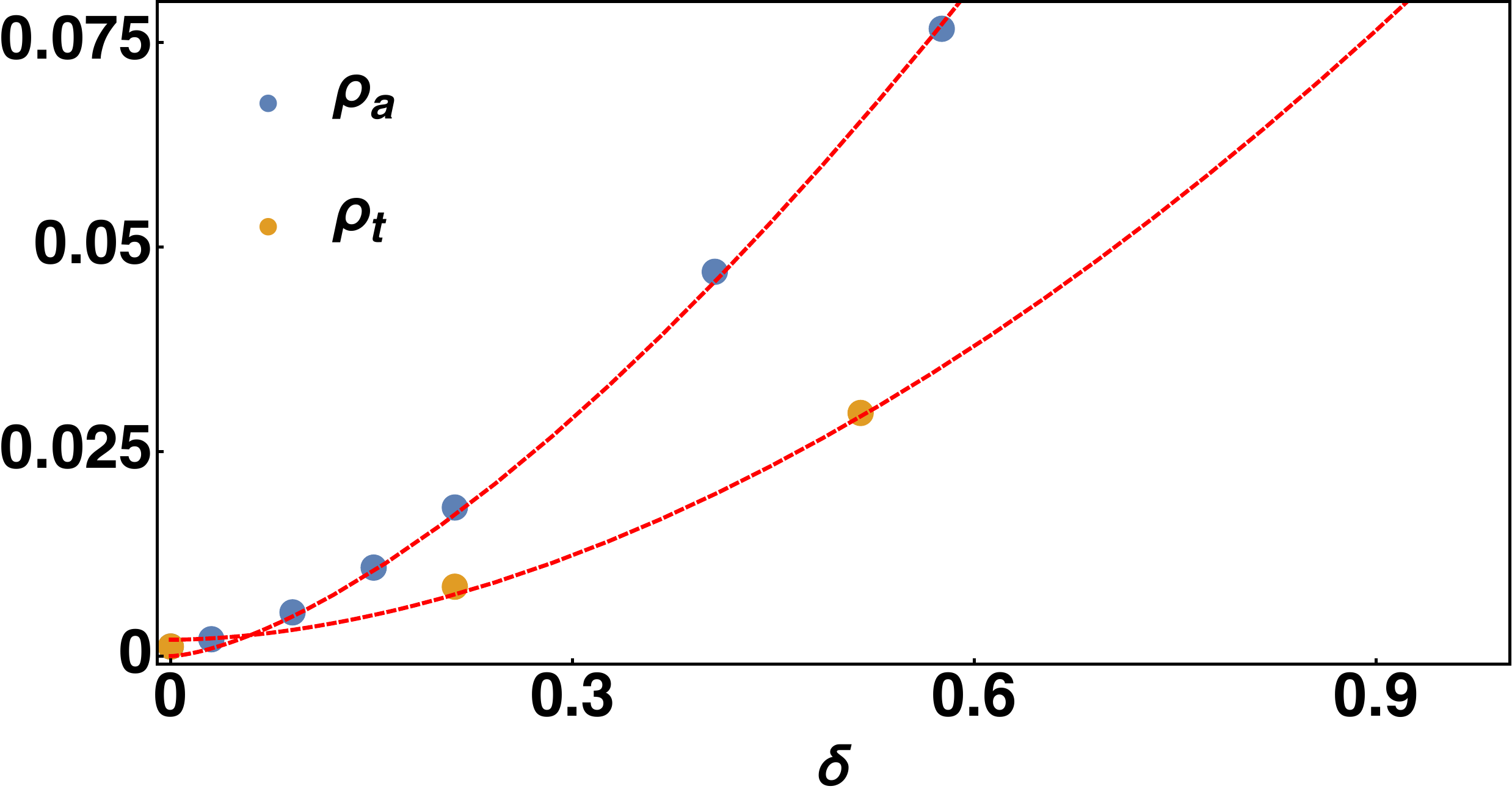}~\label{anderson:2}
}
\caption{(a) Scaling of average ($\varrho_a(0)$) and typical ($\varrho_t(0)$) density of states at zero energy as a function of disorder strength. The Weyl semimetal-metal and metal-Anderson insulator quantum phase transitions respectively takes place at $W_{c,1}=1.65 \pm 0.05$ and $W_{c,2}=9.30 \pm 0.25$. (b) Scaling of these two quantities as a function of $\delta=(W-W_{c,1})/W_{c,1}$, yielding corresponding order-parameter exponents [defined in Eq.~(\ref{OPexponents:definitions})] $\beta_a=1.50 \pm 0.05$ and $\beta_t =1.80 \pm 0.20$.
}
\end{figure}

As a final topic, we discuss the Anderson transition (AT) of a disordered diffusive Weyl metal at stronger strength of disorder. For the sake of simplicity we here focus only on the effects of random charge impurities. Possible AT in the presence of all other disorder is left for a future investigation. To study the AT we compare three different types of DOS, namely average DOS [$\varrho_a(E)$], local DOS (LDOS) [$\varrho_{L}(E)$] and typical DOS (TDOS) [$\varrho_t(E)$], respectively defined as~\cite{KPM-RMP, abrahams}
\allowdisplaybreaks[4]
\begin{eqnarray}
\varrho_a(E) &=& \Big\langle \frac{1}{2L^3} \sum^{L^3}_{i=1} \sum^{2}_{\alpha=1} \delta \left(E-E_{i,\alpha} \right) \Big\rangle, \\
\varrho^{i, \alpha}_{L}(E) &=& \sum_{k, \beta} |\langle k,\beta | i, \alpha \rangle |^2 \delta\left( E-E_{k, \beta} \right), \\
\varrho_t(E) &=& \exp \left[ \frac{1}{2 N_s} \sum^{N_s}_{j=1} \sum^{2}_{\alpha=1} \Big\langle \log \varrho^{i,\alpha}_L (E) \Big\rangle \right].
\end{eqnarray}
Here $L^3$ is the system size, $|i, \alpha \rangle$ is the eigenstate with site index $i$ and orbital index $\alpha(=1,2)$ at energy $E_{i,\alpha}$. As previously discussed, average DOS is a self-averaging quantity so to minimize statistical fluctuations we only extract the disorder-averaged smoothened data, which we carry out by computing $N_m=1024$ Chebyshev moments and performing disorder average over $20$ random disorder realizations. On the other hand, LDOS and TDOS are not self-averaging quantities. Therefore, numerical extraction of TDOS is extremely demanding for which we compute $N_m=8192$ moments and perform disorder average over $100$ random disorder realization to construct the TDOS. To further suppress statistical fluctuations in TDOS we average over a small cube of size $N_s=L^3_s \ll L^3$, and we here take $L_s=4$. Such averaging is justified since translational symmetry gets restored after disorder averaging has been performed.

\begin{figure}[t!]
\subfigure[]{
\includegraphics[width=4cm,height=3.5cm]{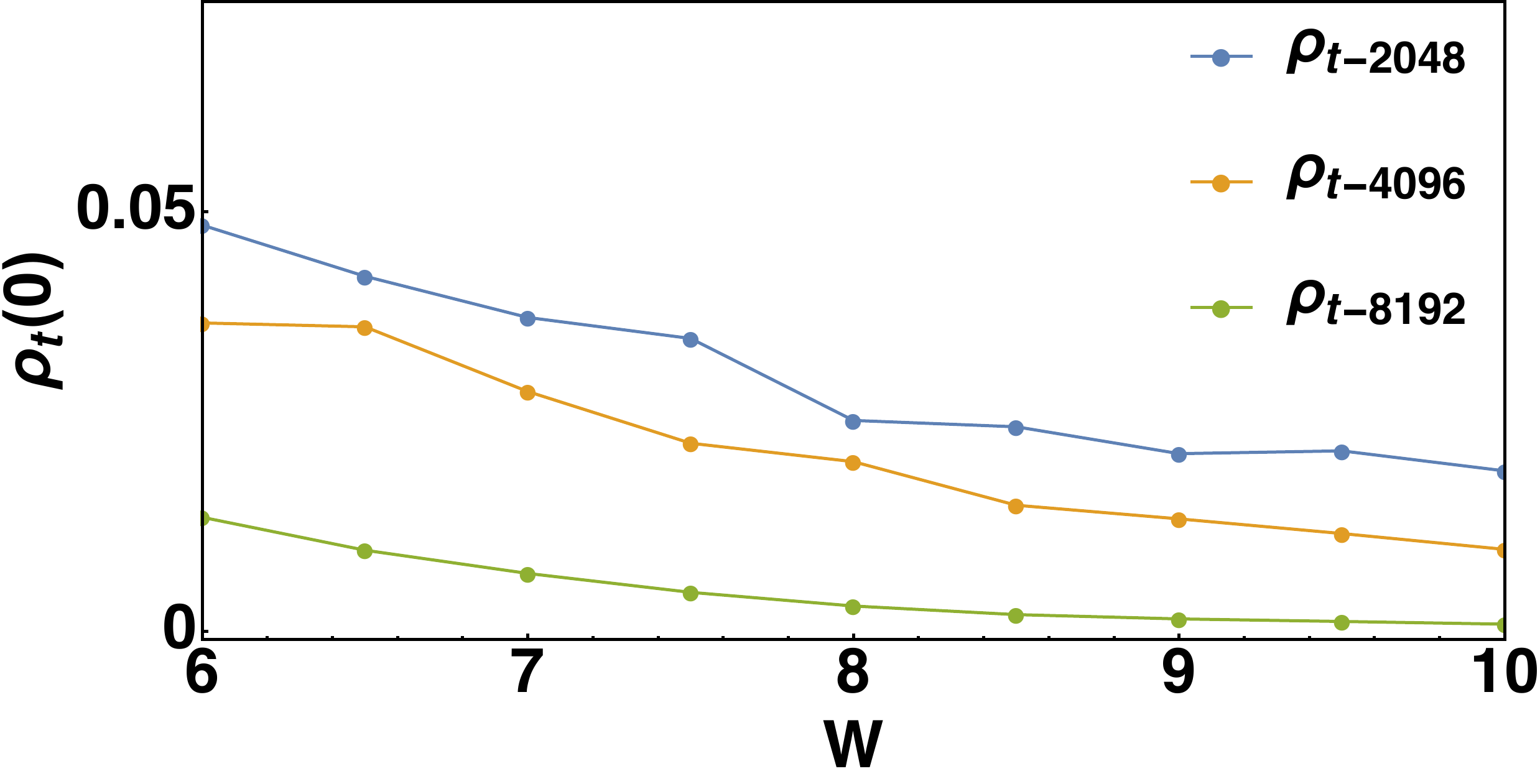}~\label{andeson:3}
}
\subfigure[]{
\includegraphics[width=4cm,height=3.5cm]{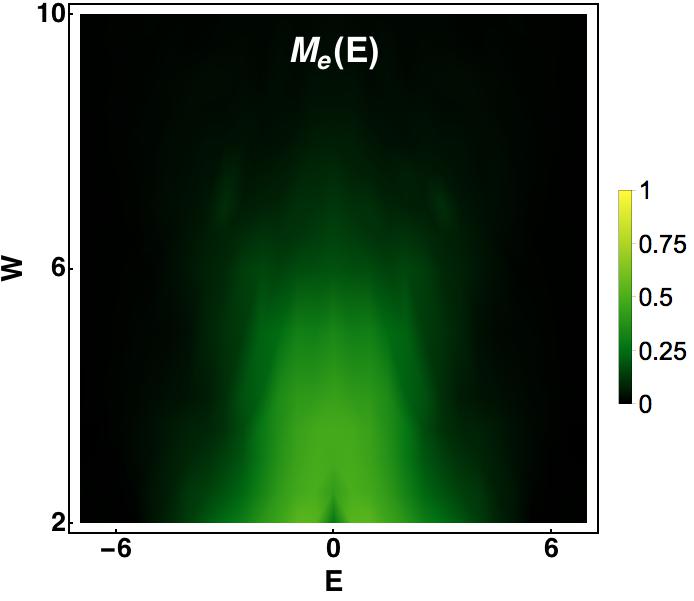}~\label{mobility-edge:Fig}
}
\caption{(a) Scaling of typical density of states at zero energy $\varrho_t(0)$ for disorder ($W$) within the range $6.0 \leq W \leq 10.0$ with the number of Chebyshev moments $N_m$. We here compute $\varrho_t(0)$ for $N_m=2048, 4096$ and $8192$. From the scaling of $\varrho_t(0)$ vs. $N_m$, we find that metal-insulator Anderson transition takes palce at $W_{c,2}=9.30$ in the $N_m \to \infty$ limit. (b) Mobility edge $M_e(E)$ [defined in Eq.~(\ref{Mob-edge:def})] as a function of energy ($E$) and disorder ($W$). Respectively, the green and the dark regions accommodates extended or metallic and localized states. Scale of $M_e(E)$ is shown in the legend. Here the system size is $L=80$.
}
\end{figure}

The scaling of average DOS and TDOS over a wide range of disorder strength is shown in Fig.~\ref{anderson:1}. Note that in the WSM phase both average DOS and TDOS at zero energy are pinned to zero, which then become finite across the WSM-metal QPT at $W_{c,1}=1.65 \pm 0.05$. Therefore, either average DOS or TDOS can be identified as a bonafide order-parameter to pin the WSM-metal QCP. Respectively these two quantities scale as
\begin{eqnarray}~\label{OPexponents:definitions}
\varrho_a (0) \sim \left(\frac{W-W_{c,1}}{W_{c,1}} \right)^{\beta_a}, \:
\varrho_t (0) \sim \left(\frac{W-W_{c,1}}{W_{c,1}} \right)^{\beta_t},
\end{eqnarray}
near the WSM-metal QCP, with
\begin{equation}
\beta_a=1.50 \pm 0.05, \: \beta_t=1.80 \pm 0.20,
\end{equation}
as shown in Fig.~\ref{anderson:2}. Even though the numerical error-bar for $\beta_t$ is quite large, in general, we expect it to be different from $\beta_a$, as their difference, $\Delta \beta=\beta_t-\beta$, is intimately tied with the \emph{multifractal dimension} of the wave-function across a disorder-driven QPT~\cite{beiltz, mirlin, janssen, foster}. However, more precise determination of $\beta_t$ requires additional extensive numerical simulation. Therefore, we leave this issue as a subject for a future investigation.

Inside the compressible diffusive metallic phase these two quantities increase monotonically and follow each each other up to a moderate strength of disorder $W_\ast \approx 3.5$. Upon further increasing strength of disorder the TDOS smoothly vanishes around $W_{c,2}=9.30 \pm 0.25$. Therefore, a metal-insulator transition (MIT) takes place at $W=W_{c,2}$, commonly known as AT. Note that the average DOS decreases monotonically across the AT, but remains non-critical, as shown in Fig.~\ref{anderson:1}. In Fig.~\ref{andeson:3} we present the scaling of TDOS with the number of Chebyshev moments ($N_m$). We explicitly compute TDOS from moderate to strong disorder regime ($6 \leq W \leq 10$), in the close vicinity of the AT, for $N_m=2048, 4096$ and $8192$. From the scaling of $\varrho_t(0)$ vs. $N_m$ we conclude that AT (identified with $\varrho_t(0) \to 0$) takes place around $W_{c,2}=9.30$ in the $N_m \to \infty$ limit. Therefore, we can conclude that a three-dimensional diffusive Weyl metal is a stable phase of matter for moderately strong disorder, which ultimately undergoes a QPT into the Anderson insulator phase for sufficiently strong disorder. Across the AT the TDOS at zero energy display single-parameter scaling 
\begin{equation}
\varrho_t(0) \sim \left(\frac{W-W_{c,2}}{W_{c,2}} \right)^{\beta},
\end{equation}
with $\beta=1.5 \pm 0.15$. Critical scaling of typical DOS across the Anderson transition strongly suggests that wave-functions at the Anderson critical point become \emph{multi-fractal} in nature~\cite{abrahams}. A detailed analysis of multi-fractal spectrum requires the notion of exact wave-function, which is numerically very time consuming. Nevertheless, analysis of multi-fractal nature of wave-functions in a time-reversal symmetry breaking topological metal is a problem of fundamental importance, which we leave for future investigation.

\begin{figure}[t!]
\subfigure[]{
\includegraphics[width=4cm,height=3.75cm]{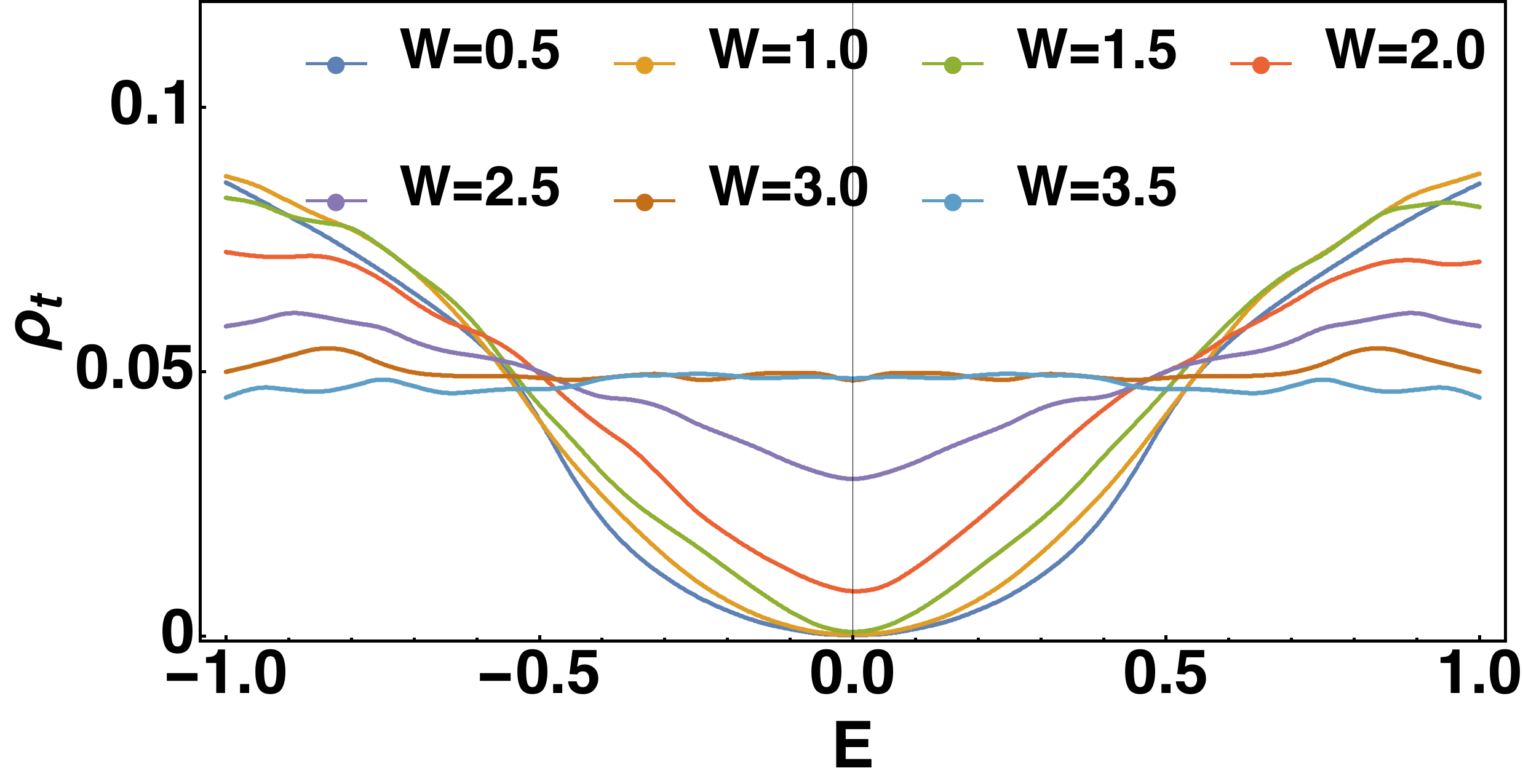}~\label{andersonfan:1}
}
\subfigure[]{
\includegraphics[width=4cm,height=3.75cm]{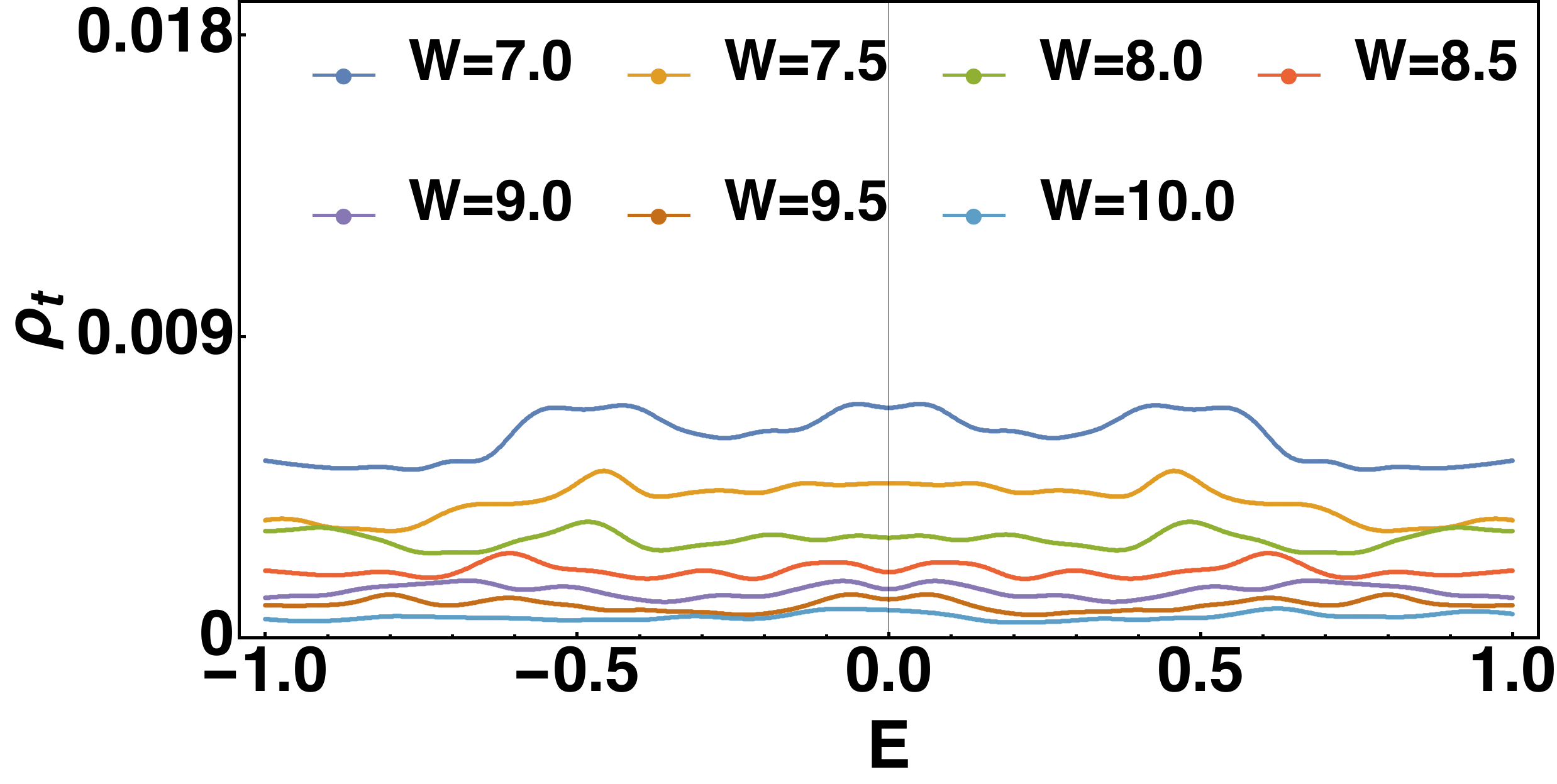}~\label{andersonfan:2}
}
\caption{(a) Scaling of typical density of states ($\varrho_t(E)$) vs. energy ($E$) from weak to moderately strong disorder, showing that $\varrho_T(E) \sim |E|^2$ of weak disorder. Also note that $\varrho_t(E) \sim |E|$ around $W=W_{c,1} \approx 1.65$, and inside the metallic phase $\varrho_t(0)$ is finite. These features are qualitatively similar to the ones for the average density of states [see Fig.~\ref{DOS_Fan_General}]. (b) Scaling of $\varrho_t(E)$ vs. $E$ for stronger disorder (close proximity to the Anderson metal-insulator transition), showing that $\varrho(0)$ smoothly vanishes across the Anderson transition, and remains pinned at zero inside an Anderson insulator. Here we compute 8192 Chebyshev moments to construct $\varrho_t(E)$ in a system with linear dimension $L=80$.
}
\end{figure}

Recall that for weak disorder average DOS $\varrho_a(E) \sim |E|^2$ and around the WSM-metal QCP it scales as $\varrho(E) \sim |E|$. Inside the metallic phase $\varrho_a(0)$ is finite. In Fig.~\ref{andersonfan:1}, we show that within the range of disorder strength $0.50$(weak) $\leq W \leq 3.5$(moderate) the TDOS also display the same scaling behavior as average DOS. This observation confirms that TDOS can also be subscribed as a bonafide order-parameter across the WSM-metal QPT. On the other hand, for strong enough disorder the TDOS $\varrho_t(E)$ decreases monotonically for any energy $E$, and ultimately $\varrho_t(0)$ becomes zero across the AT. Therefore, TDOS can serve as the order-parameter across all possible disorder-driven QPTs considered here.

Finally, we focus on the evolution of the location of the mobility edge in a dirty Weyl metal as a function of disorder strength by numerically computing the mobility edge, defined as
\begin{equation}~\label{Mob-edge:def}
M_e(E) = \frac{\varrho_t(E)}{\varrho_a(E)}.
\end{equation}
In particular, the mobility edge defines the boundary between the extended and localized states, and we here focus on this quantity in the strong disorder regime $W \geq 2 >W_{c,1}$. The results are shown in Fig.~\ref{mobility-edge:Fig}. For weak disorder the mobility edge resides at high-energy, indicating the metallic nature of a moderately dirty Weyl system. However, the mobility edge progressively slides down toward smaller energy with increasing randomness in the system. Finally, across the AT the mobility edge comes down to zero energy, indicating that all states inside the Anderson insulator are localized. Notice that the shape of the mobility edge is quite distinct in a Weyl metal in comparison to its counterpart in conventional metal~\cite{brndiar}, which however can solely be attributed to the linear dispersion of Weyl quasiparticles in the clean system.

\section{Summary and discussion}~\label{conclusions}

In this paper we have studied the role of generic disorder in a Weyl semimetal, by considering its simplest realization, comprised of only two Weyl nodes. When the system resides in the proximity of semimetal-insulator quantum phase transition, the generalized Harris criterion suggests that such critical point is stable in the presence of weak but generic disorder. By contrast, a multicritical point appears in the phase diagram for strong disorder, where the Weyl semimetal, an insulator and a metallic phase meet. Within the framework of an appropriate $\epsilon$-expansion we show that, to the leading order, the critical exponents at such multicritical point are (i) dynamic scaling exponent $z=1+\epsilon_n/2$, and (ii) correlation length exponent $\nu=1/\epsilon_n$ that controls the relevance of disorder coupling, where $\epsilon_n=1/2$ for physical system. These findings are in good agreement with the ones obtained numerically, yielding $\nu=1.98 \pm 0.10$ and $z=1.26 \pm 0.05$.

On the other hand, when the system is deep inside the Weyl semimetal phase, we have shown that the continuous global chiral $U(1)$ symmetry plays a fundamental rule in classifying the disorder-driven Weyl semimetal-metal quantum phase transitions. The simplest realization of a Weyl semimetal is susceptible to eight types of disorder, among which only four preserve such chiral symmetry. Using two different $\epsilon$-expansions, we have shown that the chiral symmetric disorder driven semimetal-metal transition takes place through either a quantum critical point or a line of quantum critical points. Irrespective of details, the critical exponents to the leading order in $\epsilon$-expansions are given by $z=1+\epsilon/2 +{\mathcal O} (\epsilon^2)$ and $\nu=\epsilon^{-1}+{\mathcal O} (\epsilon^2)$, and $\epsilon=1$ corresponds to the physical situation. Even though these exponents can receive higher order corrections ${\mathcal O} (\epsilon^2)$, presently there is no known route to compute them in a controlled fashion beyond the leading order in $\epsilon$. Such unique set of exponents in the presence of generic chiral symmetric disorder gives birth to an \emph{emergent chiral superuniversality} across the Weyl semimetal-metal quantum phase transition.

Furthermore, we have performed a thorough numerical analysis of average density of states in Weyl semimetals with chiral symmetric disorder. The emergence of chiral superuniversality has been demonstrated through numerical analysis of average density of states near zero energy. We show that for any such disorder Weyl semimetal undergoes a continuous quantum phase transition into a diffusive metallic phase. Within the numerical accuracy, we find that across this transition $z \approx 1.5$ and $\nu \approx 1$, in excellent agreement with our field theoretic predictions obtained from leading order $\epsilon$-expansions (see Table~\ref{table-exponent} for comparison). The quality as well as reliability of our numerical analysis has been anchored through two completely different types of high-quality data collapses, shown in Fig.~\ref{numeric_analysis_figure}, in the entire phase diagram of a dirty Weyl semimetal for all possible chiral disorder.

For chiral symmetry breaking disorder, the Weyl semimetal-metal quantum phase transition also takes place through a critical point or a line of critical points, but the critical exponents are significantly different from the ones reported in the presence of chiral disorder. Even though the critical exponents across such semimetal-metal transition turn out to be slightly dependent on the renormalization group scheme, we always find $z>d$ and $\nu=1/\epsilon$ from leading order $\epsilon$-expansions. Consequently, all chiral symmetric or intra-node disorder (as well as higher gradient terms that are inevitably present in a lattice) become relevant at such putative line of critical points. As a result, inter-node disorder driven semimetal-metal phase transition is ultimately always governed by the chiral symmetric disorder, yielding $\nu \approx 1$ and $z \approx 3/2$, characteristic of chiral superuniversality. We anchor these outcomes by numerically extracting the scaling of average density of states in the presence of inter-node disorder, and the results are shown in Table~\ref{Tab:CSB_exponents} and Figs.~\ref{DOS_Fan_General} (lower panel) and \ref{numeric_analysis_figure_CSB}.

Even though we promoted such classification scheme in a Weyl semimetal with only two nodes, our prescription can easily be generalized to Weyl systems with multiple flavors, as well as topological Dirac semimetals with bonafide time-reversal symmetry that has recently been found in Cd$_2$As$_3$~\cite{cdas} and Na$_3$Bi~\cite{nabi} and the ones at the quantum critical point residing between two topologically distinct insulating vacua.

We here mention that $\epsilon_d$ expansion can be problematic beyond the leading order in $\epsilon_d$, since the contribution from diagrams (c) and (d) in Fig.~\ref{laddercrossing} and their higher-loop cousins are typically ultraviolet divergent and one looses the order by order control over the perturbative calculation~\cite{roy-dassarma-erratum, carpentier-1}. For example, it was shown in Refs.~\cite{roy-dassarma-erratum, carpentier-1} that to the order $\epsilon^3_d$ the correlation length exponent is (see also Ref.~\cite{Syzranov-exponent})
\begin{equation}
\nu^{-1}=\epsilon_d+\frac{1}{2} \epsilon^2_d +\frac{3}{8} \epsilon^3_d,
\end{equation}
respectively yielding $\nu^{(2)}= 0.66$ and $\nu^{(3)}=0.53$ to the two- and three-loop order for $\epsilon_d=1$. Upon implementing the Pad\'{e} resummation~\footnote{Note that $\nu^{(2)}=0.5$ is obtained from Pad\'{e} $[1 \mid 1]$ resummation, while $\nu^{(3)}= 0.33$ and $0.375$ are respectively obtained from Pad\'{e} $[2 \mid 1]$ and Pad\'{e} $[1 \mid 2]$ resummation. See Refs.~\cite{roy-dassarma-erratum, carpentier-1} for details.}, we obtain $\nu^{(2)}=0.5$ and $\nu^{(3)}= 0.33$ or $0.375$ (both being smaller than the mean-field value of $\nu = 1/2$). Hence, $\epsilon_d$-expansion runs into serious problem of convergence beyond the leading order. Such  a class of diagrams is, however, ultraviolet finite and thus does not contribute to renormalization group flow equations in the $\epsilon_m$ expansion scheme (see Appendix~\ref{Append:Ladder_Crossing_epsilonm}). 
We, therefore, believe that higher order perturbation theory within the framework of an $\epsilon_m$-expansion should be more controlled. 
Explicit higher order calculation in $\epsilon_m$-expansion and its corroboration with a newly proposed non-perturbative approach combined with the functional renormalization group analysis~\cite{carpentier-new} is, however, left as a challenging interesting problem for future investigation. Nonetheless, we note that leading order $\epsilon_d$
and $\epsilon_m$ expansions, as well as the functional renormalization group approach from Ref.~\cite{carpentier-new}, yield identical values
for the critical exponents, namely $z = 3/2$ and $\nu = 1$.

In addition to the Weyl semimetal-metal quantum phase transition, we also establish that a compressible Weyl metal undergoes a a subsequent transition at stronger disorder into a Anderson insulator. We track the typical density of states to pin the onset of such insulating phase that only accommodates localized states. In particular, we show that across the Weyl metal-insulator transition the typical density of states at zero energy ($\varrho_t(0)$) smoothly vanishes, and thus serving as bonafide order-parameter, while the average density of states remains non-critical across this transition. In addition, we also find that $\varrho_t(0)$ remains pinned in the Weyl semimetal phase and becomes finite in the metallic phase. Therefore, typical density of states at zero energy serves as a unified order-parameter across all possible disorder-driven quantum phase transition in a Weyl semimetal.

 Finally we comment on some non-perturbative effects of disorder in Weyl semimetals, such as puddles~\cite{dassarma-puddle}, Lishiftz tail~\cite{halperin}, and rare-region states and Griffiths physics~\cite{nandkishore, pixley-3}. Puddles are inevitable in real materials as there are always density fluctuations that locally shift the chemical potential away from the Weyl nodes, while maintaining the overall charge neutrality of the system. In addition, presence of disorder can also support \emph{quasi-localized} rare states at zero-energy even for subcritical strength of disorder~\cite{nandkishore, pixley-3}. Although such effects are important and interesting, they possibly do not affect the quantum critical behavior. Also the presence of finite average DOS close to zero energy for subcritical disorder does not necessarily imply a finite typical DOS at zero energy [$\varrho_t(0)$] and a finite dc conductivity as $T \to 0$, the hallmark signatures of a metal. By contrast, we find that $\varrho_t(0)$ remains pinned to zero for weak enough disorder, see Fig.~\ref{anderson:1}. In addition, whether generic disorder (inter and intranode) accommodates rare regions, remains to be examined. Furthermore, it is not clear if the rare states can survive when they hybridize with non-rare or critical states, residing close in energy. On the other hand, a recent numerical work has demonstrated that such non-perturbative effects can be systematically suppressed with a suitable choice of the distribution of disorder, while the critical properties across the Weyl semimetal-metal quantum phase transition remain \emph{almost} unchanged~\cite{pixley-4}. Therefore, rare and critical excitations appear to be decoupled from each other (based on present numerical evidence) and these effects do not alter any physical outcome we reported in this paper.

\acknowledgements

B. R. was supported by NSF-JQI-PFC and LPS-MPO-CMTC, and partially from Welch Foundation Grant No. C-1809 and NSF CAREER Grant no.~DMR-
1552327 of Matthew S. Foster (Rice University). We thank Sankar Das Sarma, Matthew Foster, Pallab Goswami and Soumya Bera for useful discussions. B. R. and R. J. S. are thankful to Nordita for hospitality during the workshop ``\emph{From Quantum Field Theories to Numerical Methods}" where part of this work was finalized.

\appendix

\begin{figure}
\subfigure[]{
\includegraphics[width=4cm,height=4cm]{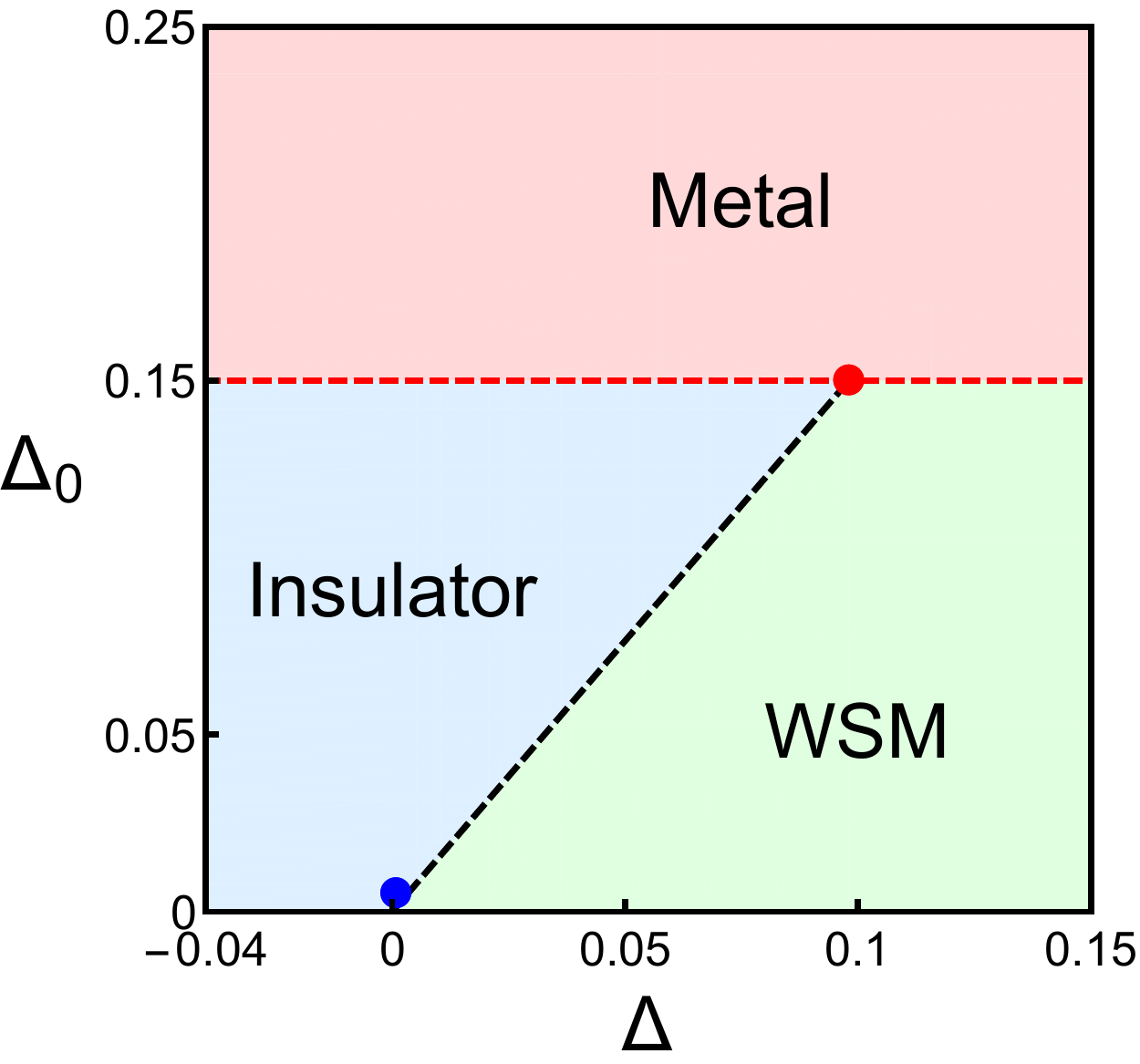}~\label{MCP_PD_Epsilonn}
}
\subfigure[]{
\includegraphics[width=4cm,height=4cm]{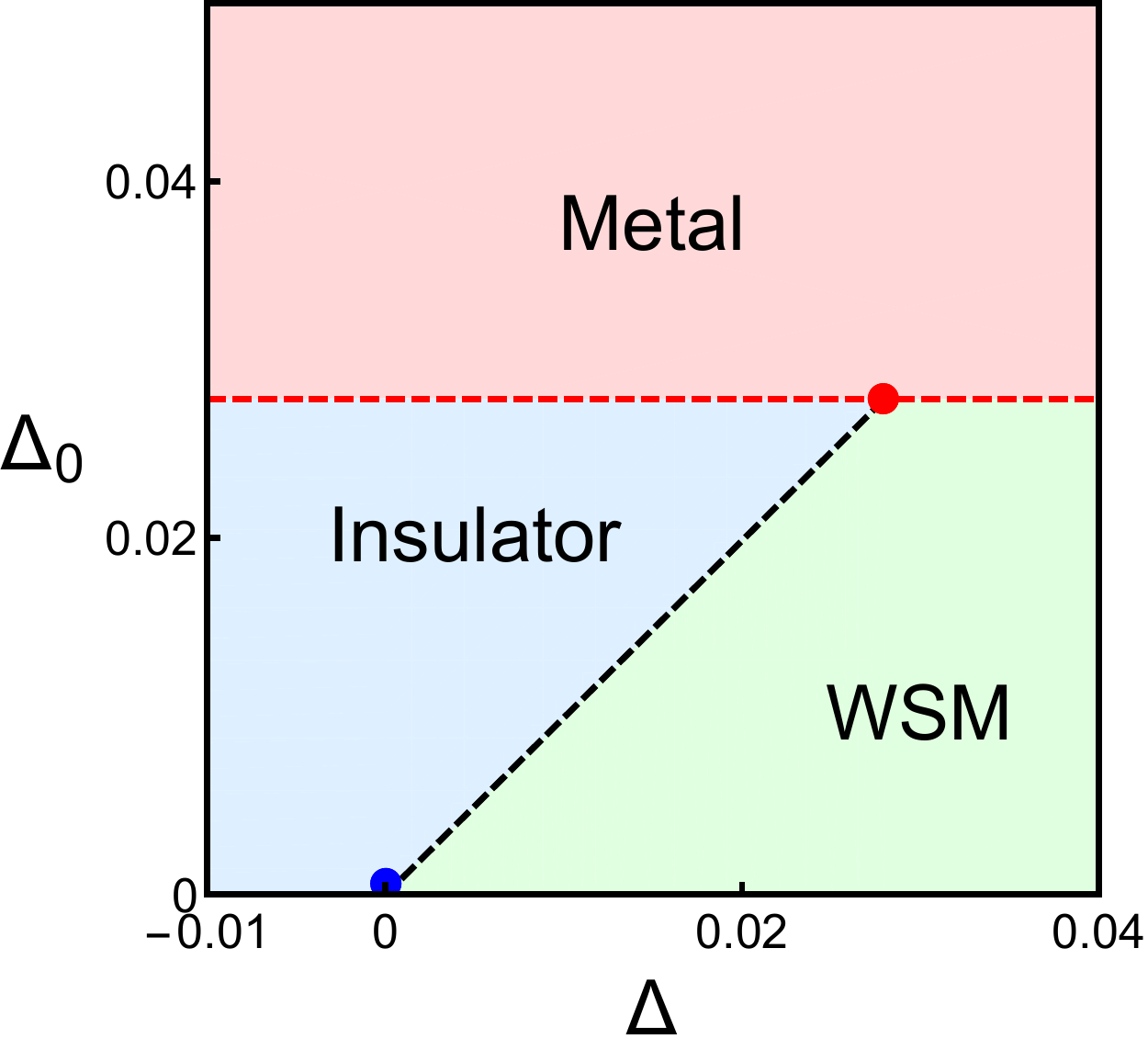}~\label{MCP_PD_Epsilond}
}
\caption[]{
The phase diagram of a dirty Weyl material residing in the close proximity to WSM-insulator QPT, obtained by solving the RG flow equations ~(\ref{RG_epsilon_n_subleading}) for (a) and ~(\ref{RG_WSM_Ins_Hard}) for (b). Here, $\Delta$ is the tuning parameter for WSM-insulator transition in the clean system and $\Delta_0$ is the strength of random charge impurities.
These two phase diagrams are qualitatively similar to the one obtained numerically, see Fig.~\ref{MCP_numerics_PD}(left), as the
WSM-insulator phase boundary shifts toward the semimetallic side with increasing (but weak) disorder.
}\label{Dirty_PD}
\end{figure}

\section{Generalized Harris criterion at WSM-insulator QCP}~\label{harris-generalized}

In this Appendix, we present a generalization of the Harris criterion applicable near the \emph{clean} WSM-insulator QCP. Let us first consider a generalized version of the Hamiltonian from Eq.~(\ref{Ins_WSM_QCP}) describing the gapless excitations residing at general WSM-insulator QCP~\cite{roy-goswami-juricic}
\begin{align}~\label{general_WSM_INS_QCP}
\hat{H}^{c}_{Q}({\bf k},\Delta) &= \alpha_c \left[ \sigma_1 k^c_\perp \cos(c \phi_{\bf k}) + \sigma_2 k^c_\perp \sin(c \phi_{\bf k}) \right] \nonumber \\
&+ \sigma_3 \left( b k^2_3 -\Delta \right),
\end{align}
where $k_\perp=\sqrt{k^2_x+k^2_y}$ and $\phi_k=\tan^{-1}\left( k_y/k_x \right)$. The above Hamiltonain for any value of $\Delta$ possesses the same symmetry, but describes two distinct phases: (i) a band insulator for $\Delta<0$ and (ii) WSM for $\Delta>0$, with $c$ representing the monopole charge of the Weyl nodes. Respectively for $c=1,2$ and $3$, single, double and triple WSMs are realized in a crystalline environment~\cite{Fang-HgCrSe, bergevig, nagaosa}. The effective dimensionality ($d_\ast$) of such critical semimetallic phase can be found from the corresponding imaginary time Euclidean action
\begin{equation}
S^{c}= \int d\tau d^2x_\perp dx_3 \; \psi^\dagger \left[ \partial_\tau + \hat{H}^{c}_{Q}({\bf k} \to -i \nabla,\Delta) \right] \psi,
\end{equation}
where $\psi$ is a two component spinor, describing the critical excitations residing at the WSM-insulator QCP. All parameters, such as $\alpha_c$ and $b$, remain invariant under the rescaling of space-time(imaginary) co-ordinates according to $\tau \to e^{l} \tau$, $(x,y) \to e^{l/c} (x,y)$, $x_3 \to e^{l/2} x_3$, when accompanied by the field normalization $\psi \to Z_\psi^{1/2} \psi$, where $Z_\psi=\exp\left[{-\left(\frac{2}{c} + \frac{1}{2}\right)l}\right] \equiv \exp[-d_\ast l]$. The spatial measure $d^2x_\perp dx_3 \to e^{d_\ast l} d^2x_\perp dx_3$, where $d_\ast=\left(\frac{2}{c} + \frac{1}{2}\right)$ is the \emph{effective dimensionality} of the system under the rescaling of spatial coordinates. Note that $\Delta$ in Eq.~(\ref{general_WSM_INS_QCP}) is the tuning (relevant) parameter at the WSM-insulator QCP, and the scaling dimension of $\Delta$, denoted by $[\Delta]$, is tied with the CLE ($\nu$) at this QCP, according to $\nu^{-1}=[\Delta]=1$. The stability of the clean WSM-insulator QCP against mass disorder [denoted by $V_z({\bf x})$ in Eq.~(\ref{generaldis_WSM_Ins})] can be assessed from the generalized Harris criterion, suggesting that such QCP is stable against mass disorder when
\begin{align}\label{harriscrit_general}
\nu > \frac{2}{d_\ast}=\frac{4 c}{4+c}.
\end{align}
Therefore, only the single ($c=1$) WSM-insulator QCP is stable against sufficiently weak mass/bond disorder. Furthermore, the stability of the WSM-insulator QCP in the presence of generic disorder, which appears similar to $V_z({\bf x})$ in Eq.~(\ref{generaldis_WSM_Ins}), can be established from the generalized Harris criterion [see Eq.~(\ref{harriscrit_general})]. Hence, a single WSM-insulator QCP is guaranteed to be stable against generic disorder. In this regard a comment is due. Our derivation of generalized Harris criterion differs from the original one in Ref.~\cite{harris}, where $d_\ast$ is replaced by the physical dimensionality of the system ($d$) and the CLE $\nu$ varies depending on the nature of the phase transition. On the other hand, within the framework of anisotropic scaling of spatial co-ordinates we always find $\nu=1$, but actual spatial dimension gets replaced by an effective dimensionality of the system ($d_\ast$) under the process of coarse graining. We believe that these two methods are complementary to each other.


\section{RG analysis near WSM-insulator QCP}~\label{Append_WSM_INS}

In this Appendix, we provide technical details of the RG calculations near the WSM-insulator QPT with disorder. First, we show the effects of subleading divergences in the RG flow equations within the $\epsilon_n$ expansion introduced in Sec.~\ref{WSM-Ins-QPT} and its consequences [see Sec.~\ref{subleading}]. Next we display the perturbative analysis of disorder near the WSM-insulator in an expansion about the lower critical dimension of the theory [see Sec.~\ref{hard_cut_off}].

\subsection{$\epsilon_n$ expansion}~\label{subleading}

Within the framework of $\epsilon_n$ expansion, discussed in Sec.~\ref{WSM-Ins-QPT}, after integrating out the fast Fourier modes within the Wilsonian shell $\Lambda e^{-l}< k_\perp < \Lambda$ and $0<k^2_3<\infty$ and accounting for subleading ultraviolet divergences, the RG flow equations read
\begin{widetext}
\allowdisplaybreaks[4]
\begin{align}~\label{RG_epsilon_n_subleading}
\beta_X&=-X\left( \Delta_0 +2 \Delta_\perp + \Delta_z \right) \; \left[ h_1(n) + h_2(n)\right)=(1-z)X, \: \:
\beta_\Delta=\Delta + \left[ \Delta \left[ h_1(n) + h_2(n) \right] - h_3(n) \right] \; \left( \Delta_0-2 \Delta_\perp+\Delta_z \right) \nonumber\\
\beta_{\Delta_0}&=-\epsilon_n \Delta_0 + 2 \Delta_0 \left( \Delta_0 + 2 \Delta_\perp + \Delta_z \right) \left[ h_1(n) + h_2(n) \right]
+4 \Delta_\perp \Delta_z h_1(n) + 4 h_2(n) \left[ \delta_{n, 2m} \Delta_0 \Delta_z + \delta_{n, 2m+1} \Delta^2_\perp \right] \nonumber \\
\beta_{\Delta_\perp}&=-\epsilon_n \Delta_\perp + 2\Delta_\perp \left(\Delta_z-\Delta_0 \right) h_2(n) + 2 \Delta_0 \Delta_z h_1(n)
+ 4 h_2(n) \Delta_\perp \left[ \delta_{n, 2m} \Delta_z + \delta_{n, 2m+1} \Delta_0 \right] \nonumber \\
\beta_{\Delta_z}&=-\epsilon_n \Delta_z  + 2 \Delta_z \left( 2 \Delta_\perp-\Delta_0-\Delta_z \right)
\left[ h_1(n) - h_2(n) \right] + 4 \Delta_0 \Delta_\perp h_1(n) + 2 h_2(n) \delta_{n,2m} \left( \Delta^2_0 + 2 \Delta^2_\perp +\Delta^2_z \right),
\end{align}
\end{widetext}
where $\delta_{n,m}$ is the Kronecker delta function, $n,m$ are  integers and $X=v,b$. Functions $h_i(n)$, $i=1,2,3$, are defined as
\allowdisplaybreaks[4]
\begin{align}~\label{fucntions_n}
h_1(n)&= \frac{\pi  (2 n-1) \csc \left(\frac{\pi }{2 n}\right)}{4 n^2} =1-\frac{1}{2 n} +{\mathcal O}\left(n^{-3} \right) \nonumber \\
h_2(n)&=\frac{\pi  \csc \left(\frac{\pi }{2 n}\right)}{4 n^2} =\frac{1}{2n} +{\mathcal O}\left(n^{-2} \right), \nonumber \\
h_3(n)&=\frac{\pi  (n-1) \sec \left(\frac{\pi }{2 n}\right)}{4 n^2} =\frac{\pi}{4 n} +{\mathcal O}\left(n^{-2} \right).
\end{align}
Therefore, as $n \to \infty$ contribution only from $h_1(n)$ survives and for any finite $n$, $h_2(n)$ and $h_3(n)$ give rise to \emph{subleading divergences}. The RG flow equations obtained by keeping only the leading divergence are shown in Eq.~(\ref{RG_WSM_INS}) of the main text. As we demonstrate below, at least to the leading order in $\epsilon_n$-expansion, inclusion of subleading
divergences affects $z$ only nominally, while leaving the CLE unchanged, and we find $\nu^{-1}=\epsilon_n=1/2$.

In Sec.~\ref{WSM-ins-RG} we neglected the quantum corrections arising from the non-trivial band dispersion in the $k_z$ direction. Note that the quantum corrections in the RG flow equations [see Eq.~(\ref{RG_epsilon_n_subleading})] can be systematically incorporated by keeping the terms to the leading order in $1/n$ from $h_j(n)$ for $j=1,2,3$ [see Eq.~(\ref{fucntions_n})], following the spirit of $1/N$ expansion, where $N$ counts number of fermion flavors~\cite{zinn-justin}. Therefore, our RG analysis is simultaneously controlled by two \emph{small} parameters $\epsilon_n$ (measuring deviation from marginal two spatial dimensions) and $1/n$ (controlling quantum corrections arising from band curvature along $k_z$) and only at the very end of the calculation we set $\epsilon_n=1/2$ and $n=2$. The resulting RG flow equations still support only two fixed points (similar to the ones reported in Sec.~\ref{WSM-ins-RG}):

1. $\left( \Delta, \Delta_0, \Delta_\perp, \Delta_z \right)=(0,0,0,0)$ representing the WSM-insulator QCP in the clean system, and 

2. The MCP where WSM, an insulator and the metal meet is now located at (obtained numerically)
\begin{equation}
\left( \Delta, \Delta_0, \Delta_\perp, \Delta_z \right) \approx \left(\frac{\pi}{8n}, \frac{1}{2}-\frac{0.40}{n}, \frac{0.195}{n}, \frac{0.185}{n} \right)\epsilon_n.
\end{equation}
The DSE at this MCP is [see Eq.~(\ref{DSE_MCP})]
\begin{equation}
z=1+\left( \frac{1}{2}-\frac{0.02}{n} \right)\epsilon_n,
\end{equation}
which for the physical relevant situation $\epsilon_n=1/2$ and $n=2$, yields $z=1.245$, extremely close to the one reported in Sec.~\ref{WSM-ins-RG}, namely $z=1.25$, obatined by neglecting quantum corrections arising from the non-trivial dispersion in the $k_z$ direction. Therefore, our proposed methodology allows to capture quantum corrections and extract the critical exponents at the MCP in a controlled fashion. The CLE, however, does not receive any $1/n$ corrections, yielding $\nu^{-1}=\epsilon_n$ as before. The resulting phase diagram after accounting for $1/n$ corrections is shown in Fig.~\ref{MCP_PD_Epsilonn}.

\subsection{$\epsilon^{\prime}_d$ expansion about lower critical dimension}~\label{hard_cut_off}

In this section we demonstrate the role of disorder in the vicinity of WSM-insulator QPT perturbatively using an $\epsilon^{\prime}_d$ expansion near the lower critical dimension $d_l=5/2$ in the theory, see Ref.~\cite{carpentier_MCP}, where $\epsilon^{\prime}_d=d-5/2$. As we will see the outcomes are qualitatively the same as in the $\epsilon_n$ regularization scheme. The exact values of the critical exponents are, however, different from the ones announced in Sec.~\ref{WSM-Ins-QPT}, although only slightly so, at least to the one-loop order. Upon integrating the fast modes within the shell $E_c e^{-l} < \sqrt{v^2 k^2_\perp+ b^2 k^4_z}< E_c$, where $E_c$ is the ultraviolet energy cutoff for critical excitations residing the WSM-insulator QCP, we arrive at the following flow equations to the leading order in $\epsilon^{\prime}_d$ expansion
\begin{align}~\label{RG_WSM_Ins_Hard}
\beta_{X}&=-5X\left( \Delta_0 +2 \Delta_\perp + \Delta_z \right)=(1-z)X, \nonumber \\
\beta_{\Delta}&= \Delta+ \left( \Delta-1 \right) \left[ \Delta_0-2 \Delta_\perp + \Delta_z \right], \nonumber \\
\beta_{\Delta_0} &=-\epsilon^{\prime}_d \Delta_0 + 10 \Delta_0 \left(\Delta_0+2 \Delta_\perp + \Delta_z \right)-16 \Delta_z \Delta_\perp, \nonumber \\
\beta_{\Delta_\perp} &= -\epsilon^{\prime}_d \Delta_\perp +2 \Delta_\perp \left(\Delta_z-\Delta_0 \right) +4 \Delta_z \left(\Delta_\perp-2 \Delta_0 \right), \nonumber \\
\beta_{\Delta_z} &= -\epsilon^{\prime}_d \Delta_z + 6 \Delta_z \left( 2 \Delta_\perp -\Delta_0 -\Delta_z \right) \nonumber \\
&+ 4 \left( \Delta^2_0 +\Delta^2+ \Delta_0 \Delta_z-4\Delta_0\Delta_\perp+2 \Delta^2_\perp \right).
\end{align}
for $X=v, b$, after defining the dimensionless disorder coupling constant as $\Delta_j \alpha \to \Delta_j$ for $j=0,\perp, z$, where $\alpha=E^{\epsilon^{\prime}_d}_c/\left(20 \pi^2 v^2 b^{1/2} \right)$ and $\Delta/E_c \to \Delta$. Then, $\beta-$function for $v$ and $b$ in the presence of disorder yields a scale dependent dynamic scaling exponent
\begin{align}
z(l)= 1+ 5 \left[ \Delta_0+ 2 \Delta_\perp+\Delta_z \right] (l).
\end{align}

The coupled RG flow equations from Eq.~(\ref{RG_WSM_Ins_Hard}) also support only two fixed points: (i) $\left(\Delta, \Delta_0, \Delta_\perp, \Delta_z \right)=(0,0,0,0)$, representing the WSM-insulator QCP in the clean limit (the blue dot in Fig.~\ref{MCP_PD_Epsilond}), and (ii) $\left(\Delta, \Delta_0, \Delta_\perp, \Delta_z \right)=(0.058,0.056,0.01, 0.02)\epsilon^\prime_d$ representing a multicritical point. The critical exponents at this multicritical point for the anisotropic critical semimetal-metal transition are
\begin{align}
\nu^{-1}=\epsilon^{\prime}_d, \: \: z=1+ 0.48 \epsilon^{\prime}_d,
\end{align}
which is extremely close to the ones reported in Sec.~\ref{WSM-ins-RG}, for $\epsilon^{\prime}_d=1/2$, leading to $z=1.24$ and $\nu=2$. Therefore, both methods produce qualitatively similar results near WSM-insulator QPT, and the obtained critical exponents for anisotropic semimetal-metal transition are extremely close to each other, at least to the leading order. The resulting phase diagram is shown in Fig.~\ref{MCP_PD_Epsilond}.


\section{Details of $\epsilon_m$ expansion}~\label{detailsfieldtheory}

In this appendix we display the detailed analysis of various one-loop diagrams, shown in Fig.~\ref{selfvertex}, within the framework of an $\epsilon_m$ expansion.

\subsection{Self-energy}

Let us first consider the self energy diagram in Fig.~\ref{selfvertex}(a). The expression for the self-energy reads
\begin{eqnarray}\label{eq:self-energy}
\Sigma(i\omega,{\bf k})&=&\sum_{N}\int\frac{d^d{\bf q}}{(2\pi)^d}N\,G_0(i\omega,{\bf k}-{\bf q})\,N\frac{\Delta_N}{q^m}\nonumber\\
&\equiv&\sum_N \Sigma_N(i\omega,{\bf k}),
\end{eqnarray}
with $d=3$, the summation is taken over all eight types of disorder (see Table~\ref{table-disorder}) and $q\equiv|{\bf q}|$.

The contribution from one-loop self-energy diagram from the disorder represented by the matrix $N$ reads
\begin{align}
\Sigma_N (i \omega, \boldsymbol k)= -i\Delta_N\int \frac{d^3 {\bf q}}{(2 \pi)^3} \; \frac{ N\left[ \gamma_0 \omega + v \gamma_j (k-q)_j\right]N}{ \left[ \omega^2 +v^2 ({\bf k}-{\bf q})^2 \right] q^m}.
\end{align}
We will evaluate the temporal and spatial components of the self-energy diagram separately. Let us first set $\boldsymbol k=0$, for which
\begin{align}
\Sigma_N (i \omega, 0) &=\Delta_N (-i \omega) \frac{N \gamma_0 N}{v^{3-m}} \; \int \frac{d^3 {\bf q}}{(2 \pi)^3} \; \frac{1}{(\omega^2+q^2) q^m} \nonumber \\
&= \Delta_N (-i \omega) \frac{N \gamma_0 N}{v^{3-m}} \; \frac{\Gamma\left(1+\frac{m}{2} \right)}{\Gamma(m/2)}  \\
&\times \int^1_0 dx x^{\frac{m}{2}-1} \; \int \frac{d^3 {\bf q}}{(2 \pi)^3} \frac{1}{\left[ q^2 + (1-x)\omega^2 \right]^{1+\frac{m}{2} } }, \nonumber
\end{align}
where $x$ is the Feynman parameter. Upon completing the integrals over $q$ and $x$, and setting $m=1-\epsilon$ (for brevity, we use here shorthand notation $\epsilon_m\rightarrow\epsilon$) we obtain
\begin{align}
\Sigma_N (i \omega, 0)=\left[ iN \gamma_0\omega N \right] \; \left( \frac{\Delta_N}{2 \pi^2 v^2} \right)\frac{1}{\epsilon} + {\mathcal O}(1).
\end{align}
Next we set $\omega=0$ and the spatial component of self-energy correction is then given by
\begin{align}
&\Sigma_N (0, \boldsymbol k) =\Delta_N\left[ -i N \gamma_j N \right] \; \frac{1}{v^{3-m}} \; \int \frac{d^3 {\bf q}}{(2 \pi)^3} \; \frac{(k-q)_j}{({\bf k}-{\bf q})^2 \; q^m} \nonumber \\
&=\Delta_N \left( \frac{ -i N \gamma_j N }{v^{3-m}} \right) \; \frac{\Gamma\left( 1+\frac{m}{2}\right)}{\Gamma(m/2)} \; \int^1_0 dx x^{\frac{m}{2}-1} \nonumber \\
&\times \int \frac{d^3 {\bf q}}{(2 \pi)^3} \; \frac{(k-q)_j}{\left[ q^2-2(1-x) {\bf q} \cdot {\bf k} + (1-x)k^2 \right]^{1+\frac{m}{2}} }.
\end{align}
After shifting the momentum variable according to ${\bf q}-(1-x){\bf k} \to {\bf q}$ and setting $m=1-\epsilon$, we obtain
\begin{align}
\Sigma_N (0, \boldsymbol k)= \left[ i N \gamma_j k_j N \right] \: \left( \frac{\Delta_N}{2 \pi^2 v^2} \right) \; \frac{k^{\epsilon}}{3 \epsilon} + {\mathcal O}(1).
\end{align}
 Hence, the total self energy correction reads
\begin{align}
\Sigma_N (i \omega, \boldsymbol k) =i N\bigg[ \gamma_0 \omega+\frac{1}{3} \gamma_j k_j\bigg]N \Delta_N \frac{1}{\epsilon} +{\mathcal O}(1),
\end{align}
where we have redefined $\Delta_N k^{\epsilon}/(2 \pi^2 v^2)\rightarrow \Delta_N$, which is Eq.~(\ref{self-energy-final}) in the main text.

\subsection{Vertex}

The vertex correction for the disorder vertex shown in Fig.~\ref{selfvertex}(b) with the matrix $N$ reads
\be
V_N({\bf k})=\sum_M\int\frac{d^3{\bf q}}{(2\pi)^3}M G_0(0,{\bf k}-{\bf q}) N G_0(0,{\bf k}-{\bf q}) M\frac{\Delta_M}{q^m},
\ee
where we kept only one external momentum as an infrared regulator. The last expression can be compactly written as
\be \label{eq:vertex-correction}
V_N({\bf k})=-\sum_M [M\gamma_j N \gamma_l M]\frac{\Delta_M}{v^2}I_{jl}({\bf k}),
\ee
where
\be\label{eq:vertex-integral}
I_{jl}({\bf k})=\int\frac{d^3{\bf q}}{(2\pi)^3}\frac{(k-q)_j (k-q)_l}{({\bf q}-{\bf k})^4 q^m}.
\ee
We now present the evaluation of the above integral
\begin{align}
I_{jl}&= \int \frac{d^3 {\bf q}}{(2 \pi)^3} \; \frac{(k-q)_j \; (k-q)_l}{({\bf k}-{\bf q})^4 \; q^m} \nonumber \\
&= \frac{\Gamma\left( 2+ \frac{m}{2}\right)}{\Gamma(m/2)} \int^1_0 dx \; x (1-x)^{\frac{m}{2}-1} \nonumber \\
&\times \int \frac{d^3 {\bf q}}{(2 \pi)^3} \frac{(k-q)_j \; (k-q)_l}{\left[ x({\bf k}- {\bf q})^2 + (1-x) q^2 \right]^{2+\frac{m}{2} } }.
\end{align}
After shifting the momentum variable as ${\bf q}-x {\bf k} \to {\bf q}$, we obtain
\begin{align}
I_{jl} &= \frac{\Gamma\left( 2+ \frac{m}{2}\right)}{\Gamma(m/2)} \int^1_0 dx \; x (1-x)^{\frac{m}{2}-1} \nonumber \\
& \times \int \frac{d^3 {\bf q}}{(2 \pi)^3} \frac{ \left( q- (1-x) k \right)_j \left( q- (1-x) k \right)_l }{\left[ q^2 + x (1-x) k^2 \right]^{2+\frac{m}{2} }} \nonumber \\
&= - \frac{k^{\epsilon}}{2 \pi^2} \: \frac{\delta_{jl}}{3 \epsilon} + {\mathcal O}(1),
\end{align}
after taking $m=1-\epsilon$, since only the $q-$dependent part in the numerator of the integrand yields a divergent contribution. We use the last expression to obtain Eq.\ (\ref{eq:vertex-general}) in the main text.

\begin{figure*}[t!]
\includegraphics[width=18cm,height=7.5cm]{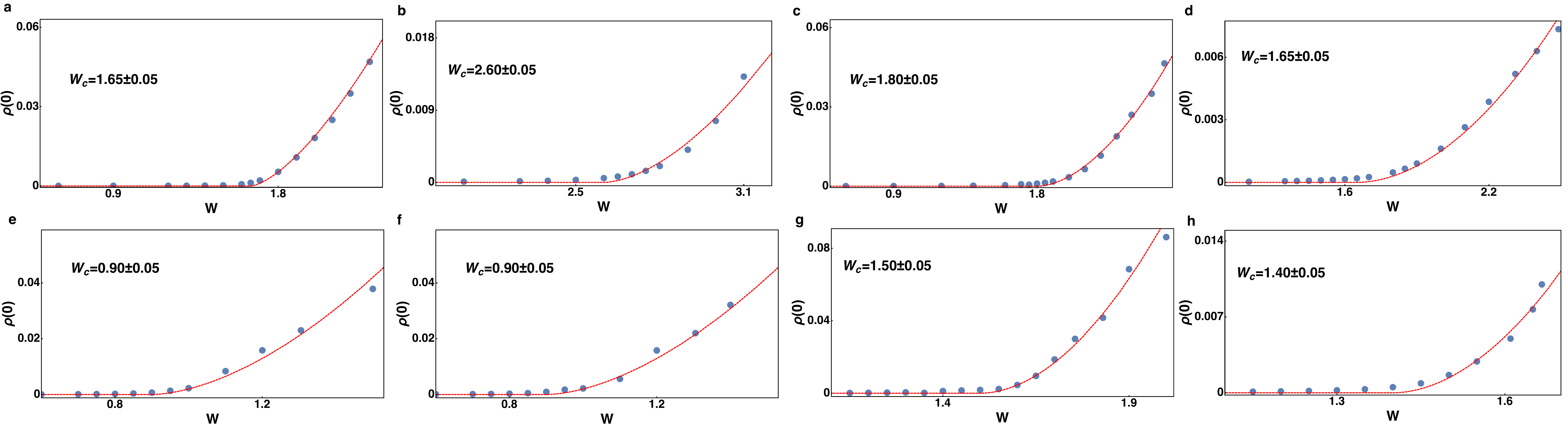}
\caption{ 
Scaling of the ADOS at zero energy [$\varrho(0)$] as a function of disorder strength W (chosen to be uniformly and independently distributed within a box $[-W,W]$) for (a) potential, (b) axial, (c) axial current (or magnetic), (d) current, (e) temporal tensor (or spin-orbit), (f) spatial tensor (or axial magnetic), (g) scalar mass and (h) pseudo-scalar mass disorder. The results are obtained by using KPM in a cubic lattice with linear dimension $L = 220$ in each direction. Note $\varrho(0)$ is pinned to zero up to a critical strength of disorder $W_c$, quoted in each panel [see also Table~\ref{table-exponent} and ~\ref{Tab:CSB_exponents}], and then it becomes finite, indicating the onset of a metallic phase. We reduce the uncertainty in determining the location of $W_c$ within the
error bar $\pm 0.05$, allowing us to minimize the \emph{fitting error} in the determination of $z$ and $\nu$ (see Appendix~\ref{Append:data_analysis} and Table~\ref{Table:Data-analysis}).
}~\label{rho0vsW}
\end{figure*}

\subsection{Ladder-crossing}~\label{Append:Ladder_Crossing_epsilonm}

We now show computation of two ladder diagrams from Fig.~\ref{laddercrossing}, in the $\epsilon_m$-expansion scheme. After setting
all the external frequencies to zero, diagram (c) from Fig.~\ref{laddercrossing} yields
\begin{eqnarray}
&&(5c)=\Delta_M \Delta_N \int\frac{d^3{\bf q}}{(2\pi)^3}\left[M\frac{i\gamma_l(p_1-q)_l}{({\bf p}_1-{\bf q})^2}N\right] \nonumber \\
&\times& \left[N \frac{i\gamma_s(p_2+q)_s}{({\bf p}_2+{\bf q})^2}M \right]\frac{1}{|{\bf q}|^m |{\bf p}_3-{\bf p}_1-{\bf q}|^m}.
\end{eqnarray}
Taking then $p_1 = p_3 = 0$ and keeping only the most singular
contribution, we obtain
\begin{equation}\label{ladder-integral}
(5c) \sim \Delta_M \Delta_N \int\frac{d^3{\bf q}}{(2\pi)^3}\frac{1}{|{\bf q}|^{2m}({\bf p}_2+{\bf q})^2}.
\end{equation}
Here, we used that $[\gamma_l,X]=0$ or $\{ \gamma_l, X\}=0$ for all $j=1,2,3$ and $X=M,N$, as well as $[M,N]=0$ or $\{M,N \}=0$.  Computation of the last integral yields
\begin{equation}
(5c) \sim \Delta_M \Delta_N ({\bf p}_2^2)^{\frac{1}{2}-m}\frac{\Gamma(\frac{1}{2})\Gamma(\frac{3}{2}-m)\Gamma(m-\frac{1}{2})}{(4\pi)^{3/2}\Gamma(m)\Gamma(2-m)},
\end{equation}
which is finite in the expansion in $\epsilon_m=1-m$ as $m \to 0$. This is also expected based on the power counting of the integral in Eq.~(\ref{ladder-integral}). The diagram Fig.~\ref{laddercrossing}(d) reads
\begin{eqnarray}
&&(5d)=\Delta_M \Delta_N \int\frac{d^3{\bf q}}{(2\pi)^3}\left[M\frac{i\gamma_l(p_1-q)_l}{({\bf p}_1-{\bf q})^2} N \right] \\
&\times& \left[M\frac{i\gamma_s(p_1+p_2-p_3-q)_s}{({\bf p}_1+{\bf p}_2-{\bf p}_3-{\bf q})^2} N \right]\frac{1}{|{\bf q}|^m |{\bf p}_1-{\bf p}_2-{\bf q}|^m}. \nonumber 
\end{eqnarray}
Taking then ${\bf p}_1={\bf p}_3=0$ and keeping only the most singular contribution, we obtain 
\begin{equation}
(5d) \sim \Delta_M \Delta_N \int\frac{d^3{\bf q}}{(2\pi)^3}\frac{1}{|{\bf q}|^{2m}({\bf p}_2-{\bf q})^2},
\end{equation}
identical to the integral from Eq.~(\ref{ladder-integral}), after substituting ${\bf q} \to -{\bf q}$. Therefore this diagram is also ultraviolet finite, confirming that both ladder diagrams are finite in the $\epsilon_m$-expansion, irrespective of the choice
of $M$ and $N$.

The reason for these two diagrams yielding ultraviolet finite contribution is the following: since disorder propagator is \emph{momentum dependent} in the $\epsilon_m$-expansion scheme (unlike the situation in $\epsilon_d$-expansion) only the self-energy and vertex diagrams [see Fig.~\ref{selfvertex}] containing only one disorder line are ultraviolet divergent and contribute RG flow equations. By contrast, each of the two ladder diagrams [see Fig.~\ref{laddercrossing}(c) and (d)] contains two disorder lines, yielding ultraviolet finite contribution and thus do not influence the RG flow equations. Now readers can convince themselves that such distinction between these two sets of Feynman diagrams persists to any order in perturbation theory. Hence, in the $\epsilon_m$-expansion scheme ladder diagrams never contribute and we do not generate any short-range disorder.

\begin{figure*}[t!]
\subfigure[]{
\includegraphics[width=4.1cm,height=3.25cm]{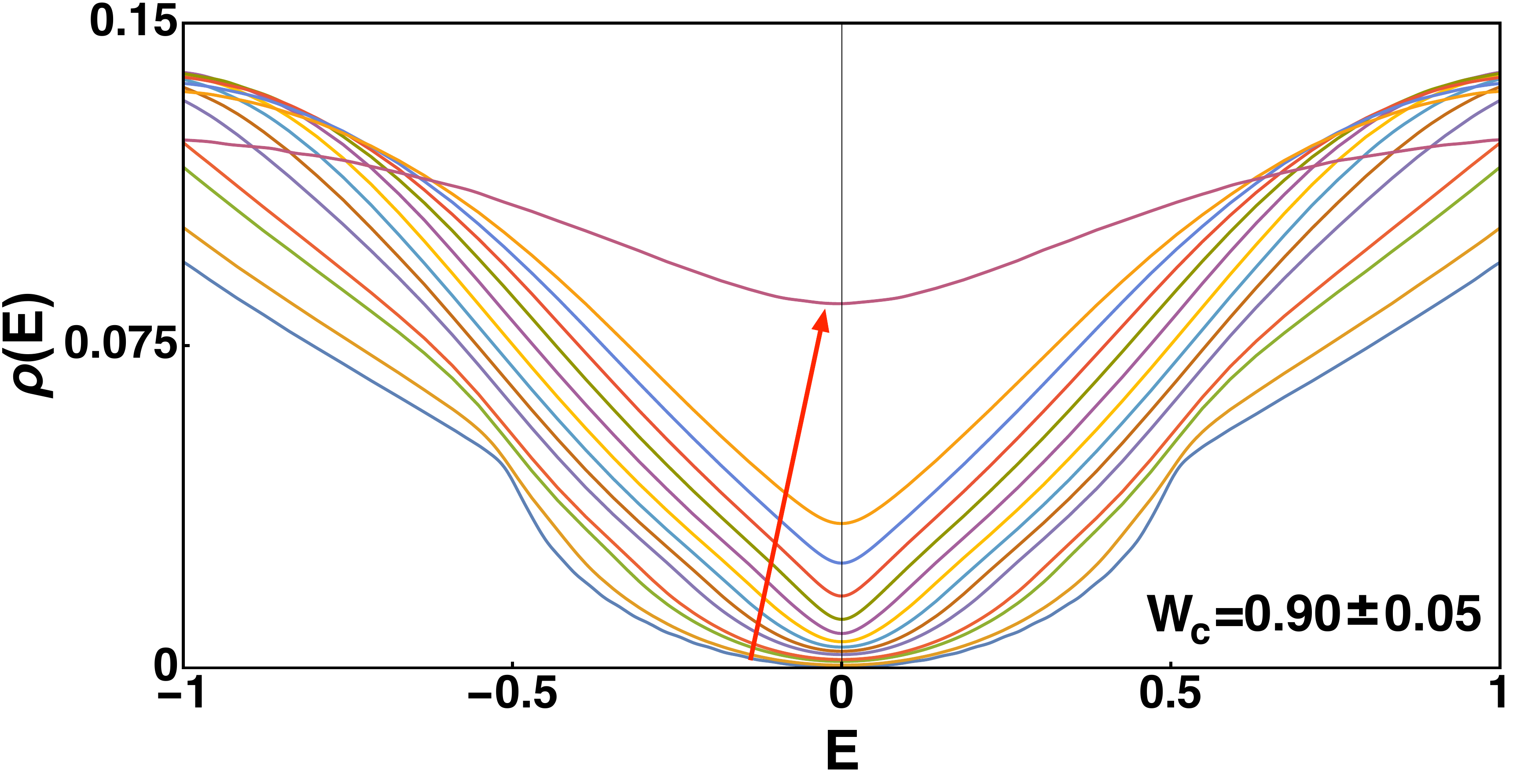}~\label{fan_correlated}
}
\subfigure[]{
\includegraphics[width=4.1cm,height=3.25cm]{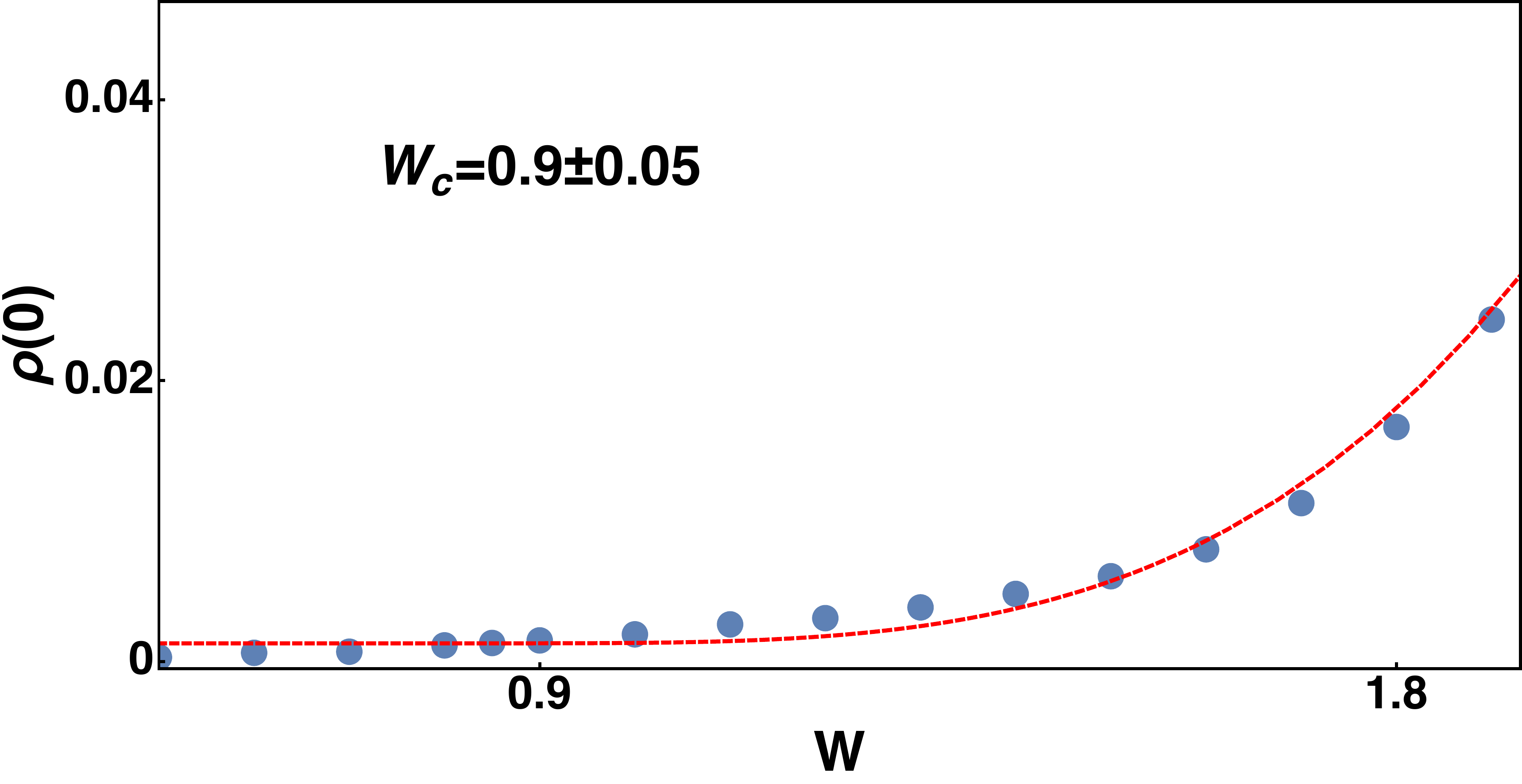}~\label{rho0_correlated}
}
\subfigure[]{
\includegraphics[width=4.1cm,height=3.25cm]{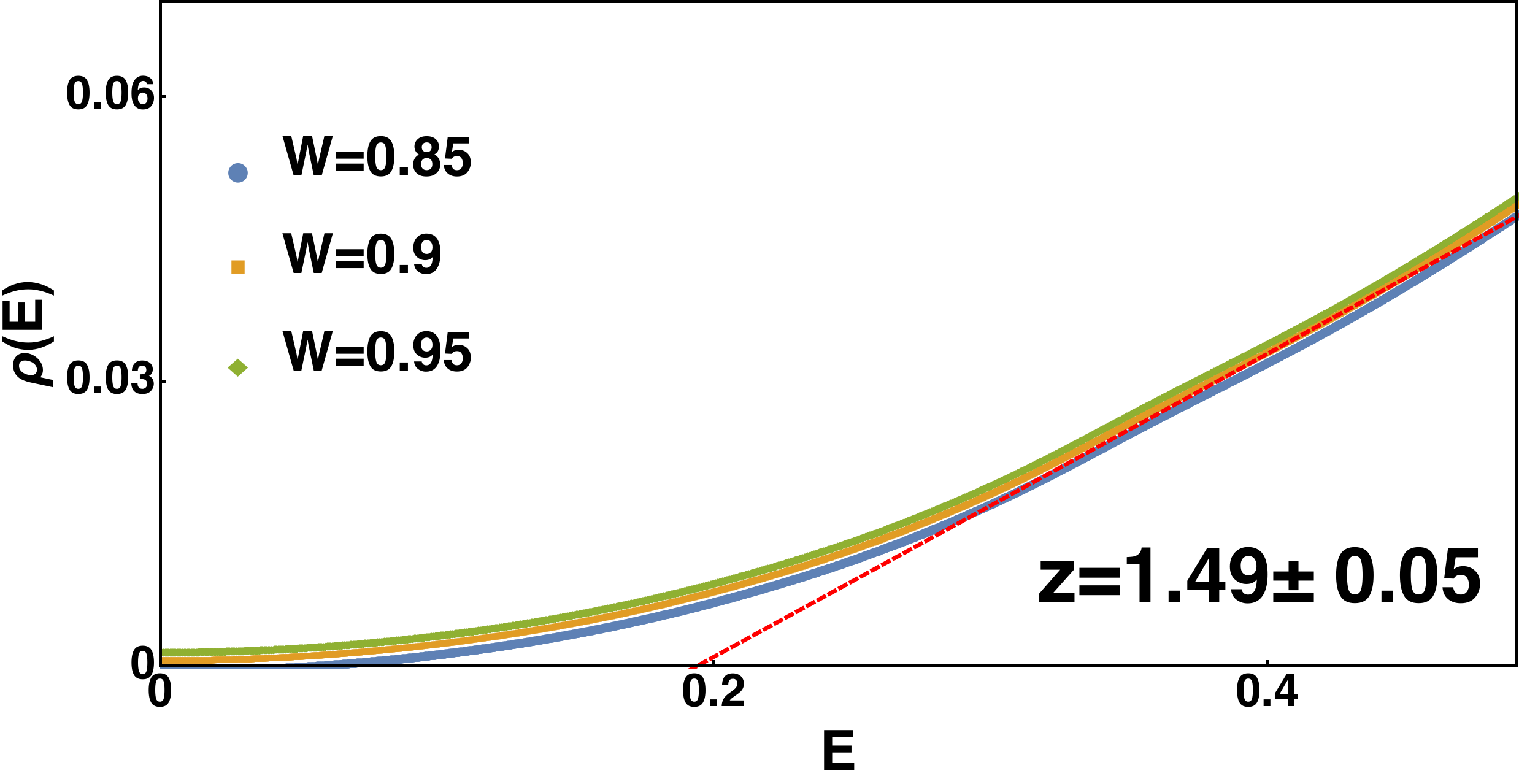}~\label{z_correlated}
}
\subfigure[]{
\includegraphics[width=4.1cm,height=3.25cm]{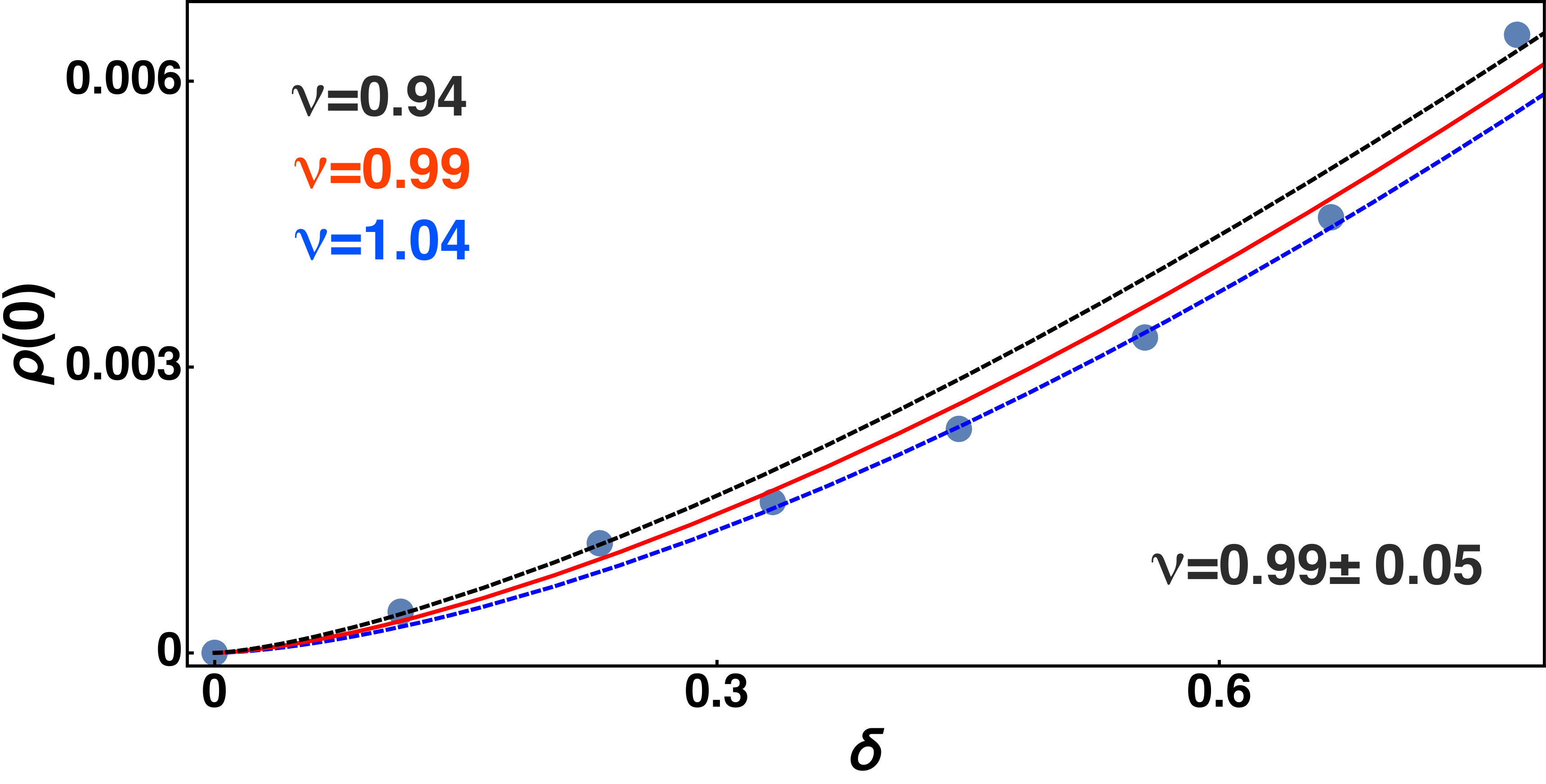}~\label{nu_correlated}
}
\caption{ Numerical analysis of average DOS in the presence of correlated potential disorder [see Append.~\ref{Append:Correlated_Disorder} and Eq.~(\ref{correlated_distribution})].
Scaling of (a) average DOS [$\varrho(E)$] with increasing strength of disorder (in the direction of the \emph{red arrow}), (b) average DOS at
zero energy [$\varrho(0)$] with increasing disorder in the system, yielding critical disorder strength $W_c = 0.90 \pm 0.05$, (c) average DOS
[$\varrho(E)$] around critical disorder $W = 0.85$ (blue), $0.90$ (yellow) and $0.95$ (green) [fitting $\varrho(E)$ with $|E|^{d/z-1}$, we obtain $z = 1.49 \pm 0.05$], (d) $\varrho(0)$ with the reduced distance from the WSM-metal critical point ($\delta$), yielding
$\nu = 0.99 \pm 0.05$ with $W_c=0.90$ and $z=1.49$. For discussion on error analysis in the determination of $\nu$, see Appendices~\ref{Append:data_analysis} and ~\ref{Append:Correlated_Disorder}, as well as Table~\ref{Table:Data-analysis} (the second last row).
}
\end{figure*}

\section{Lattice realization of generic disorder in Weyl semimetal}~\label{disorder_lattice}

In this appendix, we demonstrate the lattice realization of \emph{sixteen} possible fermionic bilinears (shown in Table~\ref{table-disorder}) from the two band tight-binding model, displayed in Eqs.~(\ref{TB_Weyl}) and (\ref{wilson}). By the virtue of the chosen tight-binding model, our construction is based on two features:

$\bullet$\; Since two Weyl nodes are located on the $k_z$ axis at $\pm k^{0}_z=\pm \pi/(2a)$, any fermionic bilinear \emph{odd} under the exchange of two Weyl nodes, can be realized by adding $h=\sum_{{\bf k}} \Psi^\dagger_{{\bf k}} \sin(k_z a) \sigma_j \Psi_{{\bf k}}$ to the tight binding model, where $j=0,1,2,3$. Such perturbation corresponds to an imaginary hopping along the $z$ direction, and does not renormalize the band width.

$\bullet$\; Any fermionic bilinear that couples two Weyl nodes, which therefore necessarily breaks  translational symmetry, can be realized through a periodic and \emph{commensurate} modulation of the nearest-neighbor hopping amplitude, but only along the $z$ direction.

With these two construction principles we can realize all sixteen fermion bilinears by adding the following terms to the tight-binding Hamiltonian.
\begin{enumerate}
\item{Regular chemical potential:
$$\sum_{{\bf r}} \Psi^\dagger_{{\bf r}} V({\bf r}) \sigma_0 \Psi_{{\bf r}},$$ }
\item{Axial chemical potential:
$$\sum_{{\bf r}} \Psi^\dagger_{{\bf r}} \left[ \frac{i V({\bf r})}{2} \sigma_0\right] \Psi_{{\bf r}+\hat{e}_3} +H.c.,$$}
\item{Abelian current:
$$ \sum_{{\bf r}} \left[ \Psi^\dagger_{{\bf r}} \left[ \frac{i V({\bf r})}{2} \sigma_3 \right] \Psi_{{\bf r}+\hat{e}_3} +H.c. + \Psi^\dagger_{{\bf r}} V({\bf r}) \left( \sigma_1+\sigma_2 \right) \Psi_{{\bf r}} \right],$$   }
\item{Abelian axial-current:
$$ \sum_{{\bf r}} \left[ \Psi^\dagger_{{\bf r}} \left[ \frac{i V({\bf r})}{2} \left(\sigma_1+\sigma_2\right) \right] \Psi_{{\bf r}+\hat{e}_3} +H.c. + \Psi^\dagger_{{\bf r}} V({\bf r}) \sigma_3 \Psi_{{\bf r}} \right],$$   }
\item{Temporal components of tensor:
 $$\sum_{{\bf r}} \sum_{j} (-1)^{j}  \Psi^\dagger_{{\bf r},j} V(r) \left[ \sigma_1+\sigma_2+i \sigma_0 \right] \Psi_{{\bf r}, \hat{e}_3,j+1} + H.c,$$  }
\item{Spatial components of tensor:
$$\sum_{{\bf r}} \sum_{j} (-1)^{j}  \Psi^\dagger_{{\bf r},j} V(r) \left[\sigma_0+ i \sigma_1+i \sigma_2 \right] \Psi_{{\bf r}, \hat{e}_3,j+1} + H.c,$$ }
\item{Scalar mass:
$$\sum_{{\bf r}} \sum_{j} (-1)^{j}  \Psi^\dagger_{{\bf r},j} \left[ V(r) \sigma_3 \right] \Psi_{{\bf r}, \hat{e}_3,j+1} + H.c,$$}
\item{Pseudo-scalar mass:
$$\sum_{{\bf r}} \sum_{j} (-1)^{j}  \Psi^\dagger_{{\bf r},j} \left[ i V(r) \sigma_3 \right] \Psi_{{\bf r}, \hat{e}_3,j+1} + H.c.$$}
\end{enumerate}
Thus, within the simplest realization of a Weyl semimetal from a tight-binding model on a cubic lattice, one can realize all possible disorder couplings by choosing $V({\bf r})$ as a random variable, and numerically study possible WSM-metal QPTs. The scaling of ADOS at zero energy for all above eight disorder is shown in Fig.~\ref{rho0vsW}.

\begin{table*}[t!]
\small
\begin{tabular}{|c|l|l|l|l|l|l|l|l|l|c|}
\hline
\multirow{3}{*}{Disorder} & \multicolumn{3}{c|}{$W_c-\delta W_c$} & \multicolumn{3}{c|}{${\bf W_c}$} & \multicolumn{3}{c|}{$W_c+\delta W_c$} & \multirow{3}{*}{$\nu$} \\ \cline{2-10}
  & \multicolumn{1}{c|}{$z-\delta z$} & \multicolumn{1}{c|}{$z$} & \multicolumn{1}{c|}{$z+\delta z$} & \multicolumn{1}{c|}{$z-\delta z$} & \multicolumn{1}{c|}{${\mathbf z}$} & \multicolumn{1}{c|}{$z+\delta z$} & \multicolumn{1}{c|}{$z-\delta z$} & \multicolumn{1}{c|}{$z$} & \multicolumn{1}{c|}{$z+\delta z$} &                  \\ \cline{2-10}
  & $\nu$ & $\nu$ & $\nu$ & $\nu$ & $\nu$ & $\nu$ & $\nu$ & $\nu$ & $\nu$ &         \\ \hline \hline
$\Delta_V$ & $0.97(0.06)$ & $1.00 (0.05)$ & $1.02(0.06)$ & $0.97(0.05)$ & ${\bf 0.99}(0.05)$ & $1.01(0.07)$ & $0.98 (0.06)$ & $0.99(0.07)$ & $1.01(0.05)$ & $1.00 (0.08)$  \\ \hline
$\Delta_A$ & $1.04(0.06)$ & $1.07(0.05)$ & $1.09(0.05)$ & 1.01 (0.05) & ${\bf 1.03}(0.05)$ & $1.08(0.08)$ & $1.04 (0.05)$ & $1.06(0.07)$ & $1.08(0.08)$  & $1.06(0.10)$  \\ \hline
$\Delta_M$ & $0.99(0.06)$ & $1.02(0.05)$ & $1.05(0.05)$ & $1.00(0.06)$ & ${\bf 1.03}(0.05)$ & 1.05(0.05) & $1.01(0.05)$ & $1.05(0.05)$ & $1.07(0.06)$  & $1.03(0.10)$  \\ \hline
$\Delta_C$ & $1.01(0.05)$ & $1.04(0.06)$ & $1.07(0.06)$ & $0.99(0.05)$ & ${\bf 1.03}(0.05)$ & 1.06(0.06) & $0.99(0.05)$ & $1.03(0.06)$ & $1.06(0.05)$  & $1.02(0.09)$  \\ \hline
$\Delta_{SO}$ & $1.01(0.05)$ & $1.03(0.06)$ & $1.06(0.05)$ & $0.97(0.07)$ & ${\bf 1.01}(0.05)$ & 1.04(0.05) & $0.99(0.06)$ & $1.02(0.06)$ & $1.05(0.05)$  & $1.01(0.10)$  \\ \hline
$\Delta_{AM}$ & $0.99(0.07)$ & $1.03(0.06)$ & $1.07(0.05)$ & $0.97(0.08)$ & ${\bf 1.01}(0.05)$ & 1.05(0.06) & $0.97(0.06)$ & $1.02(0.05)$ & $1.06(0.06)$  & $0.99(0.12)$  \\ \hline
$\Delta_{S}$ & $0.95(0.06)$ & $0.99(0.05)$ & $1.03(0.07)$ & $0.95(0.05)$ & ${\bf 0.97}(0.05)$ & 1.01(0.07) & $0.95(0.07)$ & $0.99(0.05)$ & $1.02(0.06)$  & $0.99(0.12)$  \\ \hline
$\Delta_{PS}$ & $0.99(0.08)$ & $1.02(0.05)$ & $1.05(0.06)$ & $0.99(0.05)$ & ${\bf 1.02}(0.06)$ & 1.06(0.05) & $0.97(0.05)$ & $1.01(0.05)$ & $1.05(0.06)$  & $1.01(0.11)$  \\ \hline
$\Delta^{{\rm corr}}_V$ & $0.96(0.07)$ & $1.00(0.05)$ & $1.04(0.06)$ & $0.96(0.05)$ & ${\bf 0.99}(0.05)$ & 1.03(0.07) & $0.96(0.08)$ & $1.01(0.05)$ & $1.05(0.05)$  & $0.99(0.11)$  \\ \hline
$\Delta_V$ (MCP) & $1.94(0.07)$ & $1.97(0.06)$ & $2.03(0.05)$ & $1.92(0.05)$ & ${\bf 1.97}(0.05)$ & 2.01(0.06) & $1.97(0.06)$ & $1.99(0.05)$ & $2.02(0.06)$  & $1.98(0.10)$  \\ \hline
\end{tabular}
\caption{  Details of the data analysis for the computation of the correlation length exponent $\nu$ across the WSM-metal QPT
driven by potential ($\Delta_V$), axial potential ($\Delta_A$), magnetic ($\Delta_M$), current ($\Delta_C$), spin-orbit ($\Delta_{SO}$), axial-magnetic ($\Delta_{AM}$), scalar mass ($\Delta_S$) and pseudo-scalar mass ($\Delta_{PS}$) disorder [see Table~\ref{table-disorder} for definition], where the disorder is assumed to be uniformly and independently distributed within $[-W,W]$ (first 8 rows). We here show the variation of $\nu$ with (a) the variation of the location of the WSM-metal QCP (denoted by $W_c \pm \delta W_c$) and (b) from the ``fitting error" of $z$ (denoted by $z \pm \delta z$). See Table~\ref{table-exponent} and Table~\ref{Tab:CSB_exponents} for $W_c$, $z$ and $\delta z$, and throughout we have $\delta W_c = 0.05$. The second last row represents the same analysis but in the presence of correlated potential disorder ($\Delta^{\rm corr}_V$), discussed in Appendix~\ref{Append:Correlated_Disorder}, while the last row shows similar analysis across the potential disorder driven critical anisotropic semimetal-metal transition through the multi-critical point (discussed in Sec.~\ref{WSM-Ins-QPT}). The quantities in parentheses represent corresponding ``\emph{fitting error}" for a given value of $\nu$ for the specific value of critical disorder strength and dynamic scaling exponent, see Appendix~\ref{Append:data_analysis} for discussion. Each value of $\nu$ and the corresponding ``fitting error" is determined by comparing $\varrho(0)$ with $\delta^{(d-z)\nu}$ for given values of $W_c$ and $z$, see for example, Fig.~\ref{MCP_CLE}, second column of Figs.~\ref{numeric_analysis_figure} and ~\ref{numeric_analysis_figure_CSB}, Fig.~\ref{nu_correlated}. The last column shows the corresponding values of the correlation length exponent, accompanied by \emph{maximal} fitting error. These analyses were performed in the largest system (see Secs.~\ref{MCP_numerics_analysis}, \ref{numerics_analysis} and Appendix~\ref{Append:Correlated_Disorder} for details). Data collapse in Figs.~\ref{collapse_MCP}, ~\ref{numeric_analysis_figure}, ~\ref{numeric_analysis_figure_CSB} and ~\ref{collapse_correlated} are shown with the values for $W_c$, $z$ and $\nu$ shown in bold font. 
}~\label{Table:Data-analysis}
\end{table*}


\section{Details of data analysis}~\label{Append:data_analysis}

In this Appendix, we present quintessential details of data analysis, which we employ for (a) anisotropic semimetal-metal QPT through the MCP [the blue dot in Fig.~\ref{MCP_numerics_PD}(left)] as well as (b) WSM-metal QPT [for both uncorrelated and correlated disorder (see Append.~\ref{Append:Correlated_Disorder})].

\subsection{Estimation of $W_c$}:

We determine the critical strength of disorder ($W_c$) by computing the average DOS at zero energy $\varrho(0)$. Note $\varrho(0) = 0$ in the semimetallic phase as well as at the semimetal-metal QCP. But $\varrho(0)$ is finite in a metal. Hence, by computing $\varrho(0)$ we can pin down $W_c$, as shown in Figs.~\ref{rhovsW_MCP}, ~\ref{rho0vsW} and ~\ref{rho0_correlated}. We minimize the error $\delta W_c$ in determining $W_c$ by increasing the number of data points around $W_c$, and throughout $\delta W_c = 0.05$. Note small $\delta W_c$ is the source of small ``\emph{fitting error}" in the quoted values of $z$ and $\nu$.

\subsection{Estimation of $z$} 

To determine the DSE $z$, we compare $\varrho(E)$ vs. $|E|^{d/z-1}$, for $W_c-\delta W_c$, $W_c$ and $W_c+\delta W_c$. Since continuous semimetal-metal QPT is always characterized by a unique $z$, we fit $\varrho(E)$ for a specific value of $z$. But, due to the finite-size effects (which are non-universal and also depend on the choice of disorder distribution), such a fit never goes through zero at $E = 0$, although $\varrho(0) \approx 0$ (within numerical accuracy). Hence, to find $z$ we search for its value that yields good fit with $\varrho(E)$ at finite energy (i.e., we target to fit $\varrho(E)$ with $|E|^{d/z-1}$ within the quantum critical regime, where finite size effects are nominal). For three values of $W$, namely $W_c-\delta W_c$, $W_c$ and $W_c+\delta W_c$, we obtain three values of $z$, namely $z-\delta z$, $z$ and $z +\delta z$, where $\delta z$ is the \emph{fitting error} associated with $z$ [see Table~\ref{table-exponent} and Table~\ref{Tab:CSB_exponents}, Figs.~\ref{MCP_DSE} and ~\ref{z_correlated}]. The red lines shown in Figs.~\ref{MCP_DSE}, \ref{numeric_analysis_figure} (first column), ~\ref{numeric_analysis_figure_CSB} (first column), and \ref{z_correlated} represent plot of $\varrho(E)$ vs. $|E|^{d/z-1}$ for the mean value of $z$.

\subsection{Estimation of $\nu$} 

Finally we determine CLE $\nu$, for which we compare $\varrho(0)$ in the metallic phase with $\delta = (W-W_c)/W_c$ (where $W_c$ is the mean-value of critical disorder strength), since $\varrho(0) \sim \delta^{(d-z)\nu}$. 
Within the maximally allowed range of disorder $W_c < W < W_\ast$ (due to the inevitable presence of a subsequent Anderson transition, explained below), we fit $\varrho(0)$ vs. $\delta^{(d-z)\nu}$, yielding CLE $\nu \pm \delta \nu$ with the mean value of DSE $z$, reported in Fig.~\ref{MCP_CLE}, second column of Figs.~\ref{numeric_analysis_figure}, \ref{numeric_analysis_figure_CSB}, and Fig.~\ref{nu_correlated}. The black, red and blue curves are respectively shown for $\nu \in \{ \nu-\delta \nu, \nu, \nu+\delta \nu \}$, \emph{encompassing all data points}.

To further improve our numerical analysis of $\nu$, we generate two additional data sets for $\varrho(0)$ vs. $\delta$, with $W_c \to W_c \pm \delta W_c$, but still within the range $W_c \pm \delta W_c<W < W_\ast$.  Performing the same analysis explained above, we obtain another range of CLE $\nu$. Finally, for all three sets of $\varrho(0)$ vs $\delta$, we extract the CLE taking $z \to z \pm \delta z$. With such extensive data analysis, summarized in Table~\ref{Table:Data-analysis}, we acquire maximal fitting error in the determination of $\nu$, and these values are quoted in Table~\ref{table-exponent}, Table~\ref{Tab:CSB_exponents} (for WSM-metal QPT), reported in Sec.~\ref{MCP_numerics_analysis} for anisotropic semimetal-metal QPT through the MCP and in Appendix~\ref{Append:Correlated_Disorder} for correlated potential disorder.

Finally, we highlight an important issue related to the range of disorder ($\delta$) over which we perform numerical analysis for $\nu$. Note that a three-dimensional Weyl metal undergoes a second QPT into the AI phase (discussed in Sec.~\ref{anderson}, see also Fig.~\ref{Global_PD}). Across the Anderson transition although average DOS at zero energy [$\varrho_a(0)$] remains smooth, it decreases monotonically. As shown in Fig.~\ref{anderson:1}, Anderson transition (for potential disorder) takes place at $W_{c,2} \approx 9.30$, but $\varrho_a(0)$ starts to decrease for much weaker disorder $W \geq 2.90$. On the other hand, the WSM-metal QPT takes place around $W_{c,1} \approx 1.65$. Hence, for $W > 2.9$ the Weyl metal starts to approach the Anderson fixed point, and to properly extract $\nu$ associated with the WSM-metal QPT we can only fit $\varrho_a(0)$ with $\delta^{(d-z)\nu}$ within the range $1.65 \pm 0.05 < W < 2.90$, the \emph{maximally allowed range of disorder}, mentioned earlier, with $W_\ast < 2.90$. The maximal value of $\delta$ shown in the second column of first row in Fig.~\ref{numeric_analysis_figure} is $\approx 0.60$, yielding corresponding $W \approx 2.65$ with $W_c=1.65$ (ensuring that the system is still \emph{sufficiently} far from the Anderson fixed point). Therefore, in our data analysis for $\nu$, we cover the maximally allowed range of disorder so that the system still falls outside the basin of attraction for the Anderson transition. No further variation of range of $\delta$ is permitted due to the very nature of the global phase diagram, shown in Fig.~\ref{Global_PD}. This way we acquire a \emph{maximal} ``fitting error" in determination of $\nu$ arising from the range of $\delta$ we consider, that nonetheless encompasses all data points. We follow the same strategy for the analysis of $\nu$ in the presence of arbitrary disorder driving WSM-metal QPT or the transition across the
MCP.


\begin{figure}[t!]
\subfigure[]{
\includegraphics[width=4cm,height=3.5cm]{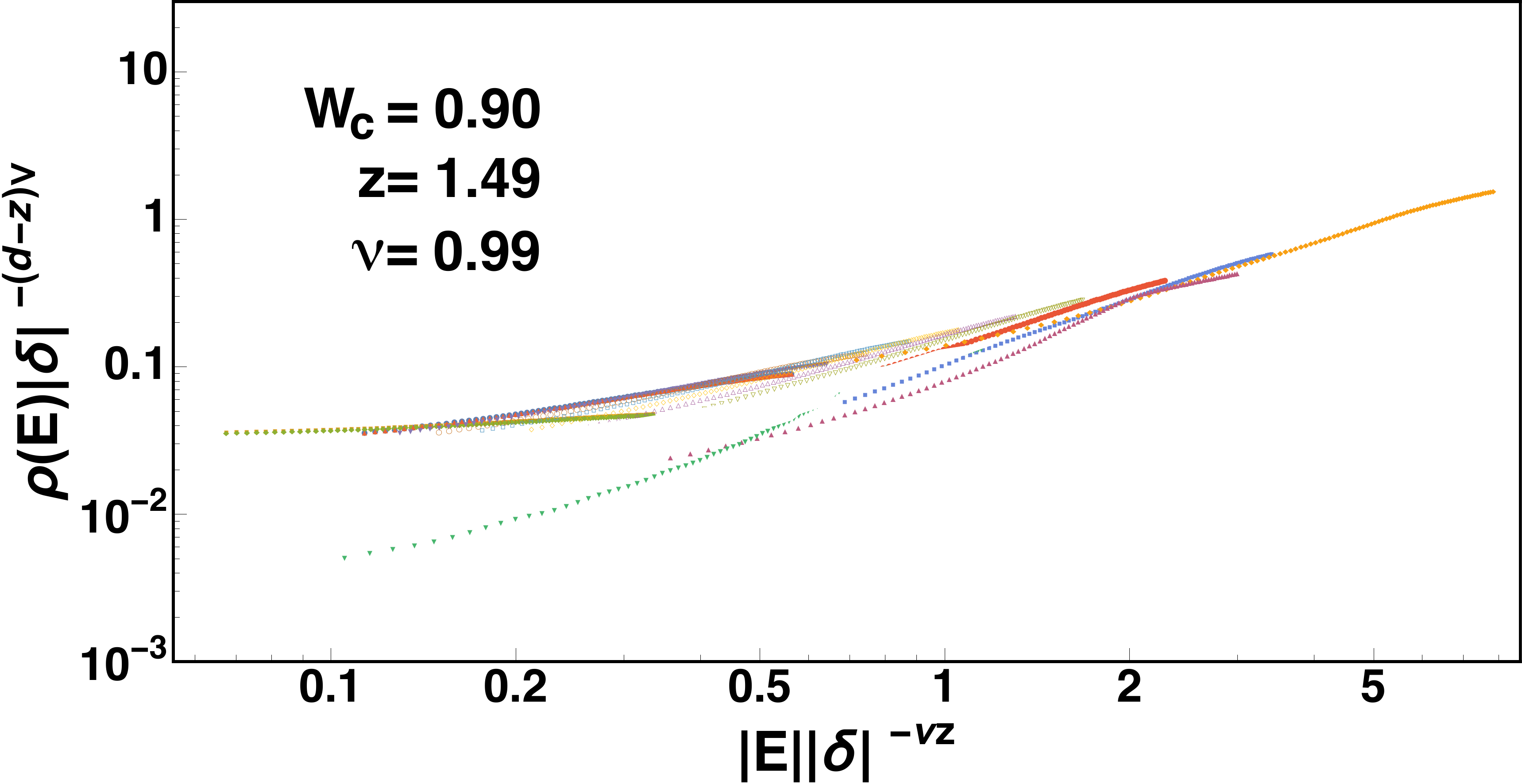}~\label{collapseE_correlated}
}
\subfigure[]{
\includegraphics[width=4cm,height=3.5cm]{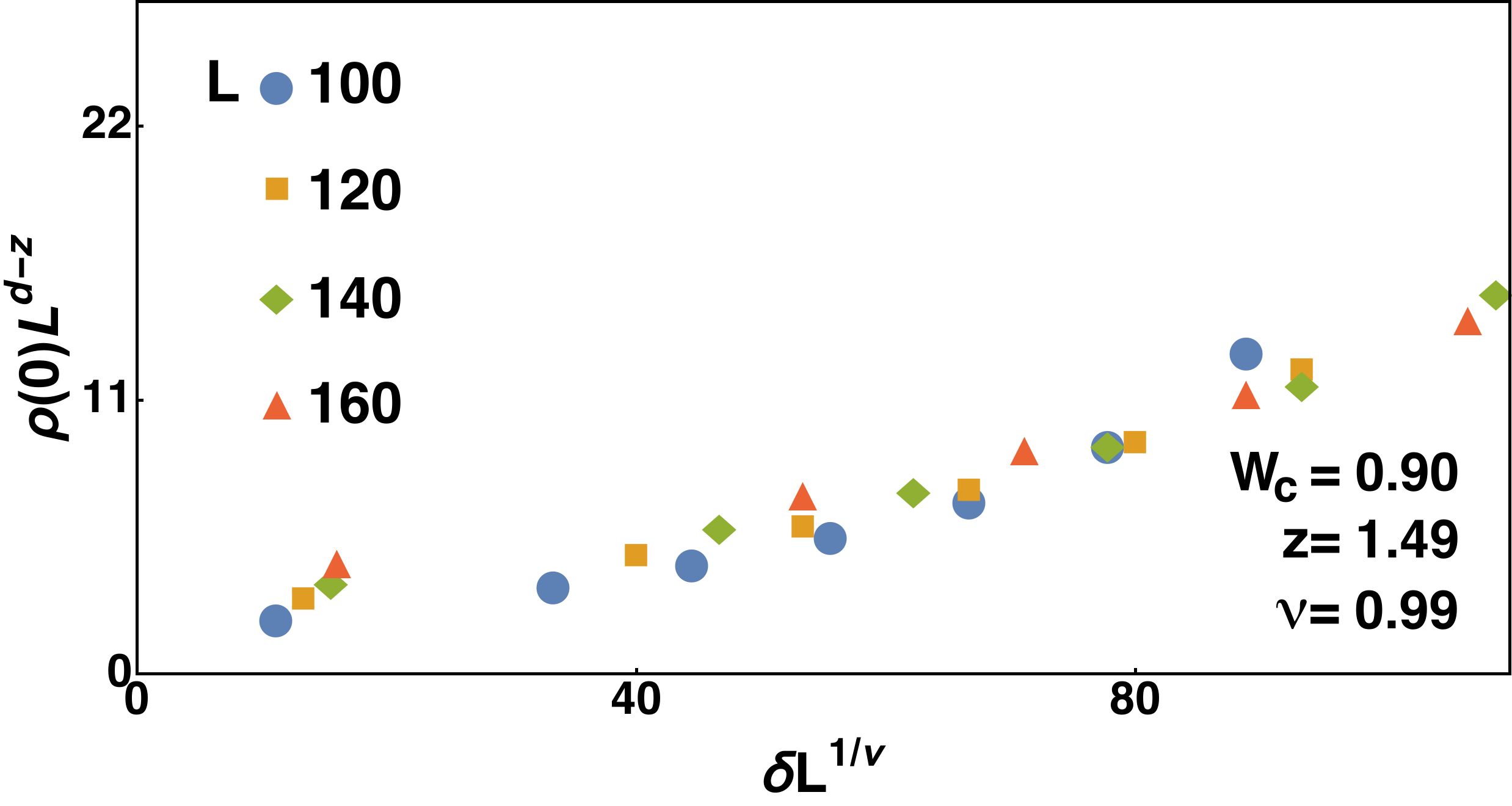}~\label{collapseL_correlated}
}
\caption{ Two types of data collapse in the presence of correlated potential disorder. (a) Finite energy collapse of $\varrho(E)|\delta|^{(d-z)\nu}$  vs. $|E| \delta^{-\nu z}$ in a system with $L = 160$. All data falls on two branches: top one corresponds to the metallic phase, while the bottom one to the semimetallic one. (b) Finite size data collapse of $\varrho(0) L^{d-z}$ vs. $\delta L^{1/\nu}$.
For these two data collapses we take $W_c = 0.9$, $z = 1.49$ and $\nu= 0.99$.
}~\label{collapse_correlated}
\end{figure}

\section{Correlated disorder}~\label{Append:Correlated_Disorder}

So far we assumed disorder to be a random variable within the range $[-W,W]$ at each site of the cubic lattice with linear dimension $L$ in each direction. Hence, disorder is uncorrelated which involves both intra-valley as well as inter-valley scattering (since in any lattice model left and right chiral Weyl points are always connected at high energies). However, our proposed scenario of the emergent superuniversality (see Sec.~\ref{chiralsuperuniversality_explanation}) suggests that in the presence of generic disorder the WSM-metal QPT is characterized by unique set of exponents, namely $z = 1.5$ and $\nu = 1$ (obtained from leading order $\epsilon$-expansions, in good agreement with numerical findings). Otherwise, such emergent superuniversality does not depend on the actual nature of the disorder (see Figs.~\ref{numeric_analysis_figure} and ~\ref{numeric_analysis_figure_CSB}, Table~\ref{table-exponent} and \ref{Tab:CSB_exponents}) \emph{nor it depends on the distribution of disorder}. To anchor the last statement we now present the numerical analysis of average DOS in the presence of \emph{correlated} random charge impurities for which inter-valley scattering is suppressed (although finite) by construction (at least when disorder is sufficiently weak). As we demonstrate the universality class of the WSM-metal transition remains unaffected (within numerical accuracy) by the choice of disorder distribution, apart from causing a non-universal shift of WSM-metal QCP ($W_c$).

We introduce a Gaussian disorder potential $W({\bf r})$, such that the mean $\langle W({\bf r}) \rangle=0$, but  
\begin{equation}~\label{correlated_distribution}
\langle W({\bf r}) W({\bf r}^\prime)= \frac{W}{\xi^2} \; \exp \left[-\frac{|{\bf r}-{\bf r}^\prime|^2}{2 \xi^2} \right].
\end{equation}
In the lattice implementation we set $\xi=4 a$, where $a$ is the lattice constant, leading to a strong suppression of inter-valley scattering by a factor $\exp \left[ -(\Delta k)^2 \xi^2/2 \right]<10^{-34}$,  where $\Delta k = \pi/a$ is the separation between two
Weyl nodes~\cite{brouwer_correlated}. Now we proceed with the numerical analysis of the average DOS using KPM in a cubic lattice
with linear dimension $L = 160$ in each direction. We average over 20 random disorder realization, compute 4096 Chebyshev moments and take trace over 12 random vectors to compute $\varrho(E)$.

First, notice that scaling of average DOS $\varrho(E)$ as a function of increasing disorder [see Fig.~\ref{fan_correlated}] is similar
to the ones found with box distribution (see Fig.~\ref{DOS_Fan_General}), showing a smooth crossover from $\varrho(E) \sim |E|^2$ to $|E|$ scaling as we approach the critical disorder strength $W_c = 0.90 \pm 0.05$ from the semimetallic side, beyond which $\varrho(0)$ becomes finite [see Fig.~\ref{rho0_correlated}] and the system enters a metallic phase. By fitting DOS $\varrho(E)$ with $|E|^{d/z-1}$, we
obtain $z=1.49 \pm 0.05$, see Fig.~\ref{z_correlated}. Finally, we compare the DOS at zero energy $\varrho(0$) with $\delta=(W-W_c)/W_c$,
and with mean-values of $W_c(= 0.90)$ and $z(= 1.49)$ and find $\nu= 0.99 \pm 0.05$ [see Fig.~\ref{nu_correlated}]. We here compare $\varrho(0)$ vs. $\delta$ up to $W=1.75$, such that $\varrho(0)$ increases smoothly within $0.90<W<1.75$ (see last paragraph of Appendix~\ref{Append:data_analysis}). Finally, performing a similar data analysis by accounting for the variation of $W_c$ and $z$, we find $\nu = 0.99 \pm 0.11$, containing maximal ``fitting error" in the determination of $\nu$ (see second last row of Table~\ref{Table:Data-analysis}).

Now with the mean values of these parameters, namely $W_c = 0.90$, $z = 1.49$ and $\nu = 0.99$, we proceed to the data collapse. The results are displayed in Figs.~\ref{collapseE_correlated} and ~\ref{collapseL_correlated}, discerning satisfactory data collapse over a large parameter space. In corroboration with the numerical results we presented for all possible disorder with uncorrelated box distribution, the present numerical analysis for correlated potential disorder supports the following fact: the universality class of the WSM-metal QPT is insensitive to the nature of disorder as well as its distribution, which in conjunction with our field theoretic predictions supports the proposed scenario of emergent superuniversality across the WSM-metal QPT.


\begin{widetext}

\section{RG analysis in the presence of generic disorder couplings}~\label{RG_8couplings}

In this Appendix we present the coupled RG flow equations for eight disorder couplings shown in Table~\ref{table-disorder}, obtained within the framework of $\epsilon_m$-expansion [defined in Sec.~\ref{epsilonm-formalism}] and $\epsilon_d$-expansion [defined in Sec.~\ref{epsilond-formalism}]. We show that under generic circumstances on the line of QCPs, defined in Eq.~(\ref{LineQCP_chiral}) [obtained from $\epsilon_m$-expansion] or Eq.~(\ref{LineQCP_chiral_hard}) [obtained from $\epsilon_d$-expansion], in the $(\Delta_V, \Delta_A)$ plane (two chiral symmetric disorders) is the legitimate solution, which provides a strong justification for the chiral superuniversality across generic disorder driven WSM-metal QPT, qualitatively discussed in Sec.~\ref{chiralsuperuniversality_explanation}.

\subsection{RG flow equations from $\epsilon_m$ expansion}~\label{Sec:epsilonm_8coupling}

The leading order coupled RG flow equations in the presence of all eight disorder couplings within the framework of an $\epsilon_m$-expansion read as
\allowdisplaybreaks[4]
\begin{eqnarray}
\beta_{\Delta_V} &=& \Delta_V \left[-\epsilon_m + \frac{4}{3} \left( 2 \Delta_A+ 5 \Delta_{AM}+4 \Delta_C+ 4 \Delta_M + \Delta _{PS}+\Delta_S+5 \Delta_{SO}+2 \Delta_V \right) \right], \\
\beta_{\Delta_A} &=& \Delta _A \left[ -\epsilon_m+ \frac{8}{3} \left(\Delta _A-2 \Delta _{\text{AM}}+2 \Delta _c+2 \Delta _M-\Delta _{\text{PS}}-\Delta _S-2 \Delta _{\text{SO}}+\Delta_V \right) \right], \\
\beta_{\Delta_M} &=& \Delta _M \left[ -\epsilon_m + \frac{4}{3} \left(\Delta_{AM}-\Delta_{PS}-\Delta_S+\Delta_{SO} \right) \right], \\
\beta_{\Delta_C} &=& -\epsilon_m \Delta_C, \\
\beta_{\Delta_{SO}} &=& \Delta_{SO} \left[ -\epsilon_m+ \frac{4}{3} \left(\Delta_{AM}-\Delta_M-\Delta _S+\Delta _V\right) \right], \\
\beta_{\Delta_{AM}} &=& \Delta _{\text{AM}} \left[-\epsilon_m-\frac{4}{3} \left(\Delta_M + \Delta_{PS}-\Delta_{SO}-\Delta_V\right) \right], \\
\beta_{\Delta_S} &=& \Delta_S \left[-\epsilon_m + \frac{4}{3} \left(2 \Delta_A - 4 \Delta_{AM} + 4 \Delta_C-5 \Delta_M + \Delta_{PS}-2 \Delta_S + 5 \Delta_{SO}-\Delta_V \right) \right], \\
\beta_{\Delta_{PS}} &=& \Delta_{PS} \left[ -\epsilon_m + \frac{4}{3} \left(2 \Delta_A + 5 \Delta_{AM} + 4 \Delta_C-5 \Delta_M - 2 \Delta_{PS}+\Delta_S - 4 \Delta_{SO}-\Delta_V \right)\right].
\end{eqnarray}
The above set of coupled flow equations only supports a line of QCPs, given by Eq.~(\ref{LineQCP_chiral}). Along the entire line of QCPs the exponents are given by $\nu^{-1}=\epsilon_m + {\mathcal O}(\epsilon^2_m)$ and $z=1+\epsilon_m/2+ {\mathcal O}(\epsilon^2_m)$ in three dimensions, to the leading order in $\epsilon_m$. Therefore, for Gaussian white noise distribution ($\epsilon_m=1$), we obtain $\nu=1$ and $z=3/2$. This outcome strongly supports the proposed emergent superuniversality across the WSM-metal QPT, driven by arbitrary disorder.

\subsection{RG flow equations from $\epsilon_d$ expansion}

The coupled RG flow equations for eight symmetry allowed disorder couplings to the leading order in the $\epsilon_d$-expansion read as
\allowdisplaybreaks[4]
\begin{eqnarray}
\beta_{\Delta_V}&=&-\epsilon_d \Delta_V +2 \Delta_V \left[\Delta_A + 3 \Delta_{AM} + 3 \Delta_C + 3 \Delta_M + \Delta_{PS} + \Delta_S + 3 \Delta_{SO} + \Delta_V \right] \nonumber \\
&+& 4 \left(2 \Delta_C \Delta_M + \Delta_{AM} \Delta_{PS}+ \Delta_S \Delta_{SO} \right), \\
\beta_{\Delta_A} &=&-\epsilon_d \Delta_A + 2 \Delta_A \left( \Delta_A- 3 \Delta_{AM} + 3 \Delta_C + 3 \Delta_M- \Delta_{PS} - \Delta_S-3 \Delta_{SO} + \Delta_V \right) \nonumber \\
&+& 4 \left( \Delta^2_{AM} + \Delta^2_C + \Delta^2_M + \Delta^2_{SO} \right), \\
\beta_{\Delta_M} &=& - \epsilon_d \Delta_M + \frac{2}{3} \Delta_M \left(-\Delta_A + \Delta_{AM} + \Delta_C + \Delta_M-\Delta_{PS}-\Delta_S
+\Delta_{SO}-\Delta_V \right) \nonumber \\
&+& \frac{4}{3} \left(2 \Delta_A \Delta_M + 7 \Delta_{AM} \Delta_{SO} + 2 \Delta_C \Delta_V + \Delta_{PS} \Delta_S \right), \\
\beta_{\Delta_C} &=& -\Delta_C + \frac{2}{3} \Delta_C \left( -\Delta_A-\Delta_{AM} + \Delta_C + \Delta_M + \Delta_{PS} + \Delta_S- \Delta_{SO}-\Delta_V \right) \nonumber \\
&+& \frac{8}{3} \left( \Delta_A \Delta_C + \Delta_{AM} \Delta_S + \Delta_M \Delta_V + \Delta_{PS} \Delta_{SO} \right), \\
\beta_{\Delta_{SO}} &=& \epsilon_d \Delta_{SO} - \frac{2}{3} \Delta_{SO} \left( \Delta_A-\Delta_{AM}-\Delta_C + \Delta_M-\Delta_{PS} + \Delta_{ S} + \Delta_{SO} - \Delta_V \right) \nonumber \\
&+& \frac{4}{3} \left( 2 \Delta_A \Delta_{SO} + 7 \Delta_{AM} \Delta_M + 2 \Delta_C \Delta_{PS} + \Delta_S \Delta_V \right), \\
\beta_{\Delta_{AM}} &=& -\epsilon_d \Delta_{AM} - \frac{2}{3} \Delta_{AM} \left( \Delta_A + \Delta_{AM}-\Delta_C + \Delta_M + \Delta_{PS}-\Delta_S-\Delta_{SO}-\Delta_V \right)  \nonumber \\
&+& \frac{4}{3} \left(2 \Delta_A \Delta_{AM} + 2 \Delta_C \Delta_S + 7 \Delta_M \Delta_{SO} + \Delta_{PS} \Delta_V \right), \\
\beta_{\Delta_S} &=& -\epsilon_d \Delta_S + 2 \Delta_S \left( \Delta_A-3 \Delta_{AM} + 3 \Delta_C - 3 \Delta_M + \Delta_{PS}-\Delta_{S}
+ 3 \Delta_{SO}-\Delta_V \right) \nonumber \\
&+& 4 \left( 2 \Delta_{AM} \Delta_C + \Delta_M \Delta_{PS} + \Delta_{SO} \Delta_V \right), \\
\beta_{\Delta_{PS}} &=& -\epsilon_d \Delta_{PS} + 2 \Delta_{PS} \left( \Delta_A + 3 \Delta_{AM} + 3 \Delta_C-3 \Delta_M-\Delta_{PS} + \Delta_S-3 \Delta_{SO}-\Delta_V \right) \nonumber \\
&+& 4 \left( \Delta_{AM} \Delta_V + 2 \Delta_C \Delta_{SO} + \Delta_{M} \Delta_S \right).
\end{eqnarray}
The above set of coupled flow equations supports only a line of QCPs, given by Eq.~(\ref{LineQCP_chiral_hard}), in the $\Delta_V-\Delta_A$ plane, shown in Fig.~\ref{Potential_Axial}. Along the entire line of QPCs, the exponents are $\nu^{-1}=\epsilon_d$ and $z=1+\epsilon_d/2$ (to the leading order in $\epsilon_d$). Therefore, in a three-dimensional WSM ($\epsilon_d=1$) the semimetal-metal QPT driven by arbitrary disorder potential is always characterized by $\nu=1$ and $z=3/2$, thus strongly supporting the proposed emergent chiral superuniversality.
Even though symmetry of a WSM is different from its two-dimensional counterpart graphene (for example, graphene does not allow presence of time-reversal symmetry breaking magnetic or current disorder), at least the coupled RG flow equations for potential ($\Delta_V$) and regular mass ($\Delta_S$) disorder (present in both WSM and graphene) are in agreement with Ref.~\cite{mirlin-2}, if we set $\epsilon_d=0$ or equivalently $d=2$.

\end{widetext}

\section{Alternative derivation of correction to optical conductivity}~\label{OC:alternative}

Direct computation of the correction to the optical conductivity (OC) due to arbitrary disorder by using the Kubo formula has already been presented in Ref.~\cite{juricic-disorder}. Specifically, we compute the disorder driven correction to the current-current correlation function (involving computation of two-loop diagrams) and then via analytic continuation we found the OC at frequency $\Omega$ in a weakly disordered WSM to be
\begin{equation}~\label{OC:disorder-correction}
\sigma(\Omega)=\frac{N e^2_0 \Omega}{12 h v} \left[ 1+ \frac{\Delta_V \Lambda}{\pi^2 v^2}\right]
\equiv \sigma_0 (\Omega) \left[ 1+ \frac{\Delta_V \Lambda}{\pi^2 v^2}\right],
\end{equation}
where $N$ is the number of Weyl nodes, $e_0$ is the electron charge in vacuum [see Eq.~(3) of Ref.~\cite{juricic-disorder}]. For concreteness, we here restrict ourselves to potential disorder or random charge impurities ($\Delta_V$), possessing Gaussian white noise distribution in three dimensions. In the absence of disorder ($\Delta_V=0$), we recover the OC in a clean WSM, $\sigma_0(\Omega)$,~\cite{chakravarty, hosur, roy-juricic}. We now present an alternative derivation of the same expression.

The OC is given by
\begin{eqnarray}
\sigma(\Omega) &=& \lim_{\Omega \to 0} \frac{1}{\Omega} \int d^Dx \;  e^{i \Omega x_0} \: \langle j_x (x) j_x (0) \rangle_R \nonumber \\
 &=& Z^2_{\Psi} \; \left[ \lim_{\Omega \to 0} \frac{1}{\Omega} \int d^Dx \;  e^{i \Omega x_0} \; \langle j_x (x) j_x (0) \rangle_0 \right], \nonumber \\
&=& Z^2_{\Psi} \; \left( \frac{N e^2_0 \Omega}{12 h v} \right),
\end{eqnarray}
where $Z_\Psi=\left[ 1+ \Delta_V \Lambda/(2 \pi^2 v^2) \right]$ is the field renormalization factor, as presented in Sec.~\ref{epsilond-formalism}, for $\epsilon_d=1$.  The same expression for the field-renormalization factor can directly be obtained by integrating over the entire Weyl-band with $0 \leq |{\mathbf k}| \leq \Lambda$, which is legitimate since we are interested in the OC of a weakly disordered WSM for which sharp quasiparticle excitations persists all the way down to zero energy or momentum. Upon substituting $Z_\psi$ in the above expression we immediately recover Eq.~(\ref{OC:disorder-correction}).

\section{$\epsilon_n$-expansion for WSM-metal QPT}~\label{epsilon_n:WSM-metal_Append}

We devote this appendix of the paper to address yet another controlled route to address the effects of disorder deep inside the WSM phase. Without any loss of generality we can express the Weyl Hamiltonian as
\begin{equation}
H_W= v_\perp \sum_{j=1,2} i \gamma_0 \gamma_j k_j + v_3 i\gamma_0 \gamma_3 k_3,
\end{equation}
and so far we have considered $v_\perp=v_3=v$. Following the spirit of ``band-flattening" method, demonstrated in Sec.~\ref{WSM-ins-RG}, we deform the above Hamiltonian to
\begin{equation}
H_W \to H^n_W= v_\perp \sum_{j=1,2} i \gamma_0 \gamma_j k_j + C_n i\gamma_0 \gamma_3 k^n_3,
\end{equation}
with the restriction that $n$ can now only take \emph{odd integer} values, so that all symmetry properties of a WSM remain unaffected. The DOS of such a deformed system is $\varrho(E) \sim |E|^{1+1/n}$. Notice in the limit $n \to \infty$ the DOS scales linearly with $E$, and disorder then become a \emph{marginal} variable (outcome from a self-consistent Born calculation). Such special limit represents a two-dimensional Weyl system (since quasiparticles do not possess any dispersion along $k_z$). Otherwise, following the same steps of coarse-graining we find that the scaling dimension of disorder couplings after performing the disorder-averaging using the replica formalism is $\left[ \Delta_j \right]=-1/n$. Therefore, we can perform a controlled RG calculation about $n \to \infty$ limit, following the spirit of an $\epsilon_n$-expansion, with $\epsilon_n=1/n$, since $\left[ \Delta_j \right]=-\epsilon_n$. For physically relevant case $\epsilon_n=1$. Otherwise, the steps are identical to the ones presented in Sec.~\ref{WSM-ins-RG} and the relevant Feynman diagrams are already shown in Fig.~\ref{laddercrossing}. For the sake of simplicity, we here focus only on the potential disorder. A detailed RG analysis within the framework of the $\epsilon_n$-expansion in the presence of generic eight disorder is left for a future investigation. The leading order RG calculation yields the following flow equations
\begin{eqnarray}
\beta_X &=& -\Delta_V H_0(n)X=(1-z)X, \nonumber \\
\beta_{\Delta_V} &=& \Delta_V \left[ -\epsilon_n +\Delta_V H_0(n) \right],
\end{eqnarray}
where $X=v_\perp, C_n$, $\hat{\Delta}_V=2 \Delta_V \Lambda^{\epsilon_n}/\left[ (2\pi)^2 C^{\epsilon_n} v^{2-\epsilon_n}_\perp \right]$ is the dimensionless disorder coupling and for brevity we have dropped the `hat' notation in the last set of equations. The function $H_0(n)$ reads as
\begin{equation}
H_0(n)=1+ \frac{\pi^2}{24} \frac{1}{n^2} + {\mathcal O} \left( n^{-4} \right).
\end{equation}
Therefore, $H_0(n)$ is a well controlled function of $1/n$. Keeping the leading order term in $H_0(n)$, the RG equations becomes
\begin{equation}
\beta_X = -\Delta_V\,X=(1-z)X, \,\,
\beta_{\Delta_V} = \Delta_V \left[ -\epsilon_n +\Delta_V \right].
\end{equation}
The DSE from the first equation reads as $z=1+\Delta_V$. The second equation supports only two fixed points: (i) the one at $\Delta_V=0$ represents the stable WSM phase, while (ii) the unstable fixed point at $\Delta_V=\epsilon_n/2$ represents the WSM-metal QCP. The DSE and the CLE at this fixed point are respectively
\begin{equation}
z=1+\frac{\epsilon_n}{2}, \: \nu^{-1}=\epsilon_n.
\end{equation}
Therefore, for physically relevant case of simple WSM ($\epsilon_n=1$), we obtain $z=3/2$ and $\nu=1$, same as the ones obtianed from $\epsilon_m$ and $\epsilon_d$ expansions, declared in Sec.~\ref{CSP_disorder}. Note that even if we chose to keep the entire function $H_0(n)$ in the RG flow equations, we obtain the same set of critical exponents.

\section{Self-consistent Born approximation at WSM-insulator QCP}~\label{appendix-Born}

In this Appendix, we present the computation of the inverse scattering life-time ($1/\tau_s$) within the framework of self-consistent Born approximation, in the presence of disorder. In this formalism the $\tau_s$ is computed from the following self-consistent equation 
\begin{equation}
\int^{E_\Lambda}_0 dE \: \frac{\varrho(E)}{\left(\hbar/\tau_s \right)^2 + E^2}=\frac{1}{W},
\end{equation}  
where $E_\Lambda$ is the ultraviolet energy cut-off up to which critical excitations separating a WSM and an insulator possess anisotropic dispersion, captured by $H_Q(0)$ in Eq.~(\ref{Ins_WSM_QCP}). Since at the WSM-insulator QCP, the average DOS scales as $\varrho(E) \sim |E|^{3/2}$ the right-hand side of the above equation displays ultraviolet divergence $\sim E^{1/2}_\Lambda$. Such divergence can be regulated by introducing a parameter 
\begin{equation}
\frac{1}{W_c}=\int^{E_\Lambda}_0 dE \: \frac{\varrho(E)}{E^2}, 
\end{equation}
where $W_c$ corresponds to the critical strength of disorder for the instability of ballistic critical fermions. The above gap equation can then be casted as 
\begin{eqnarray}
\delta=\int^{E_\Lambda}_0 dE \: \varrho(E) \; \left[ \frac{1}{E^2}-\frac{1}{\left(\hbar/\tau_s \right)^2 + E^2} \right],
\end{eqnarray}
where $\delta=W-W_c/(W W_c)$ measures the reduced disorder strength from the critical one ($W=W_c$). After regularizing the ultraviolet divergence we can take the limit $E_\Lambda \to \infty$ without encountering any divergence. The self-consistent solution of the scattering life-time is then obtained from the following universal scaling form 
\begin{equation}
\sqrt{\frac{\hbar}{\tau_s}}= \frac{\sqrt{2}}{\pi} \; \delta,
\end{equation}
which immediately implies that $\tau^{-1}_s$ is finite only when $\delta>0$ or $W>W_c$, and for $W<W_c$ we get $\tau^{-1}_s=0$. Therefore, critical fermions separating a WSM and an insulator retain its ballistic nature upto a critical strength of disorder $W_c \sim E^{1/2}_\Lambda$. Only for strong disorder $W>W_c$ a metallic phase emerges where $\tau^{-1}_s$ is finite. Therefore, conclusion from self-consistent Born approximation is in qualitative agreement with our results found by field theoretic RG analysis and numerical calculation, presented in Sec.~\ref{WSM-Ins-QPT}.


\begin{thebibliography}{10}



\bibitem{herring} C. Herring, \emph{Accidental Degeneracy in the Energy Bands of Crystals}, Phys. Rev. {\bf 52}, 365 (1937).

\bibitem{dornhaus} R. Dornhaus, G. Nimtz, and B. Schlicht, \emph{Narrow-Gap Semicounductors}, (Springer-Verlag, 1983).

\bibitem{volovik} G. E. Volovik, \emph{The Universe in a Helium Droplet} (Oxford University Press, New York, 2003).

\bibitem{balatsky} T. O. Wehling, A. M. Black-Schaffer, A. V. Balatsky, \emph{Dirac materials}, Adv. Phys. {\bf 76}, 1 (2014).

\bibitem{RyuTeo} C.-K. Chiu, J. C. Y. Teo, A. P. Schnyder, and S. Ryu, \emph{Classification of topological quantum matter with symmetries}, Rev. Mod. Phys. {\bf 88}, 035005 (2016).

\bibitem{tanmoy-RMP} A. Bansil, Hsin Lin, and Tanmoy Das, \emph{Colloquium: Topological band theory}, Rev. Mod. Phys. {\bf 88}, 021004 (2016).

\bibitem{newfermions} B. Bradlyn, J. Cano, Z. Wang, M.G. Vergniory, C. Felser, R. J. Cava, B. A. Bernevig, \emph{Beyond Dirac and Weyl fermions: Unconventional quasiparticles in conventional crystals}, Science {\bf 353}, aaf5037 (2017).

\bibitem{kane-prb} B. J. Wieder and C. L. Kane, \emph{Spin-orbit semimetals in the layer groups}, Phys. Rev. B {\bf 94}, 155108 (2016).

\bibitem{slager2016} R-J. Slager, V. Juricic, V. Lahtinen, J. Zaanen, \emph{Self-organized pseudo-graphene on grain boundaries in topological band insulators}, Phys. Rev. B {\bf 93}, 245406 (2016).


\bibitem{dirac-1} P. A. M. Dirac, \emph{The Quantum Theory of the Electron}, Proc. Roy. Soc. A {\bf 117}, 610 (1928).

\bibitem{dirac-2} P. A. M. Dirac, \emph{A theory of electrons and protons}, Proc. Roy. Soc. A {\bf 126}, 360 (1930).

\bibitem{Weyl} H. Weyl, \emph{Elektron und Gravitation. I}, Z. Physik {\bf 56}, 330 (1929).

\bibitem{burkov-review} A. A. Burkov, \emph{Chiral anomaly and transport in Weyl metals}, J. Phys.:  Condens. Matter. {\bf 27}, 113201 (2015).

\bibitem{rao-review} S. Rao, \emph{Weyl semi-metals: a short review}, arXiv:1603.02821

\bibitem{armitage-review}  N. P. Armitage, E. J. Mele, A. Vishwanath, \emph{ Weyl and Dirac Semimetals in Three Dimensional Solids},  Rev. Mod. Phys. {\bf 90}, 015001 (2018).



\bibitem{taas-1}  C. Zhang, Z. Yuan, S. Xu, Z. Lin, B. Tong, M. Z. Hasan, J. Wang, C. Zhang, S. Jia, \emph{Electron scattering in tantalum monoarsenide}, Phys. Rev. B {\bf 95}, 085202 (2017).

\bibitem{taas-2}  S-Y. Xu, I. Belopolski, N. Alidoust, M. Neupane, C. Zhang, R. Sankar, S-M. Huang, C-C. Lee, G. Chang, B. Wang, G. Bian, H. Zheng, D. S. Sanchez, F. Chou, H. Lin, S. Jia, M. Z. Hasan, \emph{Discovery of a Weyl fermion semimetal and topological Fermi arcs}, Science {\bf 349}, 613 (2015).

\bibitem{taas-3}  B. Q. Lv, H. M. Weng, B. B. Fu, X. P. Wang, H. Miao, J. Ma, P. Richard, X. C. Huang, L. X. Zhao, G. F. Chen, Z. Fang, X. Dai, T. Qian, H. Ding, \emph{Experimental Discovery of Weyl Semimetal TaAs}, Phys. Rev. X {\bf 5}, 031013 (2015).

\bibitem{nbas-1}  S-Y. Xu, N. Alidoust, I. Belopolski, C. Zhang, G. Bian, T-R. Chang, H. Zheng, V. Strokov, D. S. Sanchez, G. Chang, Z. Yuan, D. Mou, Y. Wu, L. Huang, C-C. Lee, S-M. Huang, B. Wang, A. Bansil, H-T. Jeng, T. Neupert, A. Kaminski, H. Lin, S. Jia, M. Z. Hasan, \emph{Discovery of a Weyl fermion state with Fermi arcs in niobium arsenide}, Nat. Phys. {\bf 11}, 748 (2015).

\bibitem{tap-1}  N. Xu, H. M. Weng, B. Q. Lv, C. Matt, J. Park, F. Bisti, V. N. Strocov, D. gawryluk, E. Pomjakushina, K. Conder, N. C. Plumb, M. Radovic, G. Autès, O. V. Yazyev, Z. Fang, X. Dai, G. Aeppli, T. Qian, J. Mesot, H. Ding, M. Shi, \emph{Observation of Weyl nodes and Fermi arcs in tantalum phosphide}, Nat. Commun. {\bf 7}, 11006 (2016).

\bibitem{nbp-1}  C. Shekhar, A. K. Nayak, Y. Sun, M. Schmidt, M. Nicklas, I. Leermakers, U. Zeitler, Z. Liu, Y. Chen, W. Schnelle, J. Grin, C. Felser, B. Yan, \emph{Extremely large magnetoresistance and ultrahigh mobility in the topological Weyl semimetal candidate NbP}, Nat. Phys. {\bf 11}, 645 (2015).

\bibitem{nbp-2} Z. Wang, Y. Zheng, Z. Shen, Y. Zhou, X. Yang, Y. Li, C. Feng, Z-A. Xu, \emph{Helicity-protected ultrahigh mobility Weyl fermions in NbP}, Phys. Rev. B {\bf 93}, 121112 (2016).

\bibitem{tas} G. Chang, S-Y. Xu, D. S. Sanchez, S-M. Huang, C-C. Lee, T-R. Chang, H. Zheng, G. Bian, I. Belopolski, N. Alidoust, H-T. Jeng, A. Bansil, H. Lin, M. Z. Hasan, \emph{A strongly robust type II Weyl fermion semimetal state in Ta$_2$S$_3$}, Science Advances {\bf 2}, e1600295 (2016).

\bibitem{borisenko} S. Borisenko, D. Evtushinsky, Q. Gibson, A. Yaresko, T. Kim, M. N. Ali, B. Buechner, M. Hoesch, R. J. Cava, \emph{Time-Reversal Symmetry Breaking Type-II Weyl State in YbMnBi$_2$}, arXiv:1507.04847

\bibitem{chiorescu}  J. Y. Liu, J. Hu, Q. Zhang, D. Graf, H. B. Cao, S. M. A. Radmanesh, D. J. Adams, Y. L. Zhu, G. F. Cheng, X. Liu, W. A. Phelan, J. Wei, D. A. Tennant, J. F. DiTusa, I. Chiorescu, L. Spinu, Z.Q. Mao, \emph{Discovery of a topological semimetal phase coexisting with ferromagnetic behavior in Sr$_{1-y}$MnSb$_2$ ($y \sim 0.08$)}, Nat. Mater. {\bf 16}, 905 (2017).


\bibitem{nielsen-ninomiya}  H.B.  Nielsen,  and  M.  Ninomiya, \emph{Absence of neutrinos on a lattice: (I). Proof by homotopy theory}, Nucl.  Phys.  B {\bf 185}, 20 (1981); \emph{A no-go theorem for regularizing chiral fermions}, Phys. Lett. B {\bf 105}, 219 (1981).

\bibitem{harris} A. B. Harris, \emph{Effect of random defects on the critical behaviour of Ising models}, J. Phys. C {\bf 7}, 1671 (1974).


\bibitem{fradkin} E. Fradkin, \emph{Critical behavior of disordered degenerate semiconductors. II. Spectrum and transport properties in mean-field theory}, Phys. Rev. B {\bf 33}, 3263 (1985).

\bibitem{shindou} R. Shindou, and S. Murakami, \emph{Effects of disorder in three-dimensional Z$_2$ quantum spin Hall systems}, Phys. Rev. B {\bf 79}, 045321 (2009).

\bibitem{chakravarty} P. Goswami, and S. Chakravarty, \emph{Quantum Criticality between Topological and Band Insulators in $3+1$ Dimensions}, Phys. Rev. Lett. {\bf 107}, 196803 (2011).

\bibitem{nomura-ryu} S. Ryu and K. Nomura, \emph{Disorder-induced quantum phase transitions in three-dimensional topological insulators and superconductors}, Phys. Rev. B {\bf 85}, 155138 (2012).

\bibitem{hosur} P. Hosur, S. A. Parameswaran, A. Vishwanath, \emph{Charge Transport in Weyl Semimetals}, Phys. Rev. Lett. {\bf 108}, 046602 (2012).

\bibitem{arovas} Z. Huang, T. Das, A. V. Balatsky, and D. P. Arovas, \emph{Stability of Weyl metals under impurity scattering}, Phys. Rev. B {\bf 87}, 155123 (2013).

\bibitem{nandkishore}  R. Nandkishore, D. A. Huse, S. L. Sondhi, \emph{Rare region effects dominate weakly disordered three-dimensional Dirac points}, Phys. Rev. B {\bf 89}, 245110 (2014).

\bibitem{ominato} Y. Ominato, and M. Koshino, \emph{Quantum transport in a three-dimensional Weyl electron system}, Phys. Rev. B {\bf 89}, 054202 (2014); \emph{Quantum transport in three-dimensional Weyl electron system in the presence of charged impurity scattering}, Phys. Rev. B {\bf 91}, 035202 (2015).

\bibitem{roy-dassarma} B. Roy, and S. Das Sarma, \emph{Diffusive quantum criticality in three-dimensional disordered Dirac semimetals}, Phys. Rev. B {\bf 90}, 241112(R) (2014).

\bibitem{radzihovsky} S. V. Syzranov, L. Radzihovsky, V. Gurarie, \emph{Critical Transport in Weakly Disordered Semiconductors and Semimetals}, Phys. Rev. Lett. {\bf 114}, 166601 (2015); S. V. Syzranov, V. Gurarie, L. Radzihovsky, \emph{Unconventional localization transition in high dimensions}, Phys. Rev. B {\bf 91}, 035133 (2015).

\bibitem{kim-moon} E-G. Moon, Y-B. Kim, \emph{ Non-Fermi Liquid in Dirac Semi-metals}, arXiv:1409.0573

\bibitem{altland} A. Altland, and D. Bagrets, \emph{Effective Field Theory of the Disordered Weyl Semimetal}, Phys. Rev. Lett. {\bf 114}, 257201 (2015).

\bibitem{roy-dassarma-erratum} B. Roy, S. Das Sarma, \emph{Erratum: Diffusive quantum criticality in three-dimensional disordered Dirac semimetals [Phys. Rev. B {\bf 90}, 241112(R) (2014)]}, Phys. Rev. B {\bf 93}, 119911 (E) (2016).

\bibitem{Syzranov-exponent} S. V. Syzranov, P. M. Ostrovsky, V. Gurarie, and L. Radzihovsky, \emph{Critical Exponents at the Unconventional Disorder-Driven Transition in a Weyl Semimetal}, Phys. Rev. B {\bf 93}, 155113 (2016).

\bibitem{roy-dassarma-intdis} B. Roy, S. Das Sarma, \emph{Quantum phases of interacting electrons in three-dimensional dirty Dirac semimetals}, Phys. Rev. B {\bf 94}, 115137 (2016).

\bibitem{juricic-disorder} B. Roy, V. Juri\v ci\' c, S. Das Sarma, \emph{Universal optical conductivity of a disordered Weyl semimetal}, Sci. Rep. {\bf 6}, 32446 (2016).

\bibitem{pallab-sudip2016} P. Goswami, and S. Chakravarty, \emph{Superuniversality of topological quantum phase transition and global phase diagram of dirty topological systems in three dimensions}, Phys. Rev. B {\bf 95}, 075131 (2017).

\bibitem{carpentier-1}  T. Louvet, D. Carpentier, A. A. Fedorenko, \emph{On the disorder-driven quantum transition in three-dimensional relativistic metals}, Phys. Rev. B 94, 220201(R) (2016).

\bibitem{radzihovsky-2}  S. V. Syzranov, V. Gurarie, L. Radzihovsky, \emph{Multifractality at non-Anderson disorder-driven transitions in Weyl semimetals and other systems}, Ann. Phys. {\bf 373}, 694 (2016).

\bibitem{lars} A. K. Mitchell, and L. Fritz, \emph{Signatures of Weyl semimetals in quasiparticle interference}, Phys. Rev. B {\bf 93},  035137 (2016).

\bibitem{shovkovy} E. V. Gorbar, V. A. Miransky, I. A. Shovkovy, and P. O. Sukhachov, \emph{Origin of dissipative Fermi arc transport in Weyl semimetals}, Phys. Rev. B {\bf 93}, 235127 (2016).

\bibitem{gilbert}  M. J. Park, B. Basa, M. J. Gilbert, \emph{Disorder-induced phase transitions of type-II Weyl semimetals}, Phys. Rev. B {\bf 95}, 094201 (2017).

\bibitem{carpentier-2} T. Louvet, D. Carpentier, A. A. Fedorenko, \emph{New quantum transition in Weyl semimetals with correlated disorder}, Phys. Rev. B {\bf 95}, 014204 (2017).


\bibitem{imura} K. Kobayashi, T. Ohtsuki, K-I. Imura, \emph{Disordered Weak and Strong Topological Insulators}, Phys. Rev. Lett. {\bf 110}, 236803 (2013).

\bibitem{herbut-disorder} K. Kobayashi, T. Ohtsuki, K-I. Imura, I. F. Herbut, \emph{Density of States Scaling at the Semimetal to Metal Transition in Three Dimensional Topological Insulators}, Phys. Rev. Lett. {\bf 112}, 016402 (2014).

\bibitem{brouwer-1} B. Sbierski, G. Pohl, E. J. Bergholtz, P. W. Brouwer, \emph{Quantum Transport of Disordered Weyl Semimetals at the Nodal Point}, Phys. Rev. Lett. {\bf 113}, 026602 (2014).

\bibitem{pixley-1} J. H. Pixley, P. Goswami, and S. Das Sarma, \emph{Anderson Localization and the Quantum Phase Diagram of Three Dimensional Disordered Dirac Semimetals}, Phys. Rev. Lett. {\bf 115},  076601 (2015).

\bibitem{brouwer-2} B. Sbierski, E. J. Bergholtz, P. W. Brouwer, \emph{Quantum critical exponents for a disordered three-dimensional Weyl node}, Phys. Rev. B {\bf 92}, 115145 (2015).

\bibitem{pixley-2} J. H. Pixley, P. Goswami, and S. Das Sarma, \emph{Disorder-driven itinerant quantum criticality of three-dimensional massless Dirac fermions}, Phys. Rev. B {\bf 93}, 085103 (2016).

\bibitem{ohtsuki} S. Liu, T. Ohtsuki, R. Shindou, \emph{Effect of Disorder in a Three-Dimensional Layered Chern Insulator}, Phys. Rev. Lett. {\bf 116}, 066401 (2016).

\bibitem{chen-song} C-Z. Chen, J. Song, H. Jiang, Q-F Sun, Z. Wang, X. C. Xie, \emph{Disorder and Metal-Insulator Transitions in Weyl Semimetals}, Phys. Rev. Lett. {\bf 115}, 246603 (2015).

\bibitem{roy-bera} S. Bera, J. D. Sau, and B. Roy, \emph{Dirty Weyl semimetals: Stability, phase transition, and quantum criticality}, Phys. Rev. B {\bf 93}, 201302 (2016).

\bibitem{hughes} H. Shapourian, T. L. Hughes, \emph{Phase diagrams of disordered Weyl semimetals}, Phys. Rev. B {\bf 93}, 075108 (2016).

\bibitem{pixley-3}  J. H. Pixley, D. A. Huse, S. Das Sarma, \emph{Rare-Region-Induced Avoided Quantum Criticality in Disordered Three-Dimensional Dirac and Weyl Semimetals}, Phys. Rev. X {\bf 6}, 021042 (2016).

\bibitem{roy-alavirad} B. Roy, Y. Alavirad, J. D. Sau, \emph{Global Phase Diagram of a Three-Dimensional Dirty Topological Superconductor}, Phys. Rev. Lett. {\bf 118}, 227002 (2017).

\bibitem{pixley-4} J. H. Pixley, D. A. Huse, S. Das Sarma, \emph{Uncovering the hidden quantum critical point in disordered massless Dirac and Weyl semimetals}, Phys. Rev. B {\bf 94}, 121107 (R) (2016).

\bibitem{takane2016} Y. Takane, \emph{Disorder Effect on Chiral Edge Modes and Anomalous Hall Conductance in Weyl Semimetals}, J. Phys. Soc. Jpn. {\bf 85}, 124711 (2016).

\bibitem{roy-Fermiarc} B. Roy, R-J Slager, V. Juri\v ci\' c, \emph{Dissolution of Topological Fermi Arcs in a Dirty Weyl Semimetal}, Phys. Rev. B {\bf 96}, 201401 (2017). 


\bibitem{roy-sau} B. Roy, and J. D. Sau, \emph{Magnetic catalysis and axionic charge density wave in Weyl semimetals}, Phys. Rev. B {\bf 92}, 125141 (2015).

\bibitem{model-TI} C.-X. Liu, X.-L. Qi, H.-J. Zhang, X. Dai, Z. Fang, and S-C. Zhang, \emph{Model Hamiltonian for topological insulators}, Phys. Rev. B {\bf 82}, 045122 (2010).

\bibitem{qi-anomaly} C-X. Liu, P. Ye, X-L. Qi, \emph{Chiral gauge field and axial anomaly in a Weyl semimetal}, Phys. Rev. B {\bf 87}, 235306 (2013).

\bibitem{GR-field-theory} P. Goswami, and B. Roy, \emph{ Effective field theory, chiral anomaly and vortex zero modes for odd parity topological superconducting state of three dimensional Dirac materials}, arXiv:1211.4023, and references therein.

\bibitem{li-roy2016} X. Li, B. Roy, and S. Das Sarma, \emph{Weyl fermions with arbitrary monopoles in magnetic fields: Landau levels, longitudinal magnetotransport, and density-wave ordering}, Phys. Rev. B {\bf 94}, 195144 (2016).

\bibitem{KPM-RMP} A. Wei\ss{}e, G. Wellein, A. Alverman, and H. Feshke, \emph{The kernel polynomial method}, Rev. Mod. Phys. {\bf 78}, 275 (2006).


\bibitem{KLZ1992} S. Kivelson, D.-H. Lee, and S.-C. Zhang, \emph{Global phase diagram in the quantum Hall effect}, Phys. Rev. B {\bf 46}, 2223 (1992). 
 
\bibitem{Lutken-Ross1993} C. A. L\" utken and G. G. Ross, \emph{Delocalization, duality, and scaling in the quantum Hall system}, Phys. Rev. B {\bf 48}, 2500 (1993).

\bibitem{Fradkin-Kivelson1996} E. Fradkin and S. Kivelson, \emph{Modular invariance, self-duality and the phase transition between quantum Hall plateaus}, Nucl. Phys. B {\bf 474}, 543 (1996).

\bibitem{GRV2005} I. A. Gruzberg, N. Read, and S. Vishveshwara, \emph{Localization in disordered superconducting wires with broken spin-rotation symmetry}, Phys. Rev. B {\bf 71}, 245124 (2005).


\bibitem{abrahams} For comprehensive discussion on Anderson transition see \emph{50 Years of Anderson Localization}, edited by by E. Abrahams (World Scientific Publishing Company, 1st ed., 2010).

\bibitem{slager2013} R-J. Slager, A. Mesaros, V. Juricic, J. Zaanen, \emph{The space group classification of topological band-insulators}, Nature Physics {\bf 9}, 98 (2013).

\bibitem{zinn-justin} J. Zinn-Justin, \emph{Quantum Field Theory and Critical Phenomena} (Oxford University Press, Oxford, UK, 2002).


\bibitem{mirlin-2} P. M. Ostrovsky, I. V. Gornyi, and A. D. Mirlin, \emph{Electron transport in disordered graphene}, Phys. Rev. B {\bf 74}, 235443 (2006).


\bibitem{roy-goswami-juricic} B. Roy, P. Goswami, and V. Juri\v ci\' c, \emph{Interacting Weyl fermions: Phases, phase transitions, and global phase diagram}, Phys. Rev. B {\bf 95}, 201102(R) (2017).

\bibitem{roy-foster} B. Roy, M. S. Foster, \emph{Quantum Multicriticality near the Dirac-Semimetal to Band-Insulator Critical Point in Two Dimensions: A Controlled Ascent from One Dimension}, Phys. Rev. X {\bf 8}, 011049 (2018).

\bibitem{halperin-disorder} See A. Weinrib and B. I. Halperin, \emph{Critical phenomena in systems with long-range-correlated quenched disorder}, Phys. Rev. B {\bf 27}, 413 (1983) for general discussion on correlated disorder.

\bibitem{sachdev-book} S. Sachdev, {\it Quantum Phase Transitions} (Cambridge University Press, 2nd ed., 2007).

\bibitem{HJR} I. F. Herbut, V. Juri\v ci\' c, B. Roy, \emph{Theory of interacting electrons on the honeycomb lattice}, Phys. Rev. B {\bf 79}, 085116 (2009).



\bibitem{sorella} Y. Otsuka, S. Yunoki, S. Sorella, \emph{Universal Quantum Criticality in the Metal-Insulator Transition of Two-Dimensional Interacting Dirac Electrons}, Phys. Rev. X {\bf 6}, 011029 (2016).


\bibitem{pixley-residue} J. H. Pixley, Y-Z. Chou, P. Goswami, D. A. Huse, R. Nandkishore, L. Radzihovsky, S. Das Sarma, \emph{Single-particle excitations in disordered Weyl fluids}, Phys. Rev. B  {\bf 95}, 235101 (2017). 


\bibitem{wegner}  F. Wegner, \emph{Electrons in Disordered Systems. Scaling near the Mobility Edge}, Z. Phys. B {\bf 25}, 327 (1976).


\bibitem{beiltz} D. Belitz and T. R. Kirkpatrick, \emph{The Anderson-Mott transition}, Rev. Mod. Phys. {\bf 66}, 261 (1994).

\bibitem{mirlin} F. Evers and A. D. Mirlin, \emph{Anderson transitions}, Rev. Mod. Phys. {\bf 80}, 1355 (2008).

\bibitem{janssen} M. Janssen, \emph{Statistics and scaling in disordered mesoscopic electron systems}, Phys. Rep. {\bf 295}, 1 (1998).

\bibitem{foster} M. S. Foster, \emph{Multifractal nature of the surface local density of states in three-dimensional topological insulators with magnetic and nonmagnetic disorder}, Phys. Rev. B {\bf 85}, 085122 (2012).

\bibitem{brndiar} J. Brndiar and P. Marko\ifmmode \check{s}\else \v{s}\fi{}, \emph{Universality of the metal-insulator transition in three-dimensional disordered systems}, Phys. Rev. B {\bf 74}, 153103 (2006).


\bibitem{cdas}  S. Borisenko, Q. Gibson,  D. Evtushinsky, V. Zabolotnyy, B. Buechner, and R. J. Cava, \emph{Experimental Realization of a Three-Dimensional Dirac Semimetal}, Phys. Rev. Lett. {\bf 113}, 027603 (2014).

\bibitem{nabi}  Z. K. Liu, B. Zhou, Y. Zhang, Z. J. Wang, H. M. Weng, D. Prabhakaran, S.-K. Mo, Z. X. Shen, Z. Fang, X. Dai, Z. Hussain, Y. L. Chen, \emph{Discovery of a Three-Dimensional Topological Dirac Semimetal, Na$_3$Bi}, Science {\bf 343}, 864 (2014).

\bibitem{dassarma-puddle} S. Das Sarma, S. Adam, E. H. Hwang, and E. Rossi, \emph{Electronic transport in two-dimensional graphene}, Rev. Mod. Phys. {\bf 83}, 407 (2011).

\bibitem{halperin}  B. I. Halperin, and M. Lax, \emph{Impurity-Band Tails in the High-Density Limit. I. Minimum Counting Methods}, Phys. Rev. {\bf 148}, 722 (1966).


\bibitem{Fang-HgCrSe} G. Xu, H. Weng, Z. Wang, X. Dai, and Z. Fang, \emph{Chern Semimetal and the Quantized Anomalous Hall Effect in HgCr$_2$Se$_4$}, Phys. Rev. Lett. {\bf 107}, 186806 (2011).

\bibitem{bergevig} C. Fang, M. J. Gilbert, X. Dai, and B. A. Bernevig, \emph{Multi-Weyl Topological Semimetals Stabilized by Point Group Symmetry}, Phys. Rev. Lett. {\bf 108}, 266802 (2012).

\bibitem{nagaosa} B-J. Yang, and N. Nagaosa, \emph{Classification of stable three-dimensional Dirac semimetals with nontrivial topology}, Nat. Commun. {\bf 5}, 4898 (2014).

\bibitem{roy-juricic} B. Roy and V. Juri\v ci\' c, \emph{Optical conductivity of an interacting Weyl liquid in the collisionless regime}, Phys. Rev. B {\bf 96}, 155117 (2017); \emph{Collisionless Transport Close to a Fermionic Quantum Critical Point in Dirac Materials}, arXiv:1801.03495 [Phys. Rev. Lett. (to be published)].


\bibitem{carpentier-new} I. Balog, D. Carpentier, A. A. Fedorenko, \emph{Disorder-driven quantum transition in relativistic semimetals:
functional renormalization via the porous medium equation}, arXiv:1710.07932

\bibitem{carpentier_MCP} D. Carpentier, A. A. Fedorenko, E. Orignac, \emph{Effect of disorder on 2D topological merging transition from a
Dirac semi-metal to a normal insulator}, Eur. Phys. Lett. {\bf 102}, 67010 (2013).

\bibitem{brouwer_correlated} B. Sbierski, M. Trescher, E. J. Bergholtz, P. W. Brouwer, \emph{Disordered doubleWeyl node: Comparison of
transport and density of states calculations}, Phys. Rev. B {\bf 95}, 115104 (2017).



\end{thebibliography}
\end{document}